\documentclass[aps,physrev,reprint,superscriptaddress]{revtex4-2}

\usepackage{graphicx}
\usepackage{subfigure}
\usepackage{physics}
\usepackage{mathtools}
\usepackage[version=3]{mhchem}
\usepackage{dcolumn}
\usepackage{bm}
\usepackage{xcolor}
\usepackage{siunitx}
\usepackage{url}
\usepackage{booktabs}
\usepackage{mathpazo}
\usepackage[
  font=small,
  labelfont=bf,
  justification=Justified,
  singlelinecheck=off,
  format=plain
]{caption}

\usepackage{hyperref}
\usepackage[strings]{underscore}


\begin{document}
\title{General Learning of the Electric Response of Inorganic Materials}
\date{\today}

\author{Bradley A. A. Martin}
\email[Electronic mail: ]{bradley.martin@ucl.ac.uk}
\affiliation{Department of Chemistry, University College London, London, WC1E 6BT, United Kingdom}

\author{Alex M. Ganose}
\affiliation{Department of Chemistry, Imperial College London, London, W12 0BZ, United Kingdom}

\author{Venkat Kapil}
\affiliation{Department of Physics and Astronomy, University College London, London, WC1H 0AH, United Kingdom}

\author{Tingwei Li}
\affiliation{Department of Materials, Imperial College London, London, SW7 2AZ, United Kingdom}

\author{Keith T. Butler}
\email[Electronic mail: ]{k.t.butler@ucl.ac.uk}
\affiliation{Department of Chemistry, University College London, London, WC1E 6BT, United Kingdom}


\begin{abstract}
Dielectric response governs how materials interact with electric fields and light. Still, first-principles prediction of polarisation $\mathbf P$, Born effective charges $Z^*$, and polarisability $\boldsymbol\alpha$ remains too costly for large-scale screening and long-time dynamics. We introduce \texttt{MACE-Field}, a field-aware, $O(3)$-equivariant interatomic potential that learns a single electric enthalpy functional $\mathcal F(\{\mathbf R\},\mathbf E)$ and obtains $\mathbf P$, $Z^*$, and $\boldsymbol\alpha$ by exact differentiation. A uniform field couples to latent equivariant features inside the \texttt{MACE} backbone, while the scalar energy readout preserves Maxwell reciprocity, the acoustic sum rule, and crystal tensor symmetries by construction. Because this coupling is a plug-in on top of standard \texttt{MACE}, existing energy/force foundation models can be upgraded to become field-aware. Benchmarked against semilocal DFT/DFPT reference data, a directly trained cross-chemistry ferroelectric model reproduces the same-branch Berry-phase and spontaneous polarisations across diverse inorganic crystals. Starting from the multihead foundation model \texttt{mace-mp-mh-0} and its OMAT-PBE head, joint fine-tuning on dielectric, ferroelectric, and replay data yields \texttt{MACE-Field-MH-0} foundation models, which predict $Z^*$, $\boldsymbol\alpha$, derived dielectric constants, and cross-chemistry polarisation trends with fidelity that captures branch-resolved polarisation and spontaneous-polarisation, while retaining strong force-field accuracy. Further, single-material \texttt{MACE-Field} models and \texttt{MACE-Field-MH-0} reproduce \ce{BaTiO3} hysteresis loops and $\alpha$-quartz infrared, Raman, and dielectric spectra from finite-field molecular dynamics, comparable to DFPT. These results show that a simple, physics-informed field coupling can endow atomistic foundation models with transferable dielectric and ferroelectric response, while targeted single-material training remains advantageous for the most quantitative spectroscopic predictions.
\end{abstract}

\maketitle 

Controlling how condensed matter responds to electric fields is central to chemistry, condensed-matter physics, and electronic engineering. It underpins technologies ranging from gate oxides and interconnect dielectrics in microelectronics to capacitors for power conversion, and to piezoelectrics and electro-optic crystals for sensing, imaging, and telecommunications~\cite{Scott2007,Damjanovic1998,Boyd2003,Robertson2006}. Across these settings, performance is ultimately set by how a material redistributes charge and polarisation under an applied field. Microscopically, this response is encoded in the electric polarisation $\mathbf P$, Born effective charges $Z^*$, and polarisability tensors $\boldsymbol\alpha$, which connect local bonding to macroscopic observables such as dielectric permittivity, piezoelectric coefficients, and ferroelectric response~\cite{king-smith-vanderbilt-1993,resta-1994,gonze_dynamical_1997,rabe_modern_2007}. The same quantities also govern optical and nonlinear-optical phenomena, including bulk and ferroelectric photovoltaic effects in non-centrosymmetric crystals and hybrid perovskites~\cite{Young2012,Cook2017,Dai2023,Butler2015Photoferroics,Frost2014MolecularFerroelectric}, and they shape field effects in catalysis and electrochemistry via screening and interfacial polarisation~\cite{Che2018,Huang2019,Leonard2021,Yu2022,Long2025}. A general and transferable way to predict $\mathbf P$, $Z^*$, and $\boldsymbol\alpha$ across chemistries and structures would therefore enable data-driven screening, finite-temperature simulations, and targeted optimisation of functional materials.

Achieving this goal is especially challenging in insulating crystals, where polarisation is a geometric (Berry) phase and is multivalued on a lattice of polarisation quanta~\cite{king-smith-vanderbilt-1993,resta-1994}. Only \emph{differences} in polarisation are gauge-invariant, so both dataset construction and learning objectives must contend with branch choice.

First-principles approaches based on Density Functional Perturbation Theory (DFPT) and finite-field Berry-phase methods provide rigorous access to linear and nonlinear response~\cite{gonze_dynamical_1997,nunes-gonze-2001,souza-iniguez-vanderbilt-2002}. They yield $Z^*$, electronic polarisabilities, dielectric tensors, and electro-optic coefficients. Still, at high computational cost: DFPT typically requires solving linear-response equations for all occupied bands and perturbations and can be one to two orders of magnitude more expensive than a single total-energy calculation. This limits chemical diversity, system size, and accessible time scales, leaving a gap between accurate dielectric theory and the scale required for screening, finite-temperature dynamics, and device-relevant microstructures.

Machine learning (ML) offers a route to bridge this gap, and
electric-response learning has developed along several
complementary lines. For molecular systems,
Christensen et al. introduced Operator Quantum ML (\texttt{OQML}), in which response properties are obtained by applying differential operators (e.g. $\partial/\partial\mathbf E$) to learned energy functionals, improving sample efficiency for forces and dipoles and enabling higher derivatives~\cite{Christensen2019OperatorsQML}. Symmetry-adapted kernel methods such as SA-GPR (\texttt{AlphaML} and \texttt{LODE}) with $\lambda$-SOAP kernels established how to learn covariant tensorial quantities, including molecular polarisabilities, condensed-phase dielectric response tensors and long-range Coulomb effects~\cite{Grisafi2018SymmetryAdaptedTensorial, wilkins2019, Grisafi2019}. Neural models then learned charge and dipole degrees of freedom directly from structure: \texttt{PhysNet} predicts energies/forces together with dipoles and latent partial charges (e.g. on QM9 and MD17/ISO17)~\cite{Unke2019PhysNet}, while MuML combines local atomic dipoles with a non-local partial-charge model to capture both polarisation and long-range charge separation (trained on QM7b with transfer tests)~\cite{Veit2020MuMLDipoles}. Field-explicit message passing (e.g. \texttt{FieldSchNet}) further showed that equivariance and field inputs enable direct access to response properties and vibrational/optical spectra~\cite{Gastegger2021FieldSchNet}. In materials science, analogous structure-to-property models learn dielectric constants or tensors from DFPT-labelled crystal datasets~\cite{Morita2020,anisonet-2024}.

MLIPs extend these principles to scalable atomistic simulation by reproducing \emph{ab initio} accuracy at much lower cost~\cite{behler-2007-bp,schutt-2018-schnet,batzner-2022-nequip}. $O(3)$-equivariant backbones such as \texttt{NequiP}/\texttt{Allegro} and \texttt{MACE} provide accurate, differentiable energies and forces across diverse chemistries~\cite{batzner-2022-nequip,batatia-2022-mace}. Several strategies make these models electrically aware. One approach is to augment MLIPs with explicit electronic degrees of freedom with learned charges, dipoles, or Wannier-centre proxies, to reconstruct long-range electrostatics and compute dielectric observables (e.g. \texttt{DeepWannier}/DPLR-style pipelines)~\cite{Zhang2020DeepWannier,Zhang2022DPLR}. Another is to learn a field-coupled scalar functional (electric enthalpy) and obtain $\mathbf P$, $Z^*$, and $\boldsymbol\alpha$ by exact differentiation, enforcing reciprocity by construction~\cite{allegro-pol-2025}. Complementary approaches recover polarisation and BEC-like quantities from latent long-range charge fields~\cite{les-2025} or incorporate charge response into systematically improvable local bases such as charge-augmented \texttt{ACE} variants~\cite{cace-2024}. Together, these developments suggest that symmetry-aware, differentiable MLIPs can deliver derivative-consistent dielectric response at simulation cost, provided that models capture both non-local screening physics and the gauge structure of $\mathbf P$ in periodic crystals.

From the perspective of scalable computational materials science, two limitations remain. First, most unified electric-response MLIPs are trained as \emph{single-material} models, restricting transfer across chemistries, space groups, and bonding motifs. Second, many field-aware architectures diverge from widely used MLIP backbones, limiting the reuse of foundation models pretrained on large energy/force/stress datasets. This matters because high-quality dielectric labels (Berry-phase $\mathbf P$, DFPT $Z^*$ and $\boldsymbol\alpha$) are far scarcer than energetic data; reusing and fine-tuning existing foundation models is therefore particularly valuable.

To address these challenges, we introduce \texttt{MACE-Field}~\cite{mdi_group_mace_field}, a physics-informed extension of \texttt{MACE} that learns a single \emph{electric enthalpy} functional $\mathcal F(\{\mathbf R\},\mathbf E)$ across diverse inorganic crystals and obtains dielectric observables by \emph{exact differentiation}:
\begin{equation}
\mathbf P=-\Omega^{-1}\frac{\partial \mathcal F}{\partial \mathbf E},\qquad
Z^*_{\kappa,ij}=\Omega\,\frac{\partial P_i}{\partial u_{\kappa j}},\qquad
\alpha_{ij}=\frac{\partial P_i}{\partial E_j}.
\end{equation}
A uniform field couples \emph{inside} each message-passing layer to latent equivariant features via Clebsch-Gordan tensor products with residual mixing. The readout remains scalar ($L{=}0$), and a branch-invariant loss accommodates Berry-phase multivaluedness. Because all quantities derive from one scalar, Maxwell reciprocity and the acoustic sum rule follow by construction, yielding a derivative-consistent description of dielectric response suitable for finite-field molecular dynamics and spectroscopy.

\textbf{Our two key advances are:}
\begin{enumerate}
  \item \textbf{Foundation-model inheritance.} Field coupling is implemented as a plug-in at the latent irrep level that leaves the standard \texttt{MACE} backbone and readout unchanged, enabling reuse and fine-tuning of \texttt{MACE} foundation weights in a label-scarce regime.
  \item \textbf{Cross-chemistry training.} We train and fine-tune across broad chemistries from many Materials Project structures using MP-Ferroelectric ($\sim$2.5k structures, 61 elements) and MP-Dielectric ($\sim$5.3k DFPT entries, 81 elements), rather than a single material.
\end{enumerate}
Together, these advances provide a derivative-consistent, foundation-compatible framework that scales across chemistry and predicts $\mathbf P$, $Z^*$ and $\boldsymbol\alpha$ for diverse inorganic solids at a cost suitable for large-scale simulations and screening. Unless explicitly stated otherwise, the quantitative errors reported below are measured against the underlying GGA-PBE or GGA-PBE(+U) DFPT/Berry-phase labels used to construct the datasets; they therefore quantify ML agreement with those reference workflows rather than absolute agreement with experiment.

We fine-tune a recent multihead \texttt{MACE} foundation, \texttt{mace-mp-mh-0}, using its OMAT-PBE head as the energy/force/stress prior and jointly supervising on MP-Ferroelectric, MP-Dielectric, and a 10\,000-structure OMAT-PBE replay subset associated with the \texttt{mace-mp-mh-0/1} family, to obtain the field-aware foundation model \texttt{MACE-Field-MH-0}. In Sec.~\ref{sec:results}, we show that this foundation route predicts BECs, polarisabilities, derived dielectric constants, same-branch Berry-phase polarisations, and spontaneous polarisations across chemistry. We contrast these cross-chemistry models with single-material \ce{BaTiO3} / $\alpha$-\ce{SiO2} models trained on Allegro-pol trajectories~\cite{allegro-pol-2025} for finite-field hysteresis and IR/Raman/dielectric spectra. The remainder of the paper details the relevant background theory (\S\ref{sec:theory}), architecture and design choices (\S\ref{sec:methods}), datasets and training protocols (\S\ref{sec:datasets}), quantitative results (\S\ref{sec:results}), and discussion and outlook (\S\ref{sec:discussion}).

\section{Theory}\label{sec:theory}

\subsection{Modern theory of polarisation (Berry phase)}

In periodic insulators, the macroscopic polarisation is not the naive dipole per unit volume, because the position operator is ill-defined under periodic boundary conditions. The \emph{modern theory of polarisation} resolves this by expressing the electronic contribution as a Berry phase accumulated by the occupied Bloch bands along loops in the Brillouin zone (BZ)~\cite{king-smith-vanderbilt-1993,resta-1994}. We briefly summarise the key ingredients here.

We can decompose the total polarisation into ionic and electronic parts,
\begin{equation}
\mathbf P = \mathbf P_\text{ion}+\mathbf P_\text{el},
\qquad
\mathbf P_\text{ion} = \frac{-e}{\Omega}\sum_{\kappa} Z_\kappa\,\mathbf R_\kappa,
\label{eq:P_total}
\end{equation}
where $e$ is the electron charge, $\Omega$ the cell volume, $Z_{\kappa}$ the nuclear charge for site $\kappa$, and $\vb{R}_{\kappa}$ the corresponding position vector. The electronic part of the polarisation is
\begin{equation}
\mathbf P_\text{el}
= -\frac{e}{(2\pi)^3}\sum_{n\in\text{occ}}
\int_{\text{BZ}} \! d\mathbf k\
\langle u_{n\mathbf k}\,|\, i\nabla_{\mathbf k}\,|\,u_{n\mathbf k}\rangle,
\label{eq:P_berry}
\end{equation}
where $|u_{n\mathbf k}\rangle$ are cell-periodic Bloch functions. Only \emph{changes} of $\mathbf P$ between adiabatically connected states are gauge-invariant; absolute values are defined modulo the \emph{polarisation lattice} generated by the quanta
\begin{equation}
\mathbf Q_k=\frac{e\,\mathbf a_k}{\Omega},\qquad k=1,2,3,
\label{eq:pol_quanta}
\end{equation}
with $\{\mathbf a_k\}$ the direct lattice vectors~\cite{king-smith-vanderbilt-1993,resta-1994}. This multivaluedness must be handled explicitly in learning problems (Sec.~\ref{sec:methods}).

\subsection{Finite electric fields and DFPT/finite-field formalisms}
A uniform static electric field, $\mathbf E$, couples to polarisation through the electric enthalpy (fixed-$\mathbf E$)
\begin{equation}
\mathcal F(\{\mathbf R\},\mathbf E) = E_0(\{\mathbf R\}) - \Omega\,\mathbf E\!\cdot\!\mathbf P(\{\mathbf R\}),
\label{eq:enthalpy}
\end{equation}
where $E_0$ is the zero-field energy and $\Omega$ the cell volume~\cite{nunes-gonze-2001,souza-iniguez-vanderbilt-2002}. This functional permits direct optimisation of field-polarised states and yields forces and stresses at fixed~$\mathbf E$.

Within density-functional perturbation theory (DFPT), linear-response equations for first-order wavefunctions under atomic displacements or homogeneous fields provide access to the complete set of dielectric, piezoelectric, and vibrational properties at zero field without performing finite differences~\cite{gonze_dynamical_1997}. In this work, we use DFPT quantities as labels and as a reference for assessing our learned model.

\subsection{Macroscopic response tensors as energy derivatives}
In SI units, the primary response functions used here are defined by exact derivatives of the electric enthalpy:
\begin{align}
\mathbf P(\{\mathbf R\}) &= -\frac{1}{\Omega}\,\frac{\partial \mathcal F}{\partial \mathbf E}, \label{eq:P_from_F}\\
Z^*_{\kappa,ij}(\{\mathbf R\}) &= \Omega\,\frac{\partial P_i}{\partial u_{\kappa j}}
= \frac{\partial F_{\kappa j}}{\partial E_i}, \label{eq:Zstar}\\
\alpha_{ij}(\{\mathbf R\}) &= \frac{\partial P_i}{\partial E_j}. \label{eq:alpha}
\end{align}
Here $u_{\kappa j}$ denotes the Cartesian displacement of sublattice $\kappa$, $F_{\kappa j}=-\partial \mathcal F/\partial u_{\kappa j}$ the force component,
$Z^*$ the Born effective charge tensor (dynamical charge), and $\boldsymbol\alpha$ the electronic polarisability per unit volume~\footnote{With $\mathbf D=\varepsilon_0\mathbf E+\mathbf P$ and $\mathbf P=\varepsilon_0\boldsymbol\chi^{(1)}\mathbf E$, one has $\boldsymbol\alpha=\varepsilon_0\boldsymbol\chi^{(1)}$ and $\boldsymbol\varepsilon_\infty=\varepsilon_0(\mathbf I+\boldsymbol\chi^{(1)})$.}. Maxwell reciprocity implies symmetry of mixed second derivatives (e.g.\ $\alpha_{ij}=\alpha_{ji}$). Translational invariance yields the acoustic sum rule (ASR) $\sum_\kappa Z^*_{\kappa,ij}=0$ for all $i,j$. Higher-order responses (nonlinear optics, electro-optic) follow from higher derivatives of $\mathcal F$, but are not our primary focus here.

\subsection{Ferroelectrics and spontaneous polarisation}
Ferroelectrics are insulating crystals with a switchable \emph{spontaneous} polarisation $\mathbf P_s\neq\mathbf 0$ at zero field, arising from a structural instability that breaks inversion symmetry. Near the paraelectric-to-ferroelectric transition, a soft polar phonon condenses, producing a double-well free-energy landscape as a function of the order parameter (e.g.\ the unstable mode amplitude). Within a Landau-Devonshire expansion, one writes, schematically,
$F(\mathbf P)=\tfrac{1}{2}a(T)\mathbf P^2+\tfrac{1}{4}b\,(\mathbf P^2)^2+\cdots-\Omega\,\mathbf E\!\cdot\!\mathbf P$, with $a(T)$ changing sign at the Curie temperature $T_C$.

The coupling $-\Omega\,\mathbf E\!\cdot\!\mathbf P$ tilts the double well, leading to coercive fields and hysteresis loops under cyclic fields.
Microscopically, the magnitude and direction of $\mathbf P_s$ follow from the Berry phase along an adiabatic distortion path connecting the paraelectric reference to the polar ground state, and its coupling to $\mathbf E$ is controlled by \emph{mode effective charges}, projections of $Z^*$ onto the polar eigenvectors.
Anomalously large $Z^*$ in perovskites signal strong cross-gap hybridisation and underpins large dielectric and piezoelectric responses.

\subsection{From linear dielectric response to spectra}
The frequency-dependent dielectric tensor can be decomposed into electronic and ionic parts. At high frequency (well above phonon energies and below interband transitions), the response is purely electronic, $\boldsymbol\varepsilon_\infty=\varepsilon_0(\mathbf I+\boldsymbol\alpha)$, which we compute from the electronic polarisability $\boldsymbol\alpha$. At low frequency, the \emph{ionic} contribution from infrared-active phonons adds a resonant term, schematically
\begin{equation}
\Delta\varepsilon_{ij}(\omega)\ \propto\
\sum_{m\in\text{IR}} \frac{\tilde Z_{m,i}\,\tilde Z_{m,j}}{\omega_{m}^2-\omega^2-i\gamma_m\omega},
\label{eq:ionic_eps}
\end{equation}
where $\omega_m$ are transverse-optical phonon frequencies, $\gamma_m$ linewidths, and $\tilde{\mathbf Z}_m$ the mode effective charges obtained by contracting $Z^*$ with phonon eigenvectors. In the static limit, this yields the lattice (ionic) permittivity and, together with $\boldsymbol\varepsilon_\infty$, the total static dielectric constant $\boldsymbol\varepsilon(0)$; the Lyddane-Sachs-Teller relation connects $\varepsilon(0)/\varepsilon_\infty$ to LO-TO splitting. The complex dielectric function is related to the optical conductivity by $\boldsymbol\varepsilon(\omega)=\boldsymbol\varepsilon_\infty + i\,\boldsymbol\sigma(\omega)/(\varepsilon_0\omega)$.

In molecular dynamics, infrared absorption follows from the dipole-dipole (or $\dot{\mathbf P}$-$\dot{\mathbf P}$) autocorrelation, Raman intensities from polarisability-polarisability correlations~\cite{Thomas2013, Luber2014}, and the full frequency-dependent complex dielectric function from both. This links time-domain simulations under our learned $\mathcal F$ to frequency-domain spectra.

\subsection{Multipole expansion and inspiration for equivariant field coupling}
We introduce the multipole expansion as the inspiration for the field-aware \texttt{MACE} model. The interaction of a localised charge distribution $\rho(\mathbf r)$ with a slowly varying external electrostatic potential $V_\text{ext}(\mathbf r)$ can be expanded in multipoles.
The electrostatic interaction energy is
\begin{equation}
E_\text{int}=\int d^3r\,\rho(\mathbf r)\,V_\text{ext}(\mathbf r).
\end{equation}
Expanding $V_\text{ext}$ about a reference point (e.g.\ the centre of the charge distribution)
\begin{equation}
V_\text{ext}(\mathbf r)=V_0 + r_i\,\partial_i V_0 + \tfrac12 r_i r_j\,\partial_i\partial_j V_0 + \cdots,
\end{equation}
and introducing the Cartesian multipole moments
\begin{align}
q &= \int d^3r\,\rho(\mathbf r),\qquad
p_i = \int d^3r\,r_i\,\rho(\mathbf r), \\
Q_{ij}^{(\rm T)} &= \int d^3r\,\big(3 r_i r_j - r^2\delta_{ij}\big)\,\rho(\mathbf r),
\end{align}
yields the familiar coupling to the electric field $\mathbf E=-\nabla V_\text{ext}$ and its gradients:
\begin{equation}
E_\text{int} = q\,V_0 - \mathbf p\!\cdot\!\mathbf E_0
-\frac{1}{6}\,Q_{ij}^{(\rm T)}\,\partial_i E_{0,j} \; + \; \cdots,
\label{eq:multipole}
\end{equation}
where subscripts $0$ indicate evaluation at the expansion point.
For a \emph{uniform} field ($\nabla\mathbf E=0$) only the dipole term survives.
In charge-neutral insulating crystals, the total monopole $q$ vanishes per cell; a non-zero dipole density (i.e.\ macroscopic polarisation) arises when inversion symmetry is broken (ferroelectrics)~\footnote{Different quadrupole conventions exist (traced vs \ traceless tensors, differing numerical factors). Our use of the multipole series is solely as symmetry \emph{guidance} for constructing equivariant couplings; numerical prefactors in \eqref{eq:multipole} do not enter the learned architecture.}

This multipolar structure has a direct analogue in the Atomic Cluster Expansion (ACE)~\cite{drautz-2019-ace} and its neural realisation \texttt{MACE}~\cite{batatia-2022-mace}. The local neighbour density around atom $\alpha$, $\rho_\alpha(\mathbf r)=\sum_{j} f_\text{cut}(r_{\alpha j})\,\delta(\mathbf r-\mathbf r_{\alpha j})$, admits a projection onto radial functions and spherical harmonics,
\begin{equation}
A^{(\alpha)}_{n\ell m}=\sum_{j} R_n(r_{\alpha j})\,Y_{\ell m}(\hat{\mathbf r}_{\alpha j}),
\end{equation}
which are \emph{spherical tensors} of rank $\ell$ transforming under the $\mathrm{SO}(3)$ irrep $\ell$. These coefficients play the role of \emph{generalised multipole moments} of the atomic environment. Scalar, rotation-invariant energies are then formed by contracting such tensors via Clebsch-Gordan coefficients (the spherical-tensor analogue of combining multipoles into invariants).

A homogeneous electric field is itself a rank-1 spherical tensor. By the Wigner-Eckart theorem, the symmetry-allowed linear coupling between a latent rank-$L$ feature $T^{(L)}$ (that is, one irrep block of the learned node features, with multiple channels per irrep) and the field $E^{(1)}$ decomposes as
\begin{equation}
T^{(L)} \otimes E^{(1)} \;=\; \bigoplus_{J=|L-1|}^{L+1} \big[T^{(L)} \otimes E^{(1)}\big]^{(J)} ,
\label{eq:CG-sum}
\end{equation}
with components
\begin{equation}
\big[T^{(L)} \otimes E^{(1)}\big]^{(J)}_{M}
  \;=\;
  \sum_{m,m'}
  C^{J M}_{L m,\,1 m'}\;
  T^{(L)}_{m}\, E^{(1)}_{m'} ,
\label{eq:CG-components}
\end{equation}
where $C$ are Clebsch-Gordan coefficients. This is exactly the pattern we implement in the architecture through equivariant tensor products followed by irrep-wise mixing [see Fig.~\ref{fig:architecture} and Eq.~(\ref{eq:meth_latent})]. The key idea is that the field perturbs learned latent spherical-tensor features according to angular-momentum selection rules, while the final energy remains a scalar invariant. Keeping the coupling \emph{linear} in $\mathbf E$ realises latent linear response; nonlinear field effects are nevertheless captured by stacking interaction layers and nonlinear readouts, which together generate higher-order dependence on $\mathbf E$ in the learned electric enthalpy $\mathcal F(\{\mathbf R\},\mathbf E)$.

\subsection{Physical identities enforced by a single scalar enthalpy}\label{sec:identities}
All response tensors in this work are \emph{exact derivatives} of the same twice-differentiable scalar electric-enthalpy functional $\mathcal F(\{\mathbf R\},\mathbf E)$ [Eqs.~(\ref{eq:P_from_F})-(\ref{eq:alpha})].
This immediately enforces key symmetries and sum rules, without any extra penalties, because they are nothing more than properties of mixed partial derivatives and fundamental invariances of $\mathcal F$.
\\

\paragraph*{Maxwell/reciprocity (symmetry of mixed derivatives).}
From $\mathbf P = -\Omega^{-1}\partial \mathcal F/\partial \mathbf E$ one has
\begin{equation}
\alpha_{ij}
= \frac{\partial P_i}{\partial E_j}
= -\frac{1}{\Omega}\,\frac{\partial^2 \mathcal F}{\partial E_i\,\partial E_j}
= -\frac{1}{\Omega}\,\frac{\partial^2 \mathcal F}{\partial E_j\,\partial E_i}
= \alpha_{ji},
\label{eq:maxwell-alpha}
\end{equation}
i.e.\ the electronic polarisability is \emph{exactly symmetric} at the level of the model.
Likewise, with $F_{\kappa j}=-\partial \mathcal F/\partial u_{\kappa j}$,
\begin{equation}
Z^*_{\kappa,ij}
=\frac{\partial F_{\kappa j}}{\partial E_i}
= -\,\frac{\partial^2 \mathcal F}{\partial E_i\,\partial u_{\kappa j}}
= -\,\frac{\partial^2 \mathcal F}{\partial u_{\kappa j}\,\partial E_i}
=\Omega\,\frac{\partial P_i}{\partial u_{\kappa j}},
\label{eq:maxwell-zstar}
\end{equation}
which is the standard Maxwell identity linking $Z^*$ to the field-derivative of the force and to the position-derivative of the polarisation.
\\ 

\paragraph*{Acoustic sum rule (ASR) for Born effective charges.}
$\mathcal F$ is invariant under a rigid translation of all sublattices,
$\{\mathbf R_\kappa\}\!\to\!\{\mathbf R_\kappa+\boldsymbol\delta\}$, for any $\boldsymbol\delta$ and any uniform $\mathbf E$,
hence
\begin{equation}
0 = \frac{\partial}{\partial \delta_j}\,\mathcal F(\{\mathbf R_\kappa+\boldsymbol\delta\},\mathbf E)
= \sum_\kappa \frac{\partial \mathcal F}{\partial u_{\kappa j}}
= -\sum_\kappa F_{\kappa j}.
\label{eq:rigid-shift}
\end{equation}
Differentiating \eqref{eq:rigid-shift} with respect to $E_i$ and using \eqref{eq:maxwell-zstar},
\begin{equation}
0 = \sum_\kappa \frac{\partial^2 \mathcal F}{\partial E_i\,\partial u_{\kappa j}}
= - \sum_\kappa Z^*_{\kappa,ij},
\quad \Rightarrow \quad
\sum_\kappa Z^*_{\kappa,ij}=0 \quad \forall\,i,j,
\label{eq:asr-derivation}
\end{equation}
i.e.\ the ASR holds \emph{by construction} for a uniform field and a translation-invariant $\mathcal F$.
\\

\paragraph*{Rotational covariance and crystal point-group constraints.}
Let $\mathcal R\!\in\!O(3)$ be a rigid rotation. Our construction ensures
\begin{equation}
\mathcal F(\{\mathcal R\mathbf R_\kappa\},\,\mathcal R\mathbf E) \;=\; \mathcal F(\{\mathbf R_\kappa\},\,\mathbf E),
\label{eq:rot-inv}
\end{equation}
i.e.\ the scalar output is rotationally invariant when the atomic configuration and the uniform field are co-rotated.
Taking derivatives at $\mathbf E\!=\!\mathbf 0$ yields the correct tensorial transformation laws:
\begin{align}
\mathbf P(\{\mathcal R\mathbf R\}) &= \mathcal R\,\mathbf P(\{\mathbf R\}),\\
\boldsymbol\alpha(\{\mathcal R\mathbf R\})
&= \mathcal R\,\boldsymbol\alpha(\{\mathbf R\})\,\mathcal R^\top, \label{eq:alpha-cov}\\
Z^*_{\pi(\kappa)}(\{\mathcal R\mathbf R\})
&= \mathcal R\,Z^*_{\kappa}(\{\mathbf R\})\,\mathcal R^\top,
\label{eq:zstar-cov}
\end{align}
where $\pi$ permutes sublattices according to the rotation.
\\

\paragraph*{Thermodynamic sign/definiteness.}
At fixed atomic positions, the second field-derivative of $\mathcal F$ gives
$-\Omega\,\boldsymbol\alpha = \partial^2 \mathcal F/\partial \mathbf E\,\partial \mathbf E$.
For stable insulating states at zero field, thermodynamics implies
$\mathbf v^\top \boldsymbol\alpha\,\mathbf v \ge 0$ for any vector $\mathbf v$ (passive linear response).
While symmetry \eqref{eq:maxwell-alpha} is exact in our construction, strict positive semi-definiteness is a property of the true material response and is \emph{approached} as the learned $\mathcal F$ converges; tiny negative eigenvalues may occur at inference.

\section{Methods}\label{sec:methods}

\begin{figure*}
    \centering
    \includegraphics[width=1\linewidth]{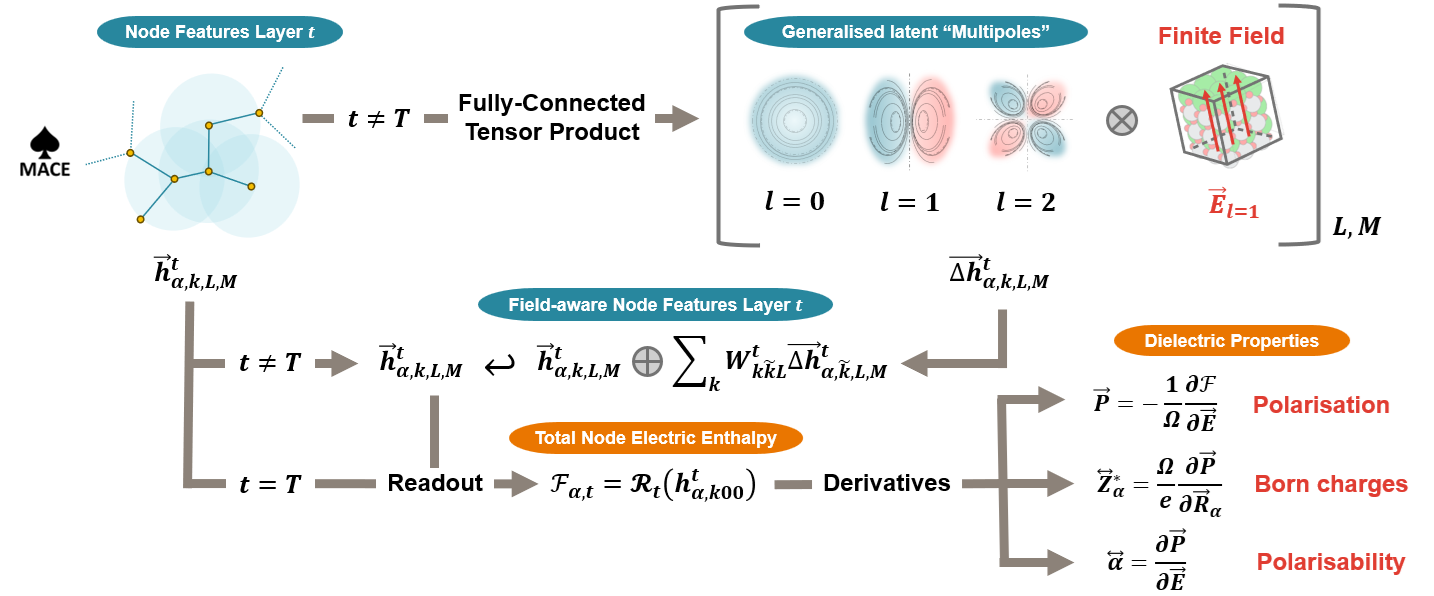}
    \caption{\textbf{MACE-Field architecture.} At message-passing layer $t$, \texttt{MACE} produces learned equivariant node features $h^{(t)}_{\alpha,kLM}$, where $L$ denotes the $O(3)$ irrep order and $k$ indexes channels. We occasionally refer to these latent tensors as ``multipoles'' only heuristically, because they transform like spherical tensors; they are not literal charge-density multipoles. A uniform external field $\mathbf E$ (irrep $l{=}1$) couples to these features via a fully-connected tensor product (Clebsch-Gordan contraction) to form $\Delta h^{(t)}_{\alpha,kLM}$; an irrep-wise linear map $W^{(t)}$ and a residual update yield field-aware features $\tilde h^{(t)}_{\alpha,kLM}$. Scalar components ($L{=}0$) are read out at each layer and summed to give a rotationally invariant electric enthalpy $\mathcal F$. All dielectric observables are obtained as exact derivatives of this single scalar: polarisation $\mathbf P=-\Omega^{-1}\partial \mathcal F/\partial \mathbf E$, Born effective charges $Z^{*}_{\alpha,ij}=\Omega\,\partial P_i/\partial R_{\alpha j}/e=\partial F_{\alpha j}/\partial E_i$, and electronic polarisability $\alpha_{ij}=\partial P_i/\partial E_j$. Hidden layers remain $O(3)$-equivariant; only the final readout in layer $t=T$ is strictly invariant.
    }
    \label{fig:architecture}
\end{figure*}

\subsection{Overview}
We learn a single scalar electric-enthalpy functional $\mathcal F(\{\mathbf R\},\mathbf E)$ that is differentiable with respect to both atomic positions and a \emph{global, uniform} electric field. The \texttt{MACE-Field} network is a field-aware variant of \texttt{MACE}: it injects the field into latent $O(3)$-equivariant features at each interaction layer while keeping the final readout scalar ($L{=}0$). Polarisation, BECs, and polarisability are then obtained by \emph{exact} automatic differentiation of $\mathcal F$
[Eqs.~(\ref{eq:P_from_F})-(\ref{eq:alpha})], ensuring derivative consistency (Maxwell symmetries and acoustic sum rules) by construction. Conceptually, this makes \texttt{MACE-Field} a physics-informed, end-to-end differentiable model: all response properties are mixed derivatives of a single learned potential, thereby avoiding finite-difference noise and naturally fitting into automatic adjoint-based optimisation and downstream simulation workflows.

\subsection{Architecture and field coupling}

A schematic of the \texttt{MACE-Field} architecture is shown in Fig.~\ref{fig:architecture}. We start from a standard \texttt{MACE} backbone with $T$ layers, $K$ hidden irreps, and message interaction and product blocks as in Ref.~\onlinecite{batatia-2022-mace}. Here, ``latent features'' refer to the learned per-atom node features in the equivariant network. The index $L$ denotes the spherical-harmonic/$O(3)$ irrep order, while $K$ counts channels within each irrep.

Let $h^{(t)}_{\alpha,kLM}$ denote the latent features at layer $t$ (atom $\alpha$, channel $k$, irrep $L$, component $M$). In the current architecture, the field input is a single, spatially uniform global vector shared across the entire graph. It is represented as an $l{=}1$ vector (odd parity $p=-1$) irrep feature $E_{1m}$ and coupled at each layer $t<T$ via a \texttt{FullyConnectedTensorProduct} followed by a linear equivariant mixing (residual update)
\begin{align}
\Delta h^{(t)}_{\alpha,kLM}
&= \sum_{l_1 m_1 m_2}
\mathcal C^{LM}_{l_1 m_1,\,1 m_2}\,
h^{(t)}_{\alpha,k l_1 m_1}\,E_{1 m_2},\nonumber\\
\tilde h^{(t)}_{\alpha,kLM}
&= h^{(t)}_{\alpha,kLM} - \sum_{\tilde k} W^{(t)}_{k\tilde k L}\,\Delta h^{(t)}_{\alpha,\tilde k LM},
\label{eq:meth_latent}
\end{align}
with Clebsch-Gordan coefficients $\mathcal C$ and learned weights $W^{(t)}$.
Message passing and products operate on $\tilde h^{(t)}$ thereafter.
At the final layer $t{=}T$, following \texttt{MACE}, we retain only $L{=}0$ scalars and employ a nonlinear readout to produce per-atom contributions, which are summed over atoms and layers to yield $\mathcal F$.
The field is identical for all atoms in a graph and does \emph{not} depend on absolute positions, preserving translation invariance. At zero field $E=0$, the residual update is identically zero, so we fall back to the normal \texttt{MACE} behaviour. Because the field coupling layers operate purely on latent features, the underlying ACE descriptors and scalar readout are unchanged, making \texttt{MACE-Field} a true plug-in: any existing \texttt{MACE} checkpoint can be upgraded to a field-aware model by inserting these layers and fine-tuning on response data. The same design choice also defines the scope of the method: it targets bulk periodic response under spatially uniform bias and does not explicitly resolve site-dependent internal fields at defects, surfaces, interfaces, or domain walls.

\subsection{Differentiable observables and units}

\subsubsection{Target properties}

We compute observables by automatic differentiation (as implemented in \texttt{PyTorch}'s \texttt{autograd} functionality) on the \emph{interaction} part of the energy (the atomic baselines are constant):
\begin{align}
\mathbf P &= - \frac{1}{\Omega} \frac{\partial \mathcal F}{\partial \mathbf E}, \qquad
Z^*_{\kappa,ij} = \frac{\partial(\Omega P_i)}{\partial u_{\kappa j}},\qquad
\alpha_{ij} = \frac{\partial P_i}{\partial E_j}.
\label{eq:meth_derivs}
\end{align}
Internally, $\mathbf P$ is stored in $e/\text{\AA}^2$, $Z^*_{\kappa,ij}$ in units of $e$, and $\alpha_{ij}$ in $\varepsilon_0$ (vacuum permittivity). We report polarisations in units of $\mu$C\,cm$^{-2}$ where convenient.

All training losses use consistent internal units. To obtain $Z^*$ and $\boldsymbol\alpha$, we differentiate $\Omega\,\mathbf P$ with respect to atomic positions and $\mathbf P$ with respect to $\mathbf E$, respectively, using nested automatic differentiation. During training, we retain and create graphs for higher-order derivatives only when those targets are present in the batch. We compute losses on the \emph{interaction} energy (the atomic baselines are constant), so that derivatives reflect the learned $\mathcal F$.

\subsubsection{Derived properties: Dielectric constants}
\label{sec:dielectric_math}

We can relate the polarisability to the \emph{relative} high-frequency dielectric tensor by
\begin{equation}
\boldsymbol\varepsilon_\infty^{(r)} = \mathbf I + \frac{\boldsymbol\alpha}{\varepsilon_0}.
\label{eq:eps_from_alpha}
\end{equation}
Likewise, we obtain the polycrystalline dielectric constant $\varepsilon^{(\text{poly})}_\infty$ at long wavelengths (ionic contributions vanish at high frequencies) from the eigenvalues ($\lambda_i$, $i\in\{1,2,3\}$) of the high-frequency dielectric tensor~\cite{Petousis2017},
\begin{equation}
\varepsilon^{(\text{poly})}_\infty \equiv \frac{\lambda_1 + \lambda_2 + \lambda_3}{3}.
\label{eq:eps_poly}
\end{equation}
We can then obtain an estimate of the refractive index, $n$, at optical frequencies and far from resonance effects using
\begin{equation}
    n = \sqrt{\varepsilon^{(\text{poly})}_\infty}.
\label{eq:refractive_index}
\end{equation}

For a given relaxed structure, the real-space Hessian is
\begin{equation}
  H_{\kappa\alpha,\kappa'\beta} = \frac{\partial^2 \mathcal{F}}{\partial u_{\kappa\alpha}\,\partial u_{\kappa'\beta}} ,
\end{equation}
where $\kappa,\kappa'$ index atoms in the unit cell, and $\alpha,\beta\in\{x,y,z\}$ are Cartesian components of the displacements $u_{\kappa\alpha}$. From this we construct the $\Gamma$-point dynamical matrix
\begin{equation}
  D_{\kappa\alpha,\kappa'\beta} = \frac{1}{\sqrt{M_\kappa M_{\kappa'}}} H_{\kappa\alpha,\kappa'\beta},
\end{equation}
with $M_\kappa$ the mass of atom $\kappa$. Diagonalising $D$ yields phonon eigenfrequencies and eigenvectors,
\begin{equation}
  \sum_{\kappa'\beta} D_{\kappa\alpha,\kappa'\beta} e^{(m)}_{\kappa'\beta} = \omega_m^2 e^{(m)}_{\kappa\alpha} ,
\end{equation}
where $m$ labels the (zone-centre) normal modes, the $\omega_m$ are the phonon frequencies, and the eigenvectors $e^{(m)}_{\kappa\alpha}$ are chosen orthonormal in the mass-weighted sense,
\begin{equation}
  \sum_{\kappa\alpha} e^{(m)}_{\kappa\alpha}\,e^{(m')}_{\kappa\alpha} = \delta_{mm'} .
\end{equation}
Given the Born effective charge tensors $Z^{\ast}_{\kappa,i\alpha}$ predicted by \texttt{MACE-Field} (with Cartesian indices $i,\alpha\in\{x,y,z\}$), we can form the mode effective charges by projecting $Z^*$ onto the phonon eigenvectors,
\begin{equation}
  \tilde{Z}_{m,i} = \sum_{\kappa\alpha} \frac{Z^{\ast}_{\kappa,i\alpha}}{\sqrt{M_\kappa}} e^{(m)}_{\kappa\alpha} .
\end{equation}
In density-functional perturbation theory, the lattice (ionic) contribution to the dielectric tensor in the static limit $\omega\to 0$ (cf.\ Eq.~\eqref{eq:ionic_eps}) is then given by
\begin{equation}
  \varepsilon^{\mathrm{ion}}_{ij} = \frac{4\pi}{\Omega} \sum_{m\,\in\,\mathrm{IR}} \frac{\tilde{Z}_{m,i}\,\tilde{Z}_{m,j}}{\omega_m^2} ,
  \label{eq:eps_ion_static}
\end{equation}
where $\Omega$ is the unit-cell volume and the sum runs over the IR-active, non-acoustic modes at $\Gamma$. In practice, we discard the three translational acoustic modes (for which $\omega_m\to 0$ and $\tilde{Z}_{m,i}\to 0$) and any modes with numerically ill-defined $\omega_m$ arising from slight residual stresses.

Finally, combining the electronic and ionic contributions yields the full static dielectric tensor
\begin{equation}
  \boldsymbol{\varepsilon}_0 = \boldsymbol{\varepsilon}_\infty + \boldsymbol{\varepsilon}_{\mathrm{ion}} .
\end{equation}

\subsection{Branch-invariant polarisation supervision}
\label{sec:pol-loss}

Polarisation is multivalued modulo the polarisation lattice,
$\mathbf Q_k=e\,\mathbf a_k/\Omega$ [Eq.~(\ref{eq:pol_quanta})].
As such, we follow the approach taken by \texttt{Allegro-pol}~\cite{allegro-pol-2025}: given a reference $\mathbf P_\mathrm{ref}$ and prediction $\mathbf P_\mathrm{pred}$, we compare the \emph{folded} difference $\Delta\mathbf P_\mathrm{fold}$:
\begin{align}
\Delta\mathbf P &= \mathbf P_\mathrm{pred}-\mathbf P_\mathrm{ref},\nonumber\\
B &= 
\begin{bmatrix}
\mathbf a_1, 
\mathbf a_2,
\mathbf a_3
\end{bmatrix}^\mathsf{T},
\qquad
Q_\mathrm{pol}=B/|\Omega|,\nonumber\\
Q_\mathrm{pol}^{\mathsf T}\mathbf c^{\mathsf T} &= \Delta\mathbf P^{\mathsf T},
\qquad
\mathbf n^\star=\arg\min_{\mathbf n\in\mathbb Z^3}\left\|(\mathbf c-\mathbf n)Q_\mathrm{pol}\right\|_2,\nonumber\\
\Delta\mathbf P_\mathrm{fold} &= \mathbf c_\mathrm{fold}\,Q_\mathrm{pol} \qquad
\mathbf c_\mathrm{fold}=\mathbf c-\mathbf n^\star.
\label{eq:meth_fold}
\end{align}
Here $B$ is the cell matrix with lattice vectors stored as rows, so in the lattice basis the polarisation quanta becomes $Q_\mathrm{pol}=B/|\Omega|$. The fractional coefficients $\mathbf c$ are therefore obtained by solving against $Q_\mathrm{pol}^{\mathsf T}$. The fold is then defined by the \emph{exact} closest-vector problem on the polarisation lattice rather than by a component-wise modulo. This closest-image construction works for general, non-orthogonal cells and makes the polarisation loss branch-invariant while leaving $\mathbf P$ itself defined by the conservative derivative in Eq.~(\ref{eq:meth_derivs}).

\subsection{Loss function}

Following the \texttt{UniversalLoss} used by \texttt{MACE} for energy, forces and stress, our field-related per-task losses are
\begin{align}
\mathcal L_P &= \frac{1}{W_P}\sum_{g=1}^{N_b} w^P_g\;
d_{\mathrm{pol}}^2\!\left(\mathbf c^{\mathrm{fold}}_g;Q_{\mathrm{pol},g}\right), \label{eq:L_P}\\[3pt]
\mathcal L_Z &= \frac{1}{W_Z}\sum_{g=1}^{N_b}\;\frac{1}{n_g}\sum_{a\in g} w^Z_{ga}\;
\overline{\mathcal H}_{\delta_Z}\!\left(
{\hat Z}^{\,*}_{ga}- Z^{\,*}_{ga}\right), \label{eq:L_Z}\\[3pt]
\mathcal L_\alpha &= \frac{1}{W_\alpha}\sum_{g=1}^{N_b} w^\alpha_g\;
\overline{\mathcal H}_{\delta_\alpha}\!\left(
{\hat{\boldsymbol\alpha}}_{g}-{\boldsymbol\alpha}_{g}\right),
\label{eq:L_alpha}
\end{align}
where
\begin{align}
d_{\mathrm{pol}}^2(\mathbf c;Q) &= \mathbf c\,M(Q)\,\mathbf c^\mathsf{T},\nonumber\\
M(Q) &= \frac{3\,Q Q^\mathsf{T}}{\mathrm{tr}(Q Q^\mathsf{T})},
\label{eq:pol_metric}
\end{align}
is the normalised, general cell-aware polarisation loss, $\mathcal H _{\delta}$ is the Huber loss (with Huber-delta $\delta$), $\overline{\mathcal H}$ denotes the average of $\mathcal H$ over vector/tensor components, $Z^*_{ga}\in\mathbb R^{3\times 3}$ is the BEC tensor on atom $a$, and $\boldsymbol\alpha_g$ is the polarisability tensor. A batch $g$ has size $N_b$ and contains $n_g$ atoms.

Normalisation constants $W_T=\sum_g w^T_g$ (or $\sum_{g,a}w^T_{ga}$ for per-atom terms) prevent batch-size/weight drift. The $w_g$ configuration weights can be used to mask per-task losses for heterogeneous labels by setting the contribution of a particular property to zero if its label is absent for a given data point. This is particularly useful for fine-tuning existing foundation models using a replay set of energies, forces, and stresses when field-related labels are absent.

The total loss is a non-negative weighted sum of per-task losses:
\begin{align}
\mathcal L_{\mathrm{tot}} &=
\lambda_E \mathcal L_E + \lambda_F \mathcal L_F + \lambda_\sigma \mathcal L_\sigma \nonumber \\
&\quad+ \lambda_P \mathcal L_P + \lambda_Z \mathcal L_Z + \lambda_\alpha \mathcal L_\alpha , 
\label{eq:L_total}
\end{align}
where the energy, forces and stress losses are the same as in \texttt{MACE}~\cite{batatia-2022-mace}. Unless otherwise stated, we use fixed weights $\lambda$ during training (details for the different trained \texttt{MACE-Field} models are provided in the Supplementary Information).

\subsection{Implementation and training protocol}
Models are implemented in \texttt{PyTorch} with \texttt{cuEquivariance} irreps (\texttt{e3nn} option is available) and tensor products, and built atop a public \texttt{MACE} codebase.

Exact hyperparameters, seeds, split definitions, and details on public data/code regeneration are listed in the Supplementary Information and the Code and Data Availability statement.

We train with Adam (with decoupled weight decay), a cosine learning rate schedule with warmup, and gradient clipping. We retain and create graphs for autograd of $\Omega\mathbf P$ with respect to $\mathbf R$ and $\mathbf E$ when BECs and $\boldsymbol\alpha$ are requested (higher-order gradients). We use early stopping on validation $\mathbf P$ (ferroelectric split) and $\boldsymbol\alpha$/$Z^*$ (dielectric split). All models are trained with Distributed Data Parallel (DDP) across GPUs; representative dtypes and experiment-specific settings are listed in the SI and repository configuration files.
Representative runtime numbers are reported in the SI. In practice, force-only MLMD with \texttt{MACE-Field} is expected to remain close to an architecture-matched ordinary \texttt{MACE} model, because the time-integration loop only queries energies and forces. The main extra overhead enters during training and whenever $\mathbf P$, $Z^*$, or $\boldsymbol\alpha$ are explicitly requested, because these require additional nested-autograd evaluations; in the \texttt{LAMMPS}/ML-IAP-style workflows used here, these response quantities may therefore be backfilled in a separate postprocessing pass.

\subsection{Finite-field MD protocols}
\label{sec:md-protocols}

We perform finite-field ML molecular dynamics using the learned $\mathcal F$, either directly in \texttt{ASE}~\cite{larsen2017ase} or through \texttt{LAMMPS}/ML-IAP exports for the longer production trajectories:
\\

\paragraph*{\ce{BaTiO3} hysteresis.}
We run field-driven dynamics and quasi-static ionic relaxations in a small bulk-periodic supercell under a spatially uniform field applied along the polar axis. In the \texttt{LAMMPS}/ML-IAP workflow, we first relax a 135-atom periodic cell at fixed DFT lattice vectors and then equilibrate athermally at 0~\si{K}; the driven segment is subsequently propagated at fixed cell with \texttt{nve} dynamics plus viscous damping. Because the cell is periodic and contains no explicit defects or domain walls, this setup probes intrinsic, spatially homogeneous switching rather than nucleation-and-growth kinetics.
\\

\paragraph*{$\alpha$-\ce{SiO2} IR/Raman and dielectric function.}
At 300~\si{K} we run fixed-cell NVT trajectories in a 72-atom periodic $\alpha$-quartz supercell, using 2~fs timesteps and 200~ps production windows after equilibration. In the \texttt{LAMMPS}/ML-IAP workflow, energies and forces are evaluated during time integration and the derivative observables entering the spectra are reconstructed afterwards on the stored frames. IR absorption is then obtained from the $\dot{\mathbf P}$-$\dot{\mathbf P}$ autocorrelation, and Raman from the $\boldsymbol\alpha$-$\boldsymbol\alpha$ autocorrelation, both Fourier-transformed with a Hann window and modest Gaussian broadening. The complex dielectric $\boldsymbol\varepsilon(\omega)$ follows from $\boldsymbol\varepsilon_\infty^{(r)}$ and the same current-current (equivalently $\dot{\mathbf P}$-$\dot{\mathbf P}$) response.

\subsection{Comparison to other models}
\label{sec:compare}

\subsubsection*{Allegro-pol (unified electric enthalpy).}
Our work is closest in spirit to Allegro-pol, which also learns a single \emph{electric enthalpy} and obtains $\mathbf P$, $Z^*$, and $\boldsymbol\alpha$ by exact differentiation, enforcing reciprocity and the acoustic sum rule by construction~\cite{allegro-pol-2025}. Both approaches handle Berry-phase multivaluedness using a branch-invariant (minimum-image) loss on $\Delta\mathbf P$, though Allegro-pol's implementation is limited to orthogonal cells. The main differences are architectural and data-scientific. Allegro-pol injects the uniform field as a rank-1 input feature before the local equivariant layers, whereas \texttt{MACE-Field} couples $\mathbf E$ \emph{inside} each message-passing layer as a perturbation to latent irreps via Clebsch-Gordan tensor products and residual mixing, preserving the standard \texttt{MACE} backbone and enabling foundation-weight reuse. Allegro-pol is demonstrated primarily in a single-material regime (e.g.\ $\alpha$-quartz, \ce{BaTiO3}) with labels constructed from small finite-field calculations around nominal zero field~\cite{allegro-pol-2025}; here we focus on cross-chemistry training using DFPT/Berry-phase labels across thousands of Materials Project structures, enabling transfer across compositions and space groups.

\begin{figure*}[t]
    \centering
    \subfigure[]{
      \includegraphics[width=0.48\linewidth]{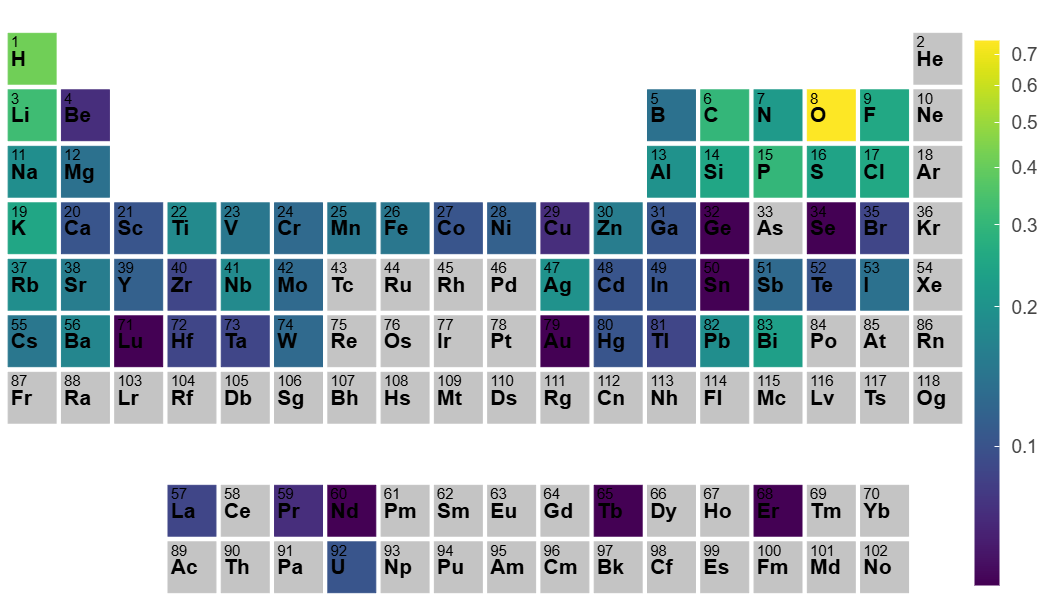}
      \label{fig:dataset_elements_ferro}
    }
    \subfigure[]{
      \includegraphics[width=0.48\linewidth]{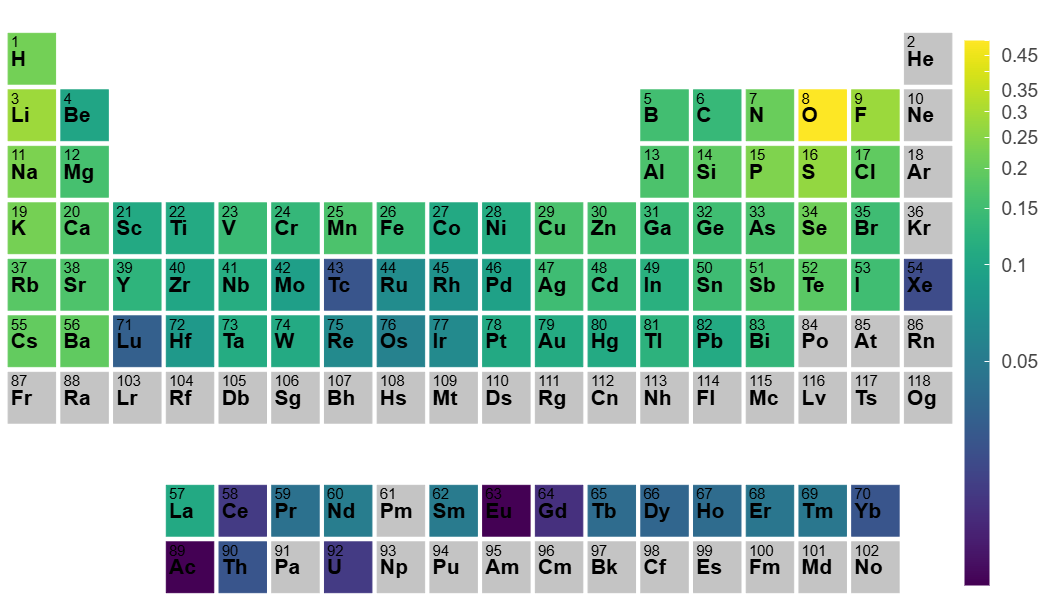}
      \label{fig:dataset_elements_dielectric}
    }
    \caption{Elemental coverage of the cross-chemistry datasets used in this work. (a) Smidt \emph{et al.} ferroelectric distortion-path set~\cite{smidt-2020-ferrodb} covering 61 elements. (b) MP-Dielectric dataset (DFPT BECs and electronic polarisabilities) covering 81 elements. Colour encodes the per-dataset normalised frequency, that is, the fraction of structures containing each element (see colour bars); grey indicates no examples.}
    \label{fig:dataset_elements}
\end{figure*}

\subsubsection*{Response-operator and symmetry-adapted learning.}
Complementary literature learns an electric response \emph{without} a force field by enforcing derivative structure and covariance at the property level. Operator QML (OQML) introduces response operators directly in the regression, so that quantities such as forces and dipoles arise as derivatives of a learned energy functional with improved data efficiency~\cite{Christensen2019OperatorsQML}. Symmetry-adapted Gaussian process regression (SA-GPR) and $\lambda$-SOAP kernels provide a general framework to learn \emph{tensorial} response with the correct $O(3)$ covariance, with demonstrations ranging from molecular and condensed-phase electrical response (e.g.\ water clusters/liquid snapshots)~\cite{Grisafi2018SymmetryAdaptedTensorial}. Similarly, the AlphaML work of Wilkins \emph{et al.} uses symmetry-adapted learning to reproduce coupled-cluster (LR-CCSD) \emph{static polarisability tensors} for 7,211 QM7b molecules and transfers to a chemically diverse showcase set~\cite{wilkins2019}. These approaches are valuable baselines for screening and for building physical inductive biases (covariant tensors, operator consistency), but they do not provide a thermodynamically consistent electric-enthalpy functional for finite-field MD in periodic solids, nor do they address Berry-phase branch structure directly.

\subsubsection*{Long-range physics and electrostatics in atomistic ML.}
A second comparison axis concerns how models represent nonlocal electrostatics and screening. Grisafi and Ceriotti proposed a general scheme to incorporate \emph{long-range physics} by building nonlocal representations (including forms with electrostatic-potential-like asymptotics) and remapping them into local, $O(3)$-equivariant features; they demonstrated the approach on electrostatic energies, charged dimers, and the electronic dielectric response of liquid water~\cite{Grisafi2019}. More MLIP-specific routes either (i) learn latent charge fields and apply Ewald-like summations (e.g.\ LES)~\cite{les-2025}, (ii) promote charges to variational degrees of freedom within systematic bases (e.g.\ ACE+Q / charge-augmented \texttt{ACE})~\cite{cace-2024}, or (iii) add explicit long-range electrostatics to neural potentials (e.g.\ \texttt{DPLR})~\cite{Zhang2022DPLR}. Compared to these, \texttt{MACE-Field} targets dielectric response by \emph{direct supervision} of $\mathbf P$, $Z^*$ and $\boldsymbol\alpha$ while retaining strict derivative consistency through a single learned enthalpy; long-range/screening physics is then represented in the learned field-coupled functional rather than reconstructed via separate electrostatic post-processing.

\subsubsection*{Empirical ferroelectric force fields and local-field limitations.}
Reactive and empirical ferroelectric force fields such as ReaxFF have already been used to study large-scale phenomena including surface chemistry, oxygen-vacancy effects, thickness dependence, and defect-assisted switching in \ce{BaTiO3} and related systems~\cite{Akbarian2019ReaxFFBaTiO3,Kelley2022VacancyFerroelectrics}. These methods are attractive for mesoscale simulations because they can be deployed in large supercells and naturally access domain formation and defect physics. \texttt{MACE-Field} has a different emphasis: it learns a single differentiable electric enthalpy and derives $\mathbf P$, $Z^*$ and $\boldsymbol\alpha$ as exact mixed derivatives, which is advantageous for derivative consistency, finite-field MD and dielectric analysis under spatially uniform bias. The present architecture, however, couples a single homogeneous external field to the entire graph and does not solve for self-consistent, spatially varying internal fields. It should therefore be viewed primarily as a model for bulk periodic response, homogeneous switching in bulk cells, and spectroscopy, rather than a complete treatment of charged defects, surfaces, interfaces, or domain walls where explicit long-range electrostatics and local depolarising fields are essential.

\subsubsection*{Molecular dipole models and charge-dipole decompositions.}
For finite systems, accurate dipoles can be obtained from learned decompositions into partial charges and local atomic dipoles. Veit \emph{et al.} (MuML) combines a nonlocal charge model with a local atomic-dipole model to capture both long-range charge separation and short-range polarisation, yielding accurate molecular dipoles on datasets such as QM7b and transfer tests beyond the training distribution~\cite{Veit2020MuMLDipoles}. Such models provide important points of comparison for the \emph{representation} of polarisation. Still, periodic insulating crystals introduce additional constraints (Berry-phase gauge/branch structure and cell-volume factors) that motivate our branch-invariant training objective and enthalpy-based formulation.

\subsubsection*{Direct tensor predictors (screening models; no force field).}
Equivariant property predictors such as AnisoNet~\cite{anisonet-2024} and graph-based dielectric models that learn scalar $\varepsilon$ or full dielectric tensors from crystal structure (e.g.\ the oxide-dielectric ML models of Morita \emph{et al.}~\cite{Morita2020} and MACE-$\mu$~\cite{Kapil2024FaradayDiscuss}) provide fast screening baselines for dielectric response. However, these approaches predict response \emph{directly} without an interatomic potential: they do not yield forces, do not define a thermodynamically consistent electric-enthalpy functional, and therefore cannot be used directly for finite-field molecular dynamics, hysteresis simulations, or spectroscopic workflows that require end-to-end derivative consistency.

\section{Datasets}\label{sec:datasets}

Our training data combine two cross-chemistry datasets that provide wide elemental coverage across the periodic table with two single-material molecular dynamics (MD) trajectory sets. Figure~\ref{fig:dataset_elements} summarises the elemental coverage of the two cross-chemistry datasets that underpin the field-aware foundation model. Unless explicitly stated otherwise, all errors reported below are with respect to the corresponding DFT/DFPT labels. Because the cross-chemistry datasets combine distinct semilocal workflows, the resulting error floor reflects both ML approximation and the intrinsic limitations and inconsistencies of the reference calculations.

\subsection{MP-Dielectric: BECs and polarisabilities of Materials Project dielectrics}
\label{sec:mp_dielectric}

For dielectric response, we assemble a broad chemistry of $\sim$6{,}000 insulating materials from the Materials Project with DFPT-computed Born effective charges and electronic polarisabilities (equivalently, the electronic dielectric tensor $\,\boldsymbol\varepsilon_\infty$). The elemental coverage and per-element frequencies are shown in Fig.~\ref{fig:dataset_elements_dielectric}. This dataset spans 81 elements and includes oxides, halides, chalcogenides, and mixed-anion chemistries.

We assemble this dataset from the public Materials Project API that queries dielectric-task provenance, extracts the DFPT tensors, and writes the filtered train/validation/test \texttt{extxyz} files used here. All DFPT calculations in this dataset were carried out within the \textbf{GGA-PBE} functional family.~\footnote{Public MP documentation describes DFPT workflows for dielectric properties (using VASP) within the semilocal GGA family; our API-assembled set consists of insulating GGA-PBE entries with dielectric tensors and Born effective charges.} As such, absolute values may differ from higher-level functionals and, as we discuss later, a small subset of entries even violate basic constraints such as the acoustic sum rule. We use $\,\boldsymbol\alpha=\partial \mathbf P/\partial \mathbf E\,$ directly as labels and supervise $\,Z^*\,$ component-wise. The DFPT forces in this dataset do not correlate to the dielectric response; therefore, we \emph{do not} train on forces or stress for this set to avoid encoding artefacts. Splits are defined at the material level (unique MP identifiers), so polymorphs of a given material may appear in different splits. Within this head, the reference labels are therefore reasonably consistent across the exchange-correlation family and basis-set class (plane-wave VASP DFPT), even though convergence settings vary across MP workflows. Together with MP-Ferroelectric (Sec.~\ref{sec:mp_ferroelectric} and Fig.~\ref{fig:dataset_elements_ferro}), this dataset covers more than 80 elements and thousands of distinct chemistries, providing a realistic test of whether a single enthalpy model can generalise dielectric response beyond single-material manifolds.

\subsection{MP-Ferroelectric: Polarisations and distortion paths of Materials Project ferroelectrics}
\label{sec:mp_ferroelectric}

To go beyond training on single-material examples, we use \texttt{MACE-Field} to predict the polarisation of a wide set of ferroelectric materials and their corresponding polarisation branches, so that we can calculate their spontaneous polarisation. We use the automatically curated first-principles MP-Ferroelectric database of Smidt \emph{et al.}~\cite{smidt-2020-ferrodb}, obtained using the \texttt{MPContribs} API~\cite{Horton2025NatMaterMP, jain-2013-mp, Huck2015eScience, Huck2016CCPE_MPContribs, AndreoniYip2020HMM}. The elemental coverage of this dataset, shown in Fig.~\ref{fig:dataset_elements_ferro}, spans 61 elements and is enriched in oxide and perovskite-like chemistries that are typical ferroelectric candidates.

The MP-Ferroelectric workflow identifies symmetry-related non-polar-to-polar pairs in the Materials Project. It computes Berry-phase polarisations along an adiabatic distortion path connecting the two end states. In this workflow, the electronic polarisation is evaluated with the Berry-phase formalism (VASP implementation) and the ionic part is added from point charges; spin-polarised GGA-PBE(+U) calculations and path-based validation (smoothness and insulating branch) are used to recover a unique spontaneous $\,\mathbf P_s\,$ for each candidate.~\footnote{See Ref.~\onlinecite{smidt-2020-ferrodb}, Secs.\ ``Identifying ferroelectricity from first principles'' and ``Post-processing spontaneous polarisation values''; 255 structure pairs satisfy the ``COMPLETED'' workflow criteria there.}

From this database, we select $\sim$250 materials for which the non-polar and polar endpoints are available and insulating along the path, and include the eight evenly spaced interpolates (fixed cell, linear in fractional coordinates) between endpoints, yielding $\sim$2{,}500 structures in this ferroelectric dataset. For each structure in the training split, we train on the DFT total energy, forces and the Berry-phase polarisation. The Berry-phase polarisations are \emph{pre-folded} onto the polar lattice branch as in Ref.~\onlinecite{smidt-2020-ferrodb}. These folded values supervise the polarisation term in our loss (Sec.~\ref{sec:methods}), while the model itself always predicts $\,\mathbf P\,$ as a derivative of the learned enthalpy. This dataset, therefore, probes a distinct semilocal workflow from MP-Dielectric: MP-Ferroelectric uses path-validated Berry-phase GGA-PBE(+U) calculations, whereas MP-Dielectric uses a separate GGA-PBE DFPT dataset assembled from the Materials Project API. 
\\

More broadly, we note that there is currently no public cross-chemistry dataset that provides energy, forces, stress, Berry-phase $\mathbf P$, $Z^*$, and $\boldsymbol\alpha$ for all materials from a single workflow, so our multihead setting should be interpreted as learning across closely related but not identical semilocal references.

\subsection{Finite-temperature MD trajectory sets: \ce{BaTiO3} and $\alpha$-\ce{SiO2}}

To validate that \texttt{MACE-Field} works as intended, and to benchmark against the unified electric-enthalpy study \texttt{Allegro-pol}, we also train two \texttt{MACE-Field} models \emph{exclusively} on finite-temperature MD trajectory datasets for tetragonal \ce{BaTiO3} (300~\si{K}) and $\alpha$-quartz (\ce{SiO2}, 300~\si{K})~\cite{allegro-pol-2025}.

Each dataset consists of time-ordered frames with per-configuration labels for total energy, forces, virial stress, Berry-phase polarisation, Born effective charges, and electronic polarisability. In contrast to the cross-chemistry setting above, these single-material datasets are internally fully consistent: all six labels are generated within one common electronic-structure workflow for a given material, following the unified-enthalpy dataset construction of Ref.~\onlinecite{allegro-pol-2025}. 

\begin{figure}[t]
  \centering
  \includegraphics[width=0.49\linewidth]{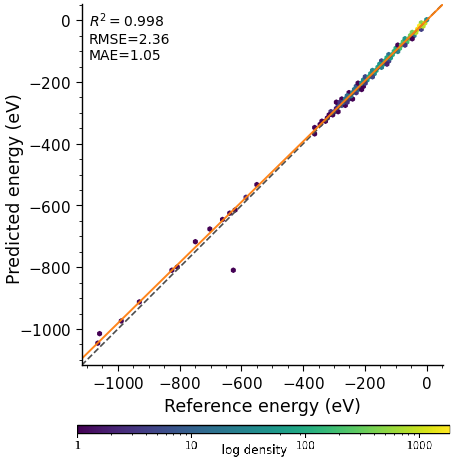}
  \hfill
  \includegraphics[width=0.49\linewidth]{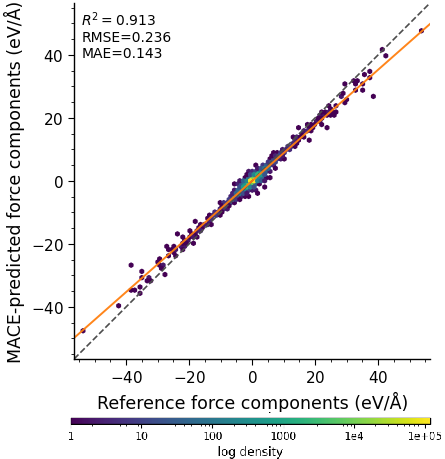}
  \\[0.01em]
  \includegraphics[width=\linewidth]{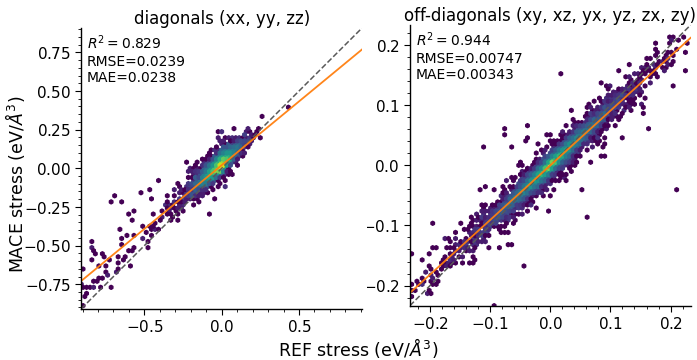}
  \caption{\textbf{OMAT-PBE replay-set parity for \texttt{MACE-Field-MH-0}.} Parities on the 10{,}000-structure OMAT-PBE replay subset used for multihead fine-tuning. Colour bars show the log number of data points. \textbf{Top-left:} Energy parity (eV). \textbf{Top-right:} Forces parity with all components combined (eV/\AA). \textbf{Bottom:} Stress parity for diagonal components (left) and off-diagonal components (right) (eV/\AA$^3$).}
  \label{fig:replay_parity}
\end{figure}

These datasets serve a dual role: (i) as a demonstration of supervised training for multi-property learning on single-material manifolds, and (ii) as the basis for our finite-field MLMD validations (Sec.~\ref{sec:results}). Specifically, for \ce{BaTiO3} we reproduce polarisation hysteresis at 0~\si{K}, extracting coercive fields and remanent $\,\mathbf P$; for $\alpha$-\ce{SiO2} we run long 300~\si{K} trajectories ($\sim$200~\si{ps}) and compute infrared absorption from the $\dot{\mathbf P}$-$\dot{\mathbf P}$ autocorrelation and Raman intensities from the $\boldsymbol\alpha$-$\boldsymbol\alpha$ autocorrelation, assembling the complex dielectric function $\boldsymbol\varepsilon(\omega)$ from time-domain response.

\section{Results}\label{sec:results}

\begin{figure}[t]
  \centering
  \includegraphics[width=\linewidth]{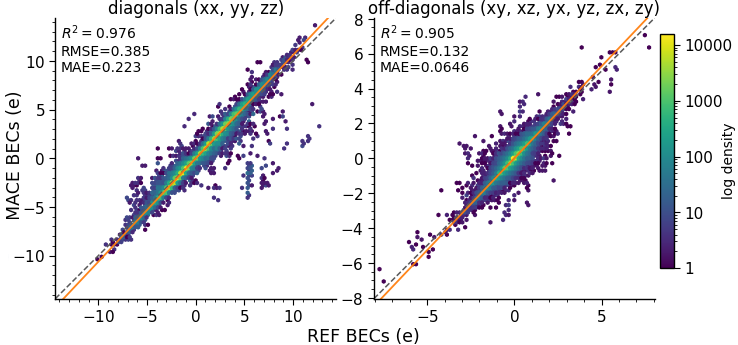}
  \\[0.01em]
  \includegraphics[width=\linewidth]{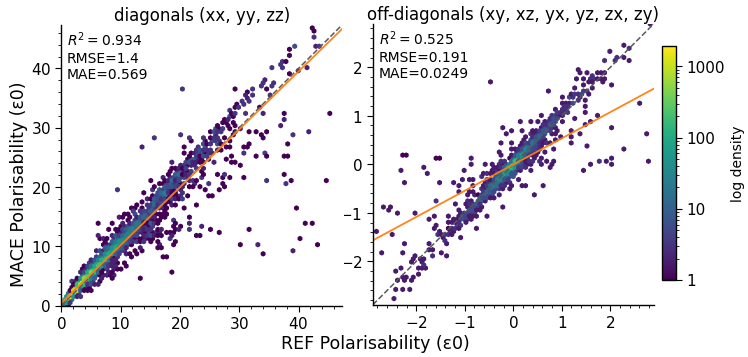}
\caption{\textbf{MP-Dielectric parity for \texttt{MACE-Field-MH-0}.} Parities for Born effective charges and electronic polarisabilities from the MP-Dielectric dataset (train, validation and test combined). Colour bars show the log number of data points. \textbf{Top:} Born effective charge tensor ($e$) parity for diagonal (left) and off-diagonal (right) components. \textbf{Bottom:} Polarisability tensor ($\varepsilon_0$) parity for diagonal (left) and off-diagonal (right) components.}
\label{fig:dielectric_parity}
\end{figure}

We organise our results in three parts. First, we show that fine-tuning a \texttt{MACE} foundation model on the MP-Dielectric and MP-Ferroelectric datasets yields a field-aware foundation, \texttt{MACE-Field-MH-0}, that preserves its original energy/force accuracy while acquiring transferable predictions of Berry-polarisation, Born effective charges, polarisabilities, and derived refractive indices across chemistry. Second, we use the MP-Ferroelectric dataset to compare a directly trained cross-chemistry polarisation model with \texttt{MACE-Field-MH-0} for predicting Berry-phase branch structure and spontaneous polarisation, and find comparable accuracy. Third, we demonstrate finite-field molecular dynamics for directly trained single-material models and \texttt{MACE-Field-MH-0} to diagnose the limits of the foundation model. Unless stated otherwise, all quantitative errors in this section are reported against the corresponding semilocal DFT/DFPT labels rather than experiment or higher-level electronic-structure benchmarks.

\begin{figure}[t]
\centering
\includegraphics[width=0.49\linewidth]{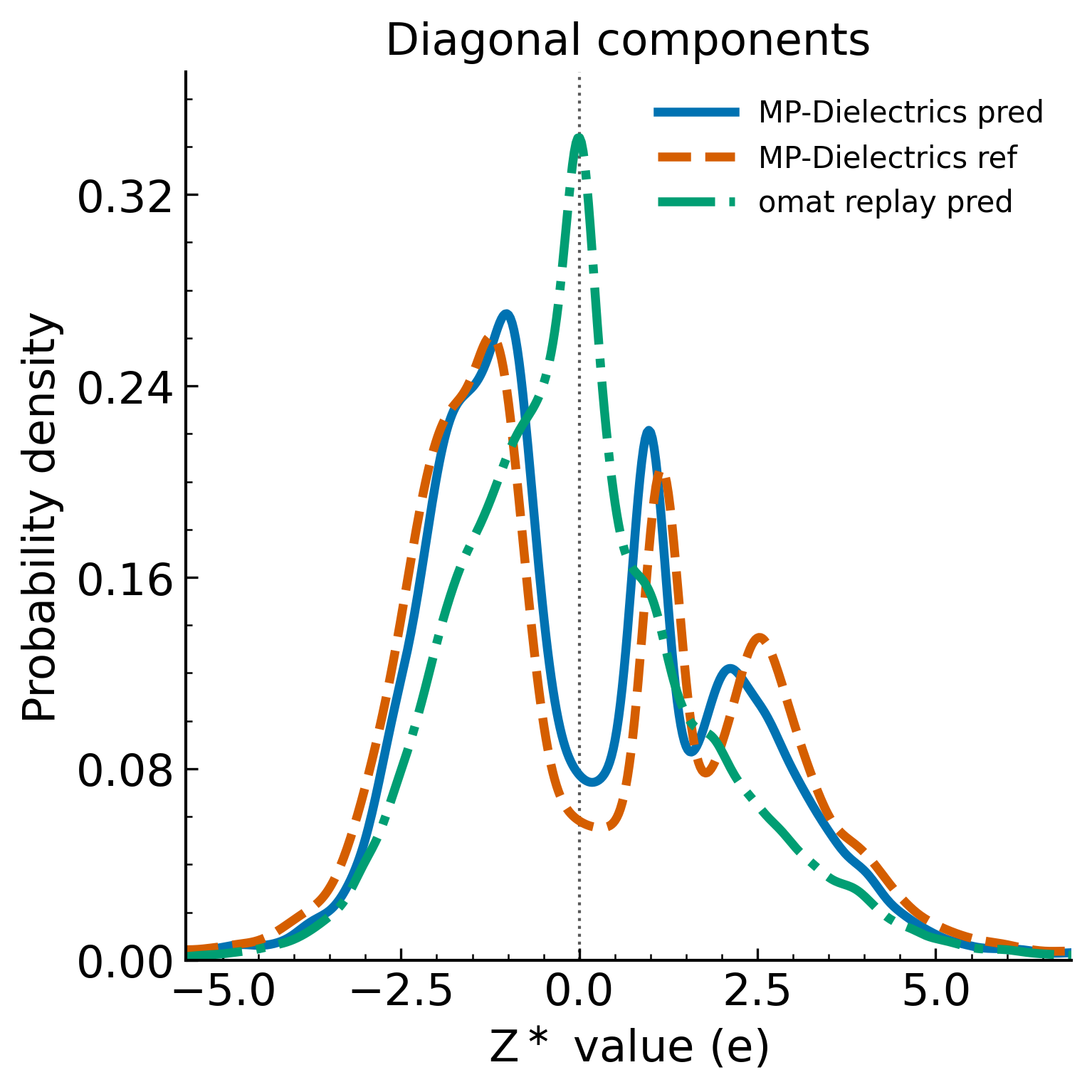}
\hfill
\includegraphics[width=0.49\linewidth]{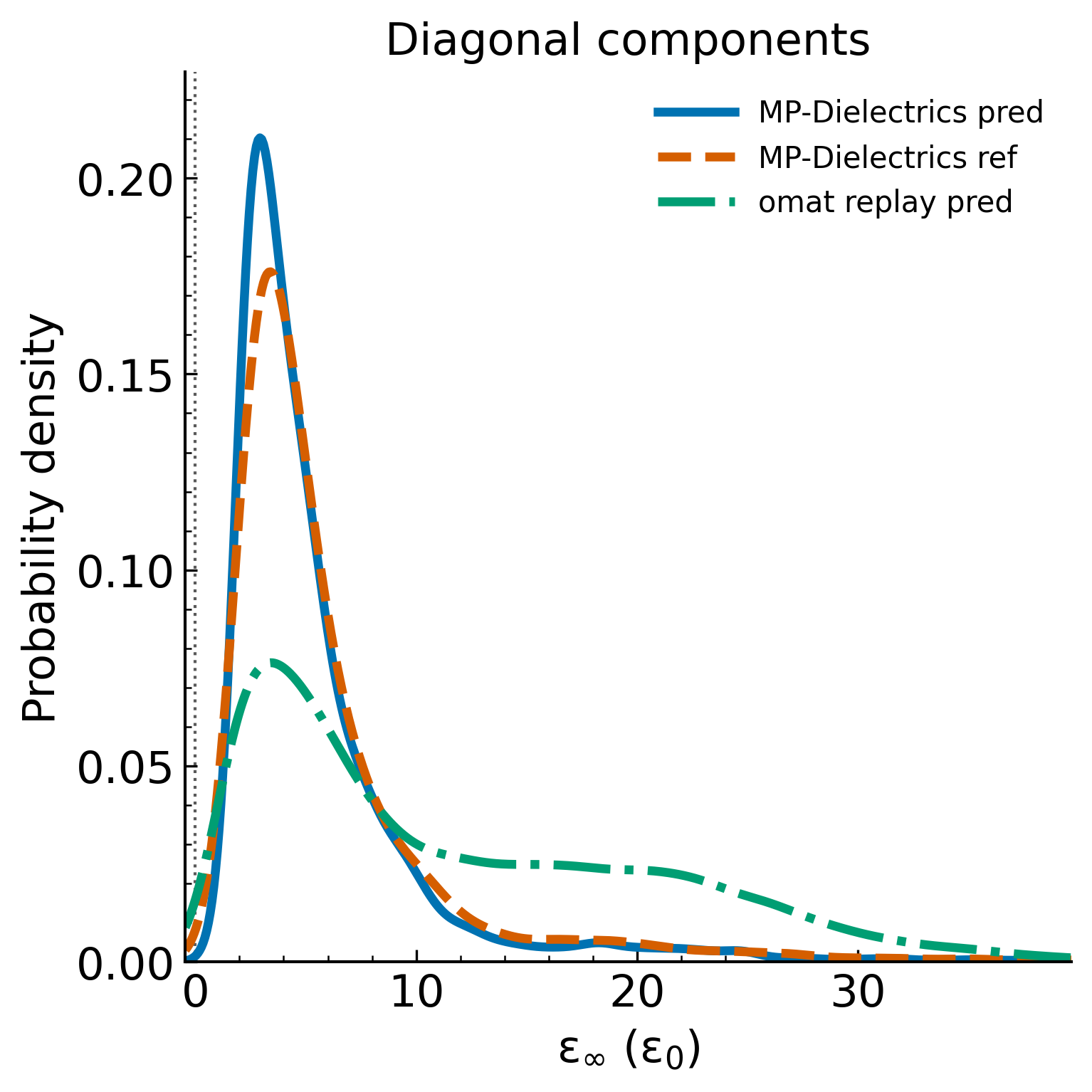}
\caption{\textbf{Density distribution of BECs and electronic dielectric constants.} KDE density curves for the Born effective charges $Z^*$ (left) and the high-frequency dielectric constants $\varepsilon_{\infty}$ (right). \textbf{Solid blue:} distribution predicted by the fine-tuned \texttt{MACE-Field} model on the MP-Dielectric dataset. \textbf{Dashed orange:} distribution from the reference DFPT MP-Dielectric dataset. \textbf{Dotted green:} distribution predicted by the fine-tuned \texttt{MACE-Field} model on the sub-selected OMAT-PBE replay subset. }
\label{fig:bec_alpha_kde}
\end{figure}

\subsection{Field-aware foundation model: \texttt{MACE-Field-MH-0}}
\label{subsec:fine-tuning_foundation}

\subsubsection{Multihead fine-tuning and parities}

We start from the recent multihead \texttt{MACE} foundation \texttt{mace-mp-mh-0} from the broader \texttt{MACE} foundation-model family~\cite{batatia2025foundationmodelatomisticmaterials}, and use its OMAT-PBE head as the energy/force/stress prior for our field-aware model. We then enable the \texttt{MACE-Field} coupling blocks while keeping the backbone and readout unchanged. For this ``\texttt{MACE-Field-MH-0}'' model, we perform joint multihead fine-tuning supervised on: a ``mp-dielectric'' head with DFPT Born effective charges and electronic polarisabilities from the MP-Dielectric dataset (Sec.~\ref{sec:mp_dielectric}); a ``mp-ferroelectric'' head with DFPT polarisations from the MP-Ferroelectric dataset (Sec.~\ref{sec:mp_ferroelectric}); and a replay/pretrained head with energies, forces and stresses from a 10{,}000-structure OMAT-PBE replay subset associated with the \texttt{mace-mp-mh-0/1} family. This joint dielectric+ferroelectric+replay protocol exploits multihead fine-tuning to combine partially heterogeneous datasets while preserving the accuracy of the underlying \texttt{mace-mp-mh-0} foundation model on its original training distribution.

Fig.~\ref{fig:replay_parity} shows parity plots on the OMAT-PBE replay subset, which remain tightly clustered around the diagonal for energies, forces (all components combined), and both diagonal and off-diagonal stress components, with some increase in variance but no visible systematic bias. In other words, adding field-coupling layers and supervising on dielectric labels does not detrimentally degrade the core potential, so \texttt{MACE-Field-MH-0} retains a high-quality general-purpose force field after fine-tuning.

Turning to dielectric response, Fig.~\ref{fig:dielectric_parity} reports parities for individual BEC tensor components and electronic polarisabilities on the MP-Dielectric dataset. \texttt{MACE-Field-MH-0} reproduces DFPT $Z^*$ and $\alpha$ across chemistry with near-diagonal trends and narrow scatter, including both diagonal and off-diagonal entries. This indicates that a single electric-enthalpy functional $\mathcal{F}(\mathbf{R},\mathbf{E})$, inherited from a foundation model and jointly fine-tuned on dielectric and ferroelectric response data, is sufficient to capture linear response over a broad chemical space.

\begin{figure}[t]
\centering
\includegraphics[width=\linewidth]{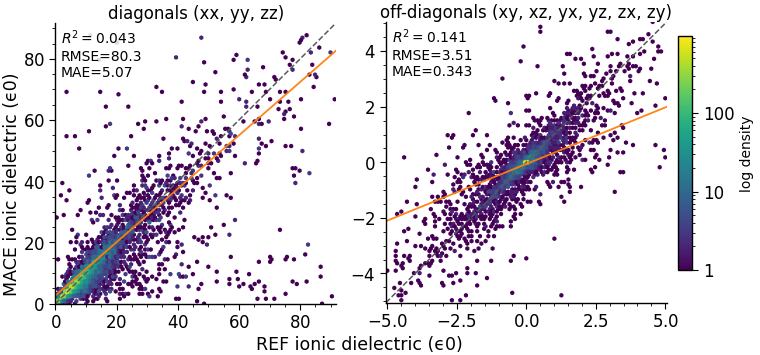}
\\[0.01em]
\includegraphics[width=\linewidth]{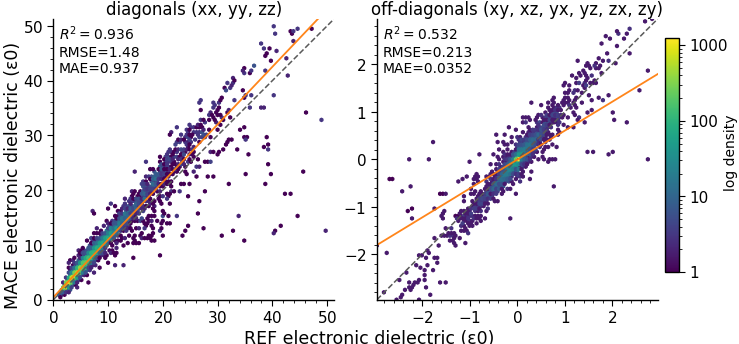}
\caption{\textbf{MP-Dielectric parity for predicted ionic and electronic dielectric constants using \texttt{MACE-Field-MH-0}.} \textbf{Top:} Parity for the diagonal (left) and off-diagonal (right) components of the ionic dielectric constant. \textbf{Bottom:} Parity for the diagonal (left) and off-diagonal (right) components of the electronic dielectric constant.}
\label{fig:eps_parity}
\end{figure}

To probe distributional behaviour, Fig.~\ref{fig:bec_alpha_kde} compares kernel-density estimates of the BEC and polarisability/electronic-dielectric components predicted on (i) the OMAT-PBE replay subset and (ii) the MP-Dielectric structures, against the reference DFPT distributions. The three curves closely overlap for both BECs and electronic dielectrics/polarisabilities, with the \texttt{MACE-Field-MH-0} model neither collapsing toward overly narrow peaks nor developing spurious heavy tails. This suggests that the inherited foundation prior, combined with moderate DFPT supervision, is enough to reconstruct realistic response distributions even on structures for which DFPT labels were never provided (the replay-only OMAT-PBE configurations). Notably, \texttt{MACE-Field-MH-0} can capture the small or zero BECs of many atoms in neutral, bulk-like insulating crystal structures present in this OMAT-PBE replay set, which are largely absent in the MP-Dielectric dataset. In practical terms, this means that a single model, trained once, can be used as a general-purpose force field that also supplies consistent dielectric response across tens of thousands of inorganic structures, enabling workflows where dielectric properties are treated on the same footing as energies and forces.

\subsubsection{Direct prediction of dielectric constants}
\label{sec:direct_eps}

As \texttt{MACE-Field-MH-0} can predict both Born effective charges and polarisability, we investigate whether we may combine this with analytical Hessians in \texttt{MACE} to evaluate the electronic and ionic dielectric constants directly, without supervising on dielectric tensors themselves.

Concretely, we first obtain the high-frequency (purely electronic) dielectric tensor directly from the predicted polarisability using Eq.~\eqref{eq:eps_from_alpha}; the electronic contribution is a straightforward algebraic post-processing of a primary model output.

The ionic contribution requires vibrational information. Here we exploit recent work on analytic Hessians for MACE-type models~\cite{Gonnheimer2025}, which provides second derivatives of the energy with respect to atomic displacements at negligible additional cost compared to ordinary force evaluation. The ionic dielectric constant is obtained from the \texttt{MACE-Field-MH-0} BECs and the corresponding \texttt{mace-mp-mh-0} foundation model Hessians following Sec.~\ref{sec:dielectric_math}.

\begin{figure}[t]
\centering
\includegraphics[width=0.8\linewidth]{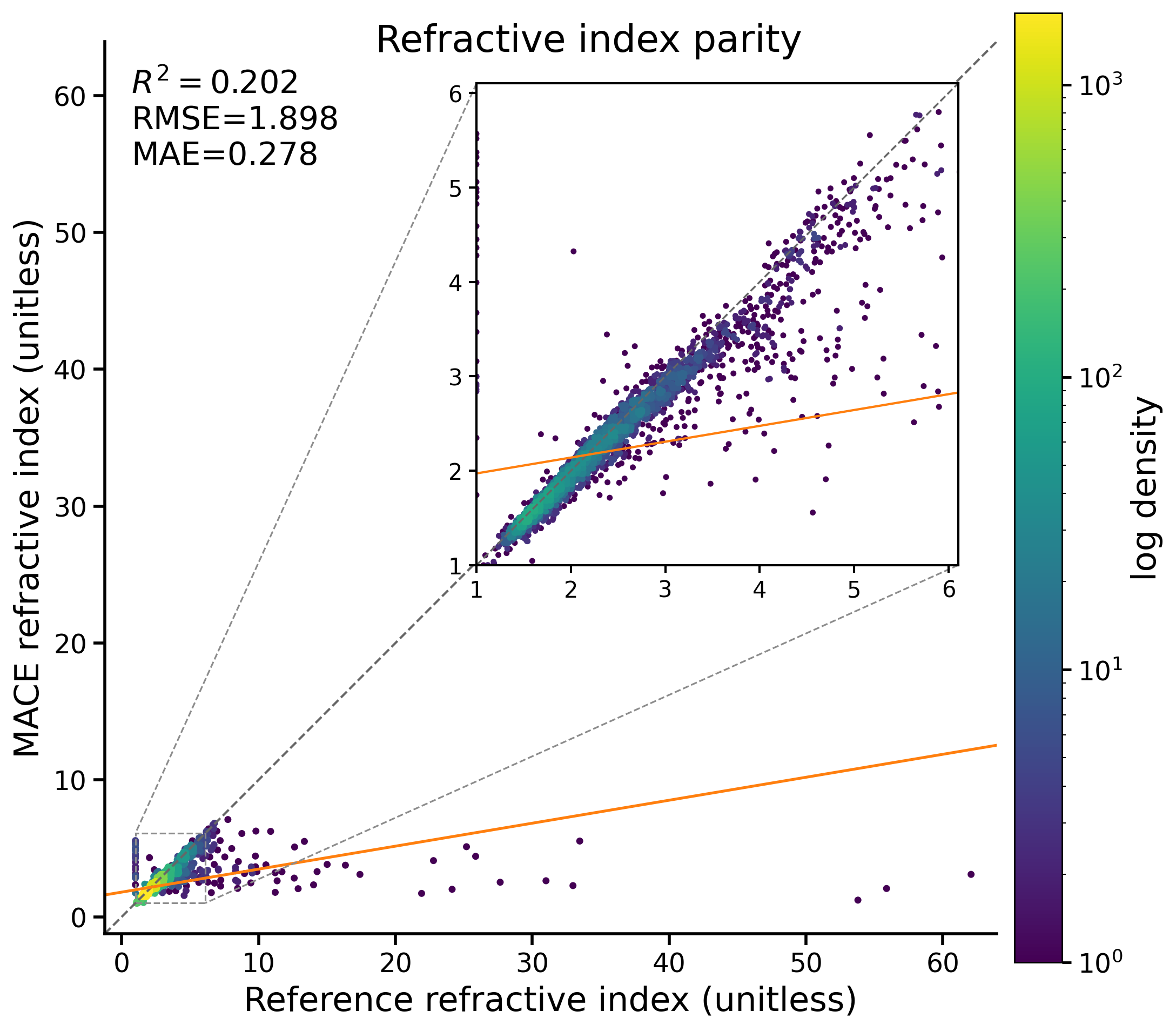}
\caption{\textbf{Matbench dielectric refractive-index benchmark for the fine-tuned \texttt{MACE-Field} foundation model.} Refractive-index parity on the Matbench v0.1 test interface, obtained by deriving $n$ from the learned \texttt{MACE-Field-MH-0} $\boldsymbol\alpha$ predictions. Because both Matbench and MP-Dielectric are derived from Materials Project data, this comparison should be interpreted with caution and is not a strict out-of-distribution benchmark. Still, it nevertheless provides a useful external point of comparison for derived optical properties. The colour bar shows the logarithm of the number of data points. \textbf{Inset:} 99\% percentile of data points.}
\label{fig:refractive_index_parity}
\end{figure}

Figure~\ref{fig:eps_parity} summarises the parity between the DFPT reference dielectric tensors and the \texttt{MACE-Field-MH-0} predictions for both ionic and electronic contributions (diagonal and off-diagonal components). The electronic dielectric constants closely follow the DFPT values, as expected from the strong agreement already observed for $\boldsymbol{\alpha}$. By contrast, the ionic dielectric constants, while still broadly following a near-diagonal trend, exhibit substantially larger scatter. This is not surprising: the ionic dielectric tensor depends on both $Z^*$ and the phonon spectrum via Eq.~\eqref{eq:eps_ion_static}, and is therefore sensitive to errors in both the BECs and the Hessian. Because the ionic dielectric response involves the \emph{inverse} of the Hessian, any residual bias in curvature inherited from the foundation prior is amplified in $\boldsymbol{\varepsilon}_{\mathrm{ion}}$.

\subsubsection{Matbench dielectric benchmark}

As an external benchmark, we test \texttt{MACE-Field-MH-0} on the Matbench dielectric task~\cite{Dunn2020}. Using the learned $\alpha$ to construct $\varepsilon_\infty$ and $n$ (see Eqs.~(\ref{eq:eps_from_alpha}), (\ref{eq:eps_poly}) and~(\ref{eq:refractive_index})), we obtain a reasonable refractive-index parity on the Matbench v0.1 test set (Fig.~\ref{fig:refractive_index_parity}), with an RMSE that remains competitive with strong direct-property predictors such as \texttt{MODNet}~\cite{DeBreuck2021} and \texttt{AnisoNet}~\cite{anisonet-2024}. Importantly, this comparison is obtained without directly training on Matbench labels: the same interatomic model is used ``as is'' to predict $\alpha$ and then derive $\varepsilon_\infty$ and $n$. At the same time, it should not be interpreted as a strict out-of-distribution optical benchmark, because \texttt{matbench\_dielectric} is itself derived from Materials Project dielectric data. Structural matching between our filtered MP-Dielectric set and the Matbench payload identifies 2{,}842 common structures, corresponding to 53.8\% of the MP-Dielectric entries used here and 60.7\% of the 4{,}764 Matbench records. We therefore interpret Fig.~\ref{fig:refractive_index_parity} as a useful external comparison point for derived refractive indices, while noting that the substantial overlap and the sensitivity of this derived quantity to dataset filtering and label-protocol differences warrant caution in over-interpreting the benchmark.

\subsubsection{Error and outlier analysis}

\begin{figure}[t]
\centering
\includegraphics[width=\linewidth]{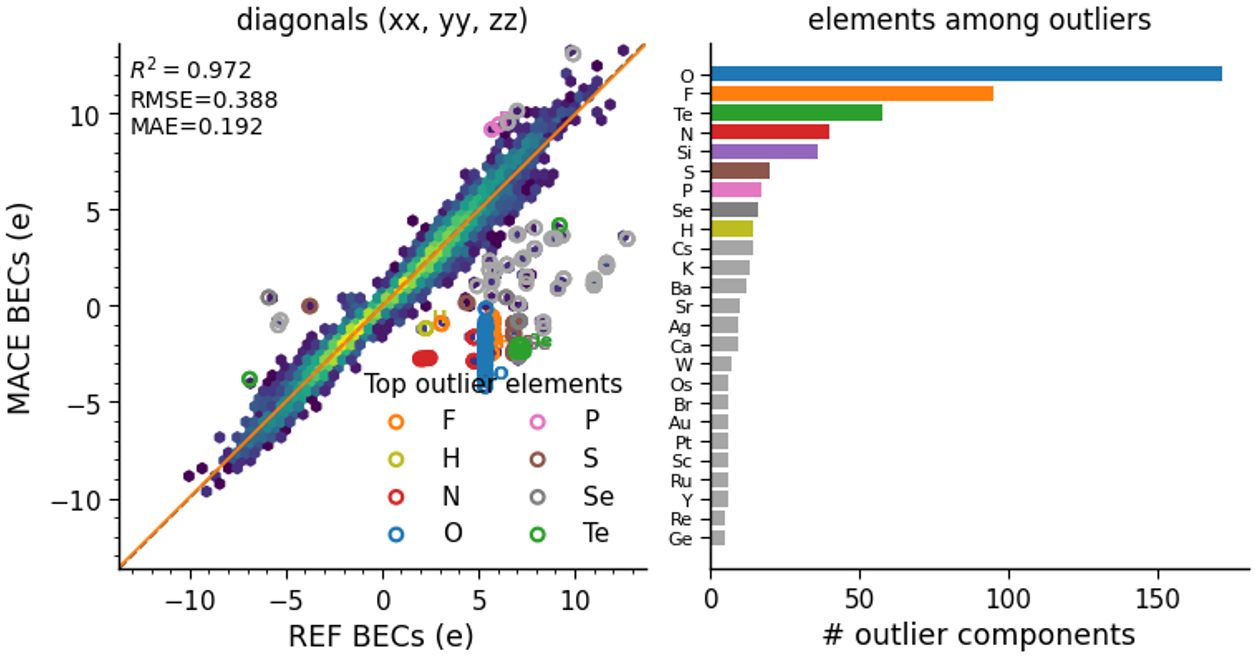}
\caption{\textbf{Error analysis of outliers in BEC predictions.} \textbf{Left:} Top 1\% most erroneous BEC predictions and their corresponding chemical species. \textbf{Right:} Bar chart showing which elements most frequently appear in these outliers.}
\label{fig:bec_error_dist}
\end{figure}

In Fig.~\ref{fig:bec_error_dist}, we summarise the error distribution for BEC components, highlighting that large deviations are confined to a small number of outliers, predominantly reactive non-metallic elements such as oxygen and metalloids like tellurium, which typically exhibit negative BECs but are positive in the MP-Dielectric dataset. In Table~\ref{tab:bec-outliers} we list the top ten most erroneous BEC predictions by \texttt{MACE-Field-MH-0} compared to the DFPT reference. Notably, the majority of mismatches involve elements that would typically have negative BECs. Worse, the DFPT BECs violate the acoustic sum rule and are clearly incorrect. Despite these outliers, \texttt{MACE-Field-MH-0} obeys the ASR by construction and so predicts sensible BECs for these structures, which explains the disparity. This illustrates a pragmatic benefit of the physics-informed setup: by enforcing charge-neutrality and the ASR at the model level, we can, in some cases, ``repair'' inconsistent training labels and obtain more physically plausible predictions than those from the raw DFPT dataset.

\begin{table}[t]
  \centering
  \begingroup
\makeatletter
\@ifundefined{DIFaddFL}{}{\renewcommand{\DIFaddFL}[1]{#1}}
\@ifundefined{DIFdelFL}{}{\renewcommand{\DIFdelFL}[1]{}}
\@ifundefined{DIFaddbeginFL}{}{\renewcommand{\DIFaddbeginFL}{}}
\@ifundefined{DIFaddendFL}{}{\renewcommand{\DIFaddendFL}{}}
\@ifundefined{DIFdelbeginFL}{}{\renewcommand{\DIFdelbeginFL}{}}
\@ifundefined{DIFdelendFL}{}{\renewcommand{\DIFdelendFL}{}}
\makeatother
\begin{tabular}{l l c c l c c l c c}
\hline\hline
\multicolumn{1}{c}{} & \multicolumn{3}{c}{$1^{st}$ Element} & \multicolumn{3}{c}{$2^{nd}$ Element} & \multicolumn{3}{c}{$3^{rd}$ Element} \\
\hline
Material & El & Ref & Pred & El & Ref & Pred & El & Ref & Pred \\
\hline
\ce{Y2Ag2Te4} & Te & 7.171 & -2.366 & Y & 8.915 & 3.486 & Ag & 5.692 & 1.247 \\
\ce{K2Au2Se4} & Se & 7.119 & -2.615 & K & 7.515 & 1.524 & Au & 7.082 & 3.707 \\
\ce{Sm4Ta4O16} & O & 5.362 & -2.751 & Sm & 9.361 & 3.656 & Ta & 7.648 & 7.349 \\
\ce{Ca2SnS4} & S & 6.824 & -2.402 & Ca & 7.356 & 2.498 & Sn & 6.491 & 4.610 \\
\ce{La4Ta4O16} &  La & 12.74 & 3.486 & O & 5.364 & -2.937 & Ta & 7.648 & 8.261 \\
\ce{CdPt3O6} & O & 5.392 & -1.831 & Pt & 6.941 & 2.572 & Cd & 6.143 & 3.270 \\
\ce{MgPt3O6} & O & 5.385 & -2.032 & Pt & 6.947 & 3.148 & Mg & 3.821 & 2.748 \\
\ce{Bi2O2F2} & O & 5.389 & -3.607 & F & 5.731 & -1.786 & Bi & 6.588 & 5.393 \\
\ce{Tm4Ta4O16} & O & 5.359 & -2.933 & Tm & 4.936 & 3.356 & Ta & 7.648 & 8.376 \\
\ce{ZnPt3O6} & O & 5.392 & -1.837 & Pt & 6.947 & 2.744 & Zn & 3.840 & 2.791 \\
\hline\hline
\end{tabular}
\endgroup

  \caption{\textbf{Top 10 BEC outliers by material.} Largest erroneous BEC predictions (unit of \si{e}) by the fine-tuned \texttt{MACE-Field} model. The first listed element has the largest difference between the MP-Dielectric reference values and the \texttt{MACE-Field} prediction. The second and third elements' BECs in the material are then listed. Notably, all of these outliers correspond to clearly incorrect BECs in the MP-Dielectric dataset; this is most evident from the fact that they do not obey the acoustic sum rule (ASR). The \texttt{MACE-Field} predictions do obey the ASR and give sensible BECs.}
  \label{tab:bec-outliers}
\end{table}

\subsection{Polarisation across distortion paths and spontaneous $P_s$}
\label{subsection:polarisation_branches}

In this experiment, we assess the capability of \texttt{MACE-Field} to learn ferroelectric polarisation branches from the MP-Ferroelectric distortion-path dataset (Sec.~\ref{sec:mp_ferroelectric}). 

\begin{figure}[t]
  \centering
    \includegraphics[width=0.48\linewidth]{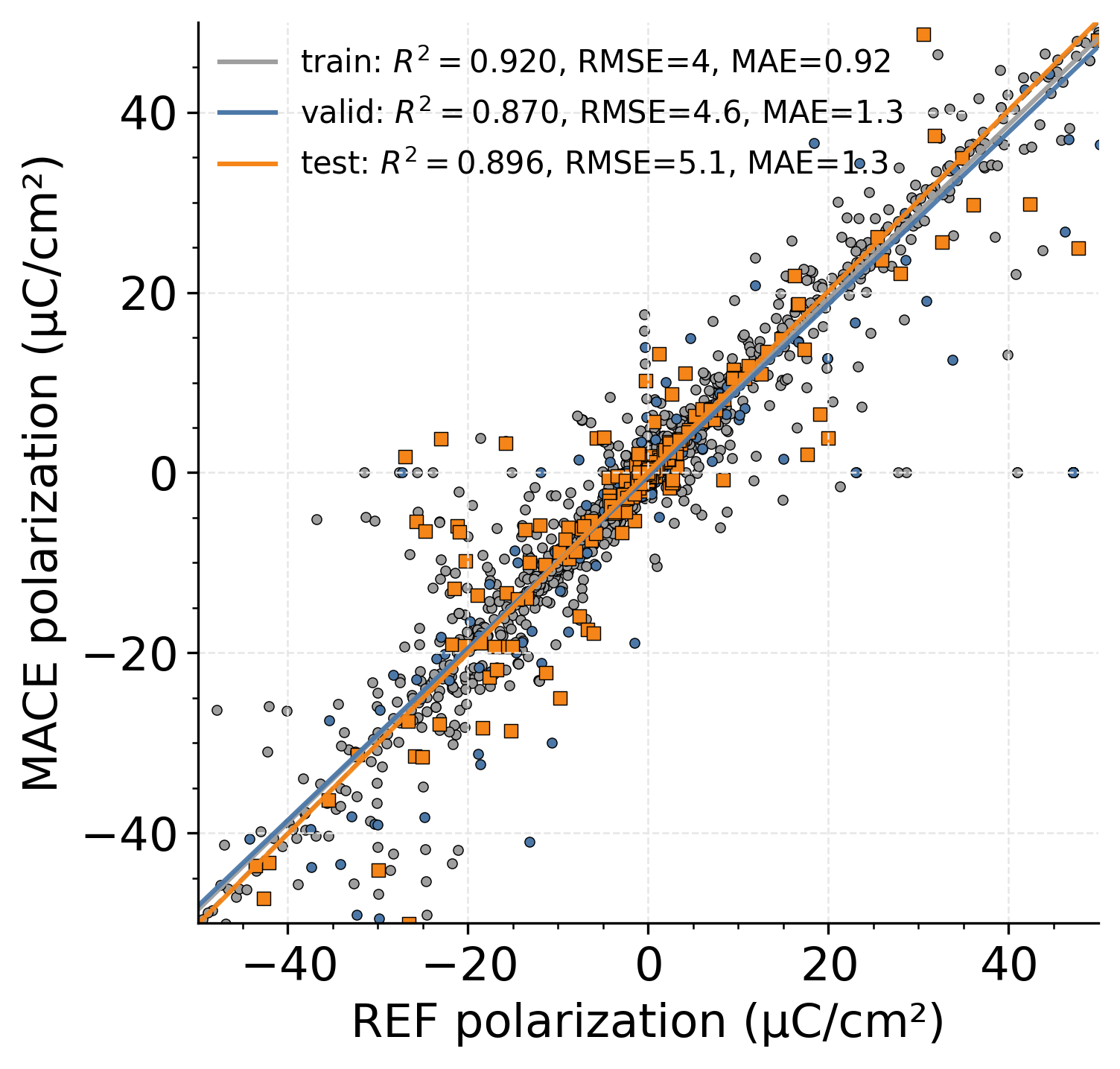}
    \hfill
    \includegraphics[width=0.48\linewidth]{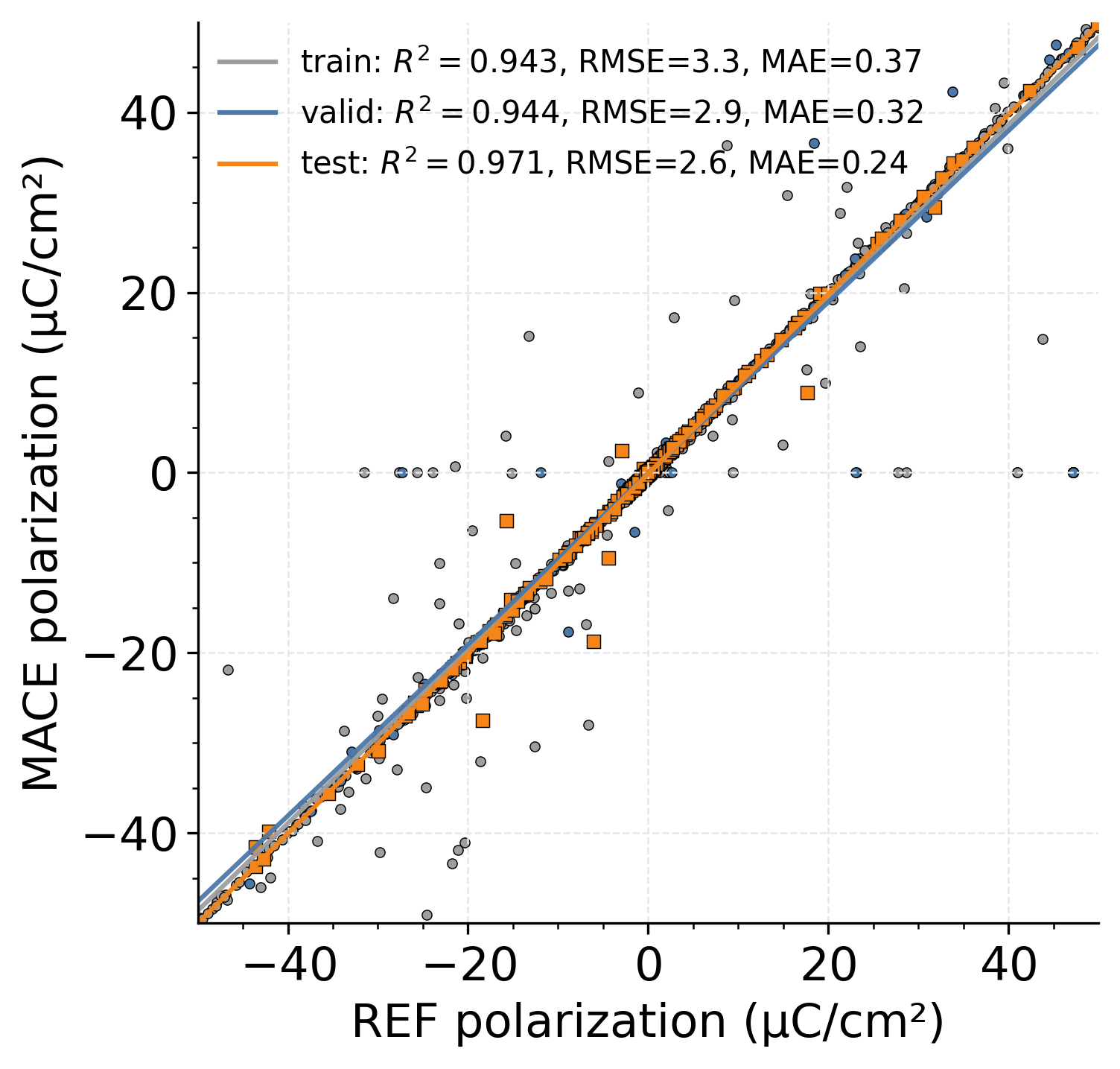} 
  \\[0.01em]
    \includegraphics[width=0.48\linewidth]{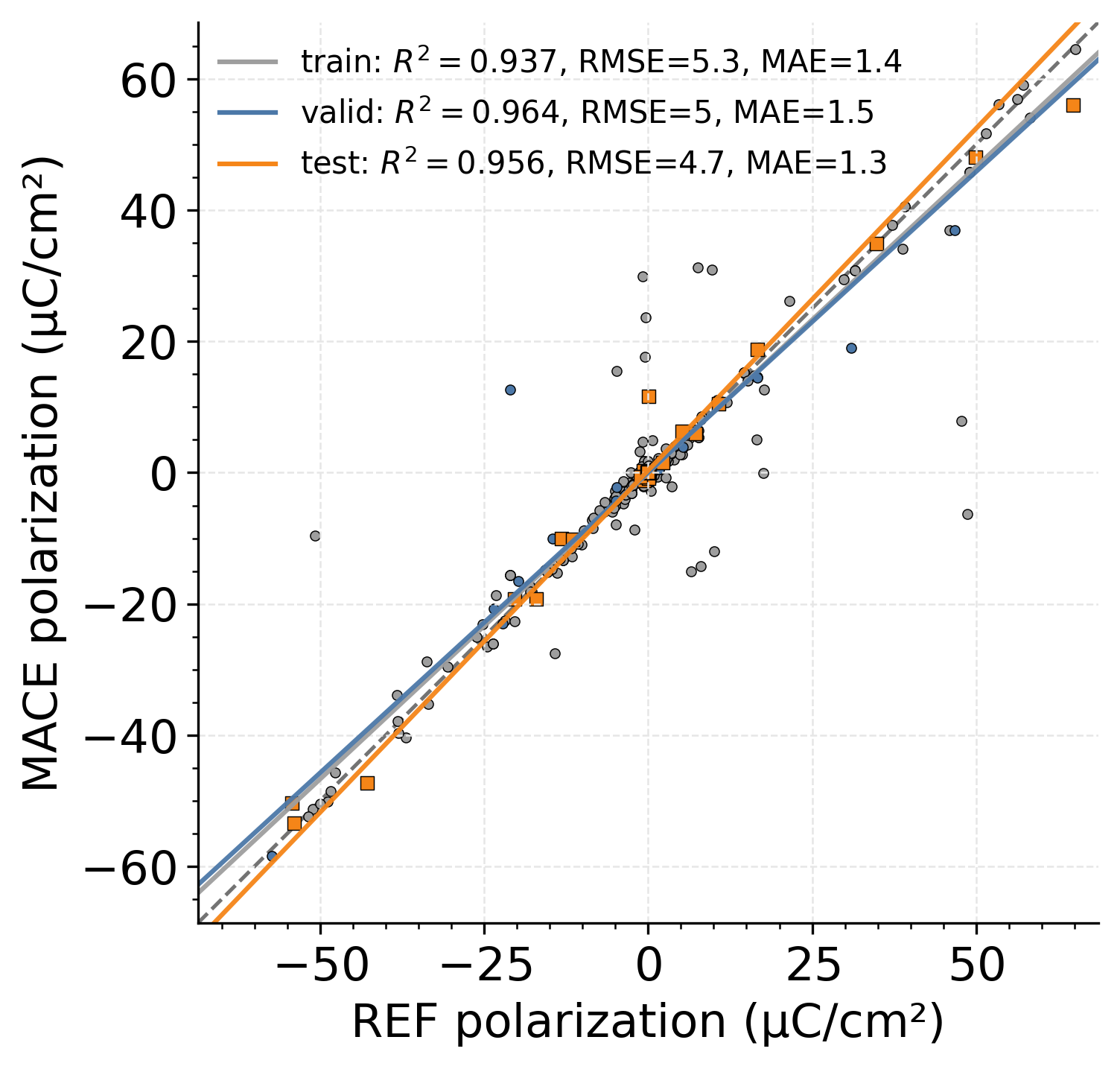}
    \hfill
    \includegraphics[width=0.48\linewidth]{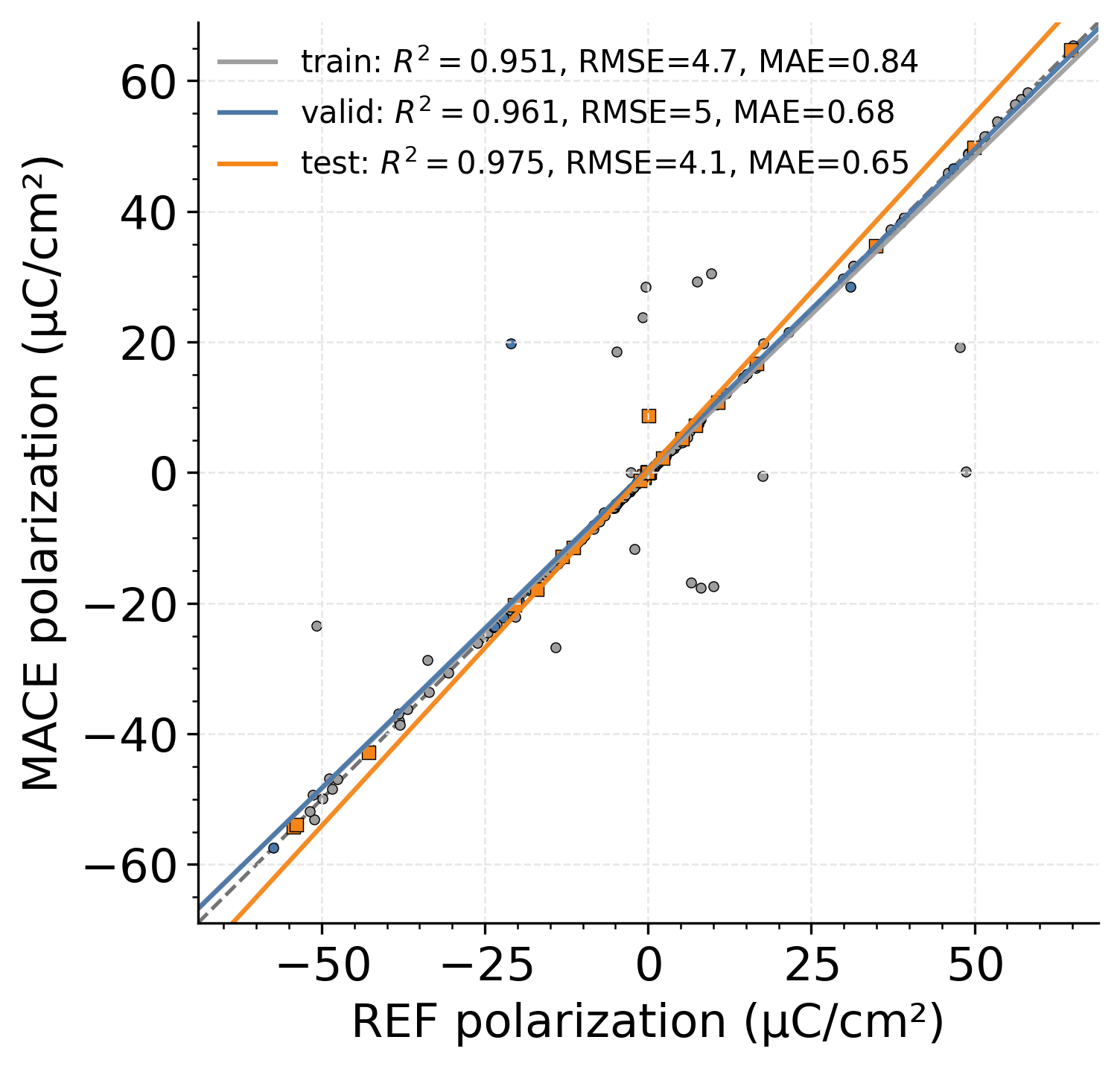}
  \caption{\textbf{Polarisation parity comparisons.}
  \textbf{Top:} Component-wise parity between DFPT Berry-phase polarisation and \texttt{MACE-Field} predictions over all materials and path frames; per-panel $R^2$, RMSE, and MAE are annotated. (Left) Trained-from-scratch model. (Right) \texttt{MACE-Field-MH-0}.
  \textbf{Bottom:} Parity of spontaneous polarisation $P_s$ (folded endpoint difference) on the training (grey circles), validation (blue circles) and test set (orange squares); each point is one material. (Left) Trained-from-scratch model. (Right) \texttt{MACE-Field-MH-0}.}
  \label{fig:pol-parity}
  \label{fig:spol-parity}
\end{figure}

Figure~\ref{fig:spol-parity} compares DFPT Berry-phase labels with the \emph{learned} polarisations obtained as 
$\mathbf P=-\Omega^{-1}\partial\mathcal F/\partial\mathbf E$ from (left) \texttt{MACE-Field} trained directly on polarisation labels and (right) from \texttt{MACE-Field-MH-0} (see Sec.~\ref{subsec:fine-tuning_foundation}). $R^2$, RMSE and MAE are reported for each model for the training (grey circles), validation (blue circles) and testing (orange squares) sets. Both models capture polarisations from all materials and distortion paths well, but the jointly fine-tuned OMAT-based \texttt{MACE-Field-MH-0} model is slightly stronger on the folded branch-resolved task overall.

\begin{figure}[t]
  \centering
    \includegraphics[width=0.48\linewidth]{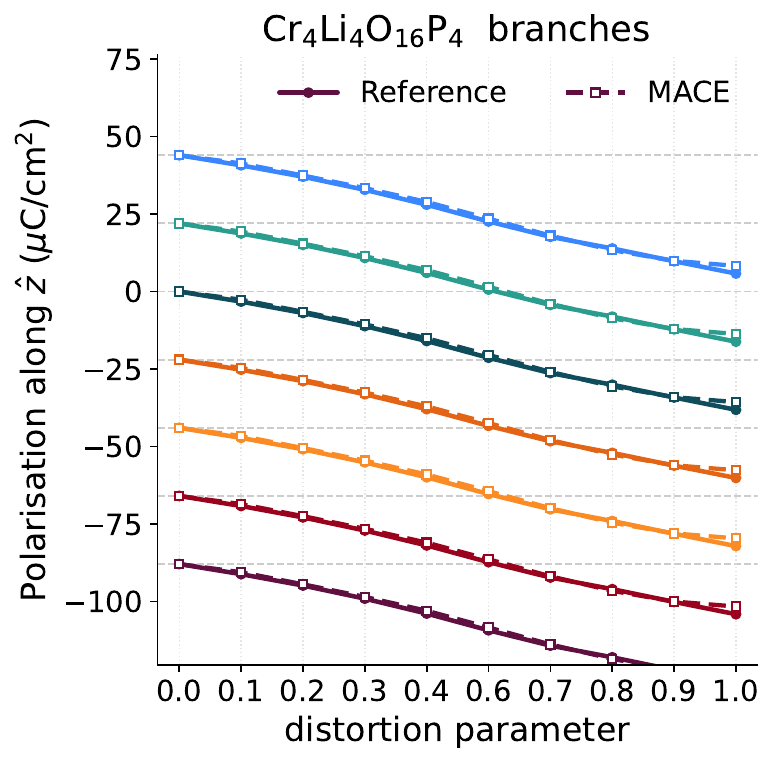}
    \label{fig:pol-branches:a}
  \hfill
    \includegraphics[width=0.48\linewidth]{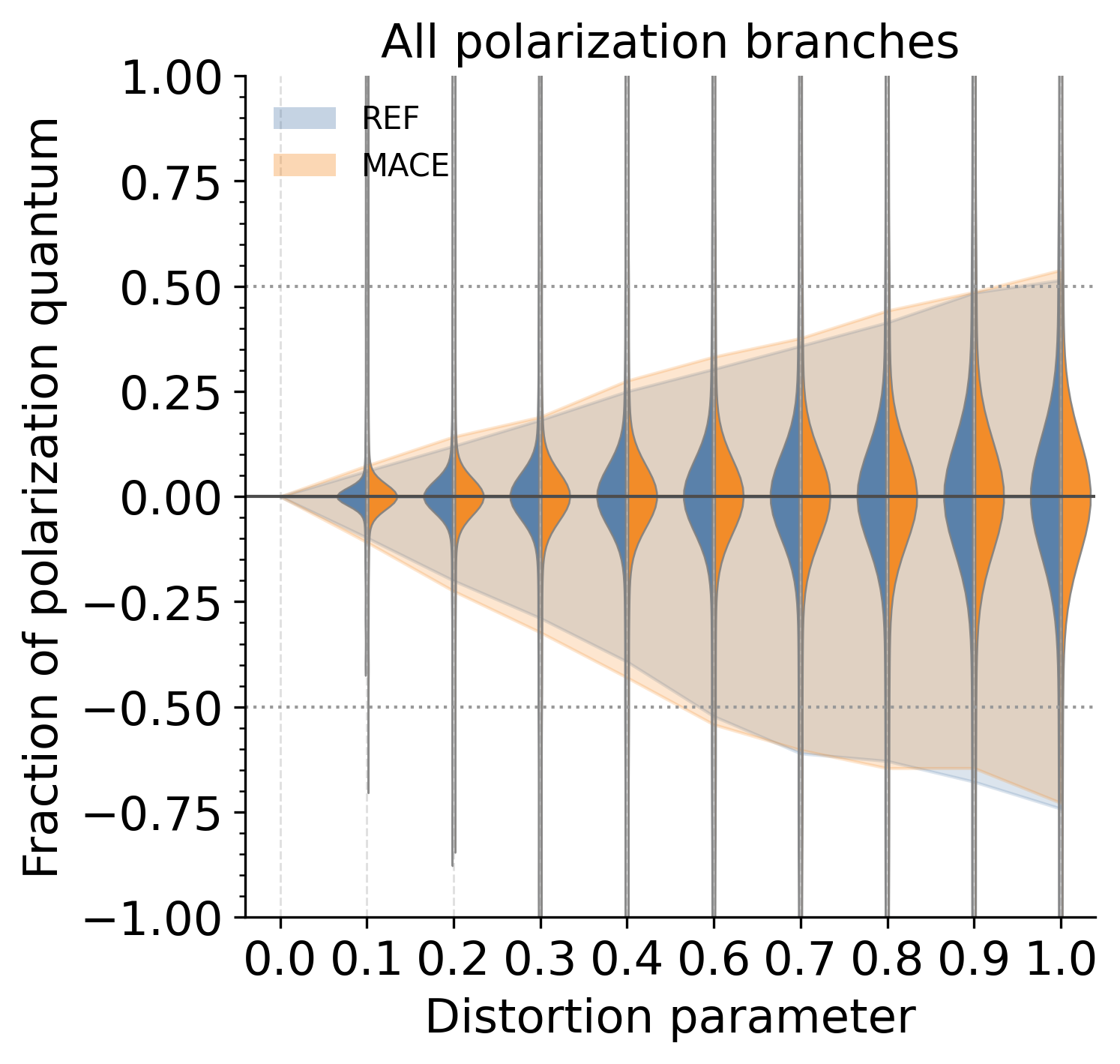}
    \label{fig:pol-branches:b}
  \caption{\textbf{Polarisation across distortion paths.}
  \textbf{Left:} Example polarisation branches for \ce{Cr4Li4O16P4} from the test set; reference (solid) and \texttt{MACE-Field-MH-0} (dashed) curves align across the path.
  \textbf{Right:} Branch-invariant ``fan'' plot: for each Cartesian component, polarisation as a fraction of the corresponding polarisation quantum vs distortion parameter. Reference DFPT distributions (blue) and \texttt{MACE-Field-MH-0} (orange) show close agreement; shaded regions indicate 95\% percentiles.}
  \label{fig:pol-branches}
\end{figure}

The spontaneous polarisation is obtained as the difference in polarisation between the polar and non-polar reference structures at the endpoints of a given distortion path. Because the polarisation is multivalued modulo a quantum, all polarisations along a path must first be mapped onto a common branch to define this difference unambiguously. The parities in Fig.~\ref{fig:spol-parity} (one point per material; grey circles for the training set, blue circles for the validation set, and orange squares for the test set) compare DFPT reference values with \texttt{MACE-Field} predictions from (left) the directly trained model and (right) \texttt{MACE-Field-MH-0}. Consistent with the pointwise parities, the OMAT-based foundation model reproduces the spontaneous-polarisation distribution at least as well as, and slightly better than, the directly trained cross-chemistry model.

To illustrate the branch structure, Fig.~\ref{fig:pol-branches} (left) shows the $\hat{z}$ Cartesian polarisation branches for \ce{Cr4Li4O16P4} from the test set. The \texttt{MACE-Field-MH-0} foundation model reproduces both the magnitude and slope of the Berry-phase branches and does so without any spurious branch hopping. Figure~\ref{fig:pol-branches} (right) then provides a global view of branch behaviour across the full dataset (train, validation, and test combined): it shows a combined violin and ``fan'' representation (shaded bands indicating the 95\% percentile range) of the polarisation, normalised by the corresponding polarisation quantum, as a function of the distortion parameter. At each distortion step, the predicted polarisation distributions closely overlap with the DFPT reference, demonstrating that the model captures realistic polarisation branches even in the presence of $\mathbf{P} $'s multivalued nature.

From a methodological standpoint, the small but consistent advantage of \texttt{MACE-Field-MH-0} on the folded branch-resolved and spontaneous-polarisation parities is informative. Joint multihead fine-tuning on MP-Ferroelectric, MP-Dielectric, and replay data does not merely preserve the ferroelectric signal from direct supervision; in these benchmarks, it slightly improves upon the directly trained cross-chemistry model. This suggests that foundation-model pretraining, together with simultaneous supervision on local dielectric response, provides a useful inductive bias for the global branch structure of $\mathbf P$. The transfer is therefore strong not only for local linear response but also for branch-consistent ferroelectric polarisation across chemistry, which is encouraging for future universal multi-task materials models.

Overall, these results demonstrate \texttt{MACE-Field-MH-0} as a field-aware foundation model that (i) preserves its original accuracy on energies, forces and stresses, and (ii) acquires transferable predictions of Born effective charges, polarisabilities, derived refractive indices, and useful ferroelectric-response trends across diverse inorganic materials through joint multihead fine-tuning of a single enthalpy functional.

\subsection{Finite-field MLMD for \ce{BaTiO3} and $\alpha$-\ce{SiO2}}
\label{subsec:finite-field-mlmd}

\begin{figure}[t]
    \centering
    \includegraphics[width=0.49\linewidth]{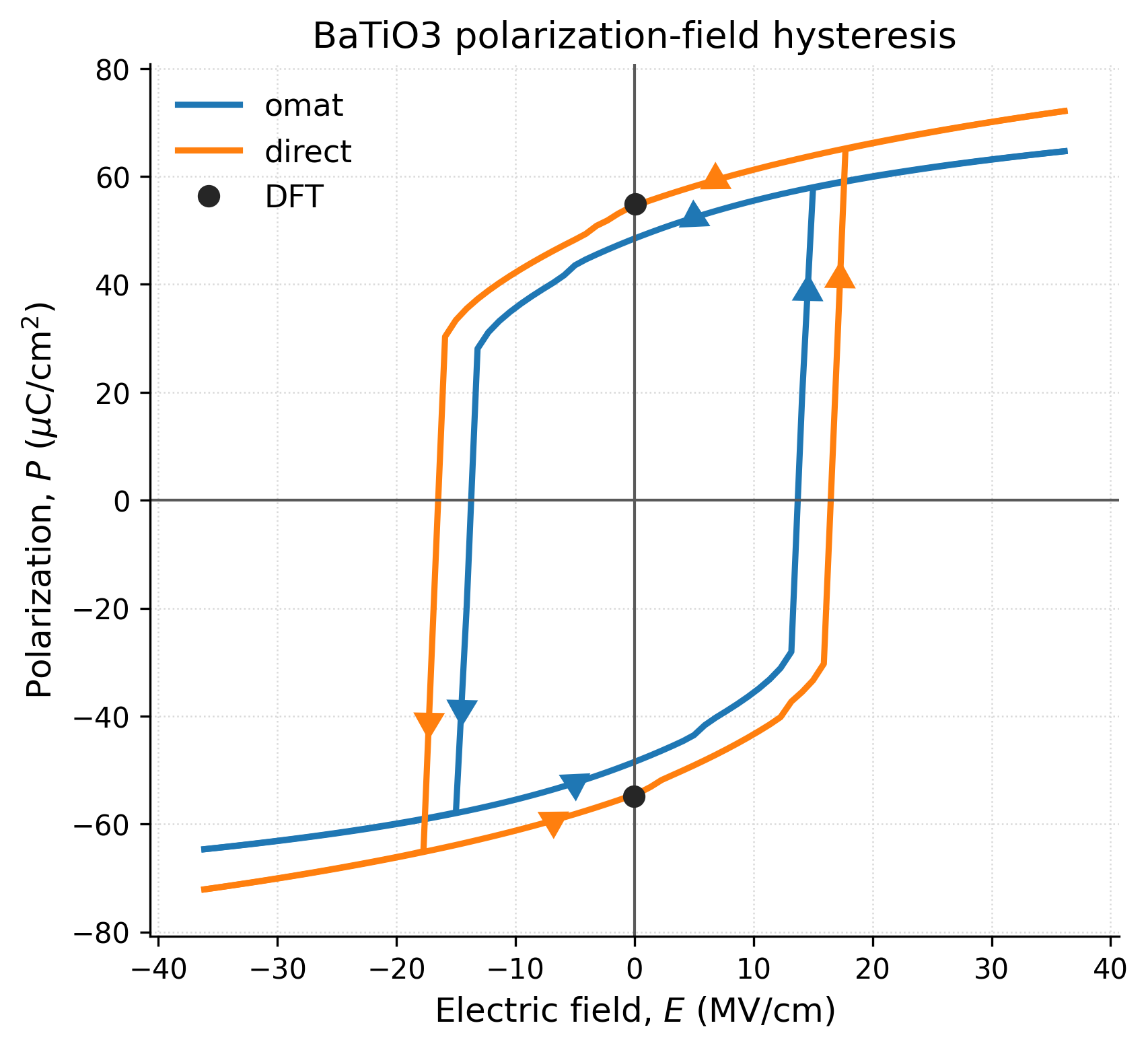}
    \hfill
    \includegraphics[width=0.49\linewidth]{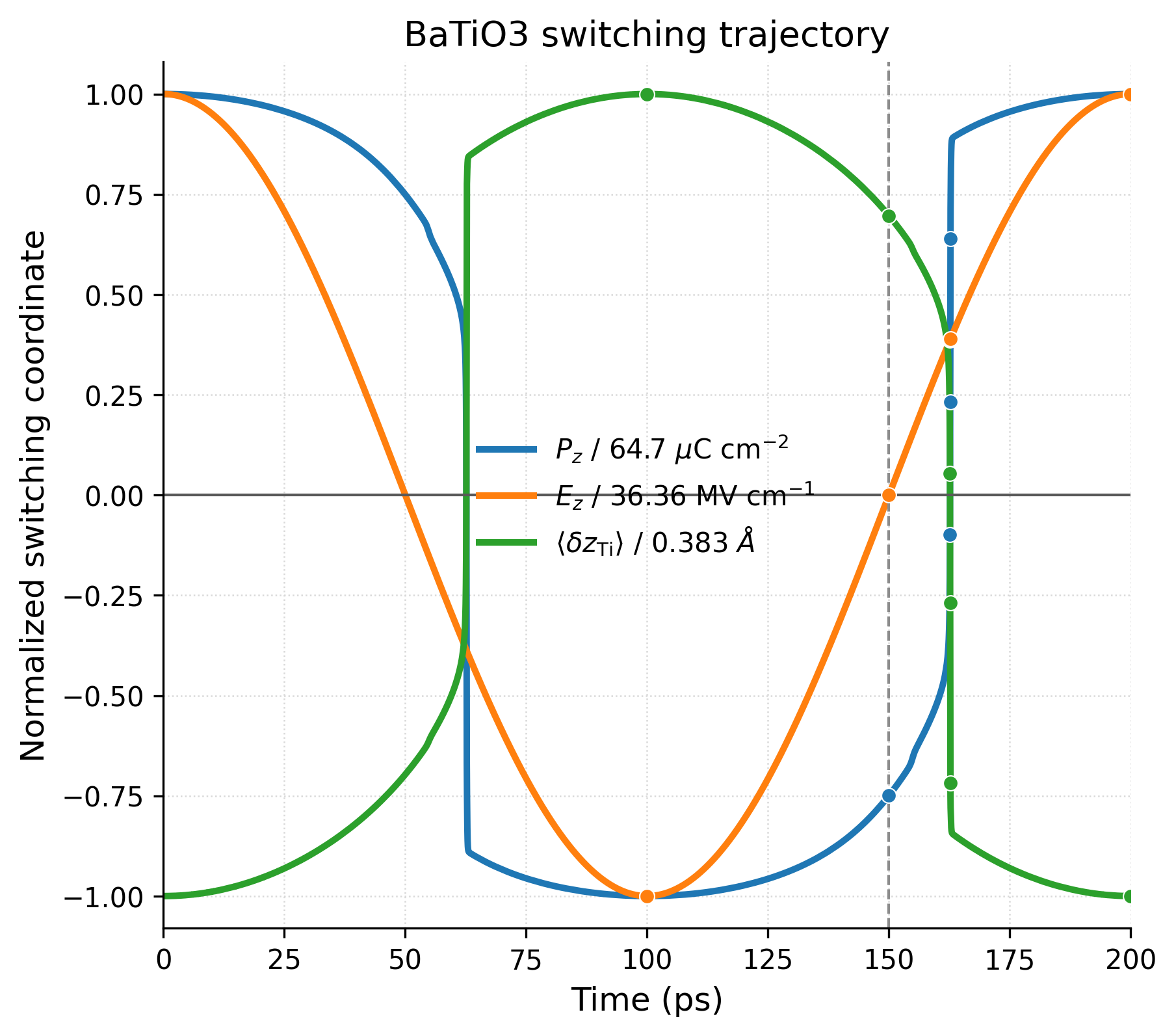}
    \caption{\textbf{BaTiO3 ferroelectric hysteresis.}
    \ce{BaTiO3} finite-field MLMD at 0~\si{K} in a 135 atom supercell under a single 5~GHz cosine field cycle applied along the polar axis. Left: overlaid polarisation-field hysteresis loops from \texttt{MACE-Field-MH-0} (blue), the directly trained specialist (orange), and DFT branch points at zero field (black). Right: representative trajectory diagnostics for the \texttt{MACE-Field-MH-0} loop, showing the evolution of $P_z(t)$, the applied field $E_z(t)$, and the average Ti off-centring along the tetragonal $c$ axis during one cycle. The circular markers indicate the six representative frames shown in the Supplementary Information. These simulations probe intrinsic homogeneous switching rather than domain nucleation and growth.}
    \label{fig:bto_hyst_ft}
\end{figure}

\begin{figure}[t]
    \centering
    \includegraphics[width=\linewidth]{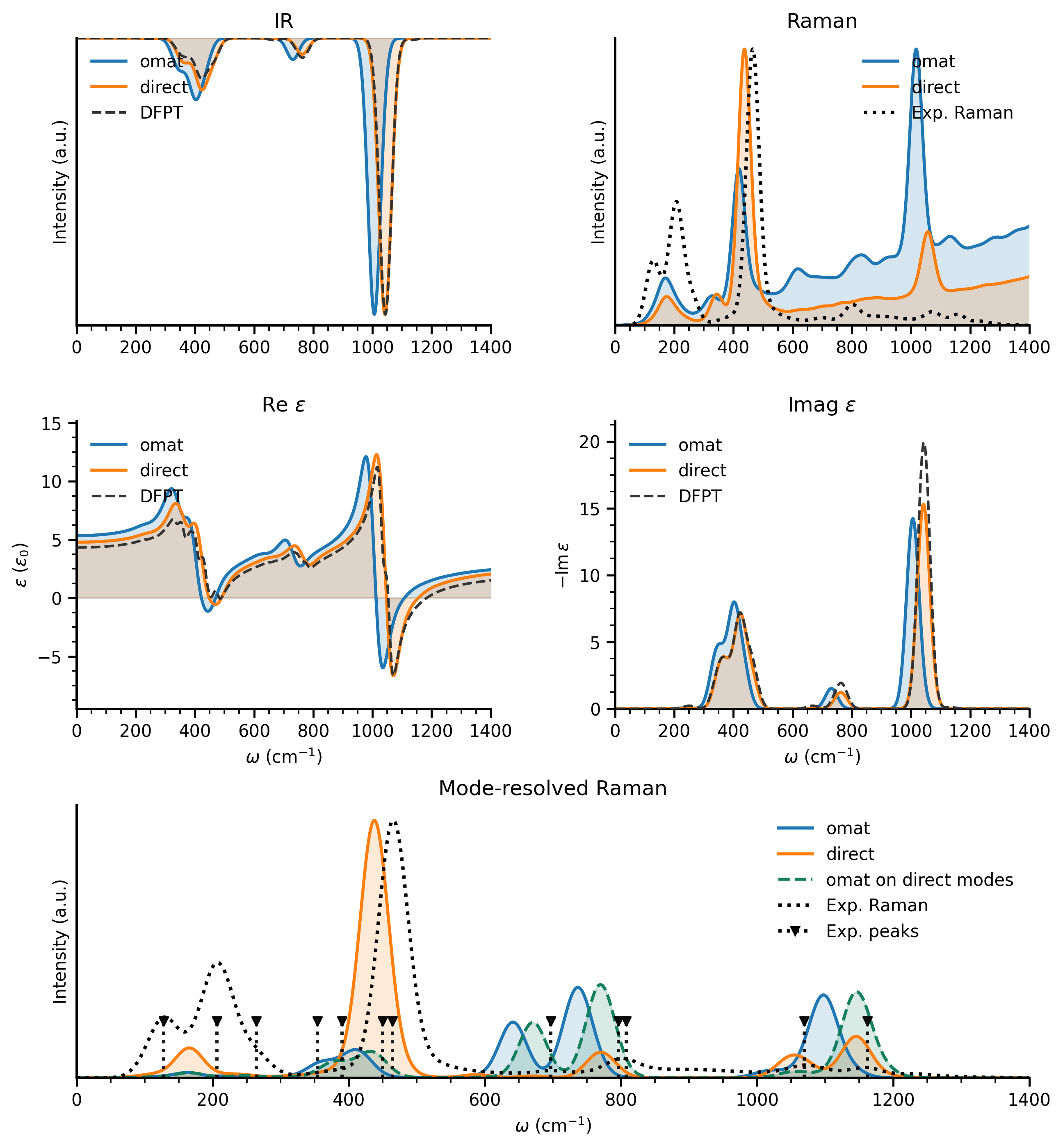}
    \caption{\textbf{$\alpha$-\ce{SiO2} spectroscopy.}
    $\alpha$-quartz spectroscopy simulations at 300~\si{K} from \texttt{MACE-Field-MH-0} (blue) and the directly trained specialist (orange). \emph{Top row:} IR spectra (left; from $\dot{\mathbf P}$-$\dot{\mathbf P}$), compared with the DFPT reference response (dashed black), and MLMD Raman spectra (right; from $\boldsymbol\alpha$-$\boldsymbol\alpha$) compared with experimental Raman~\cite{ENSLyonQuartzRaman} (dotted black). \emph{Middle row:} real and imaginary parts of the dielectric function $\varepsilon(\omega)$ assembled from time-domain response, again with the DFPT reference shown as dashed black where available. \emph{Bottom row:} mode-resolved Raman spectra reconstructed from finite-difference polarisability derivatives along optical normal modes for the direct model (orange), the foundation model in its own native mode basis (blue), the foundation response evaluated on the direct-model normal-mode basis (green dashed), and for experimental Raman (black dotted); black triangle markers denote experimental quartz Raman peak positions~\cite{ENSLyonQuartzRaman}. Spectra use Hann windows and Gaussian broadening $\sigma=20$~\si{cm^{-1}}; intensities are in arbitrary units.}
    \label{fig:sio2_spectra}
\end{figure}

To assess whether \texttt{MACE-Field} can drive field-dependent dynamics and reproduce time-domain observables, we follow the same broad validation strategy as \texttt{Allegro-pol}~\cite{allegro-pol-2025}. Using the same training signals (energy, forces, stress, Berry-phase $\mathbf{P}$, $Z^*$, and $\boldsymbol{\alpha}$), we perform fixed-cell MLMD under a spatially uniform field or at zero field, depending on the observable of interest (Sec.~\ref{sec:md-protocols}), comparing two models: a \texttt{MACE-Field} instance trained directly on MD trajectories and our fine-tuned \texttt{MACE-Field-MH-0} foundation model.

For \ce{BaTiO3}, we drive a 135-atom periodic cell through a single 5~GHz cosine field cycle of amplitude $E_0 = 0.3636$~$\mathrm{V/\AA}$ (about $36$~\si{MV/cm}) along the polar axis, corresponding to a 200~ps trajectory at 2~fs timestep. The trajectory is propagated at 0~\si{K} with fixed-cell \texttt{nve} dynamics plus viscous damping, and the polarisation-field hysteresis is reconstructed afterwards from the stored frames. These field scales are large compared with experimental coercive fields. Still, the present setup is a small bulk-periodic cell under a spatially uniform bias. It therefore targets the intrinsic homogeneous-switching regime rather than defect- and domain-mediated switching in macroscopic samples~\cite{dawber-2005-physics}. Moreover, the supervised labels for these single-material models constrain the response near zero applied field, so the large-amplitude loop should be viewed as an extrapolative test of the learned enthalpy rather than an interpolation benchmark in field magnitude. Both models generate robust hysteresis loops (Fig.~\ref{fig:bto_hyst_ft}). The directly trained model produces a nearly square, symmetric loop with an average absolute remanent polarisation $|P_r| \approx 54.5\,\mu\mathrm{C/cm}^2$ and a coercive field $|E_c| \approx 16.5$~\si{MV/cm}. The OMAT-based \texttt{MACE-Field-MH-0} foundation model yields $|P_r| \approx 48.5\,\mu\mathrm{C/cm}^2$ and $|E_c| \approx 13.7$~\si{MV/cm}, giving a somewhat softer and narrower loop. Table~\ref{tab:bto_hysteresis_compare} places these values alongside a DFT branch reference. In this intrinsic, homogeneous-switching regime, the foundation model gives a credible finite-field ferroelectric response. The right panel of Fig.~\ref{fig:bto_hyst_ft} further shows that the polarisation reversal tracks both the applied field and the sign change of the average Ti off-centring along $c$, consistent with homogeneous tetragonal switching; the six circular markers pick out the representative atomistic snapshots discussed in the Supplementary Information.

\begin{table}[t]
\centering
\small
\setlength{\tabcolsep}{6pt}
\begin{tabular}{lcc}
\hline\hline
Model/reference & $|P_{s/r}|$ (\si{\micro C / cm^2}) & $|E_c|$ (\si{MV/cm}) \\
\hline
Direct & 54.5 & 16.5 \\
OMAT & 48.5 & 13.7 \\
DFT (PBE)~\cite{allegro-pol-2025} & 54.8 & 18.1 \\
\hline\hline
\end{tabular}
\caption{\textbf{\ce{BaTiO3} intrinsic switching metrics from finite-field MLMD and reference DFT branches.} For the dynamic MLMD loops (direct \texttt{MACE-Field} and OMAT foundation \texttt{MACE-Field-MH-0}) we report the average absolute remanent polarisation $|P_r|$ and coercive field $|E_c|$ extracted from the loop. For the DFT reference from Fig.~3a of Ref.~\cite{allegro-pol-2025}, we report the zero-field branch polarisation and the terminal field magnitude of the stable upper/lower branches, so the listed $|E_c|$ should be interpreted as an intrinsic branch-limit field rather than a dynamic coercive field. In the present 0~K homogeneous-switching setup, $|P_r|$ and $|P_s|$ are numerically close, so the table provides a compact comparison of the intrinsic ferroelectric scale across models.}
\label{tab:bto_hysteresis_compare}
\end{table}

\begin{table}[t]
\centering
\small
\setlength{\tabcolsep}{4pt}
\resizebox{\columnwidth}{!}{%
\begin{tabular}{lccccc}
\hline\hline
Quantity & Allegro-pol & Direct & MFMH-0 & DFPT (PBE) & Expt. \\
\hline
$\omega_{1}$ (cm$^{-1}$) & 1041 & 1042 & 1007 & 1044 & 1072 \\
$\omega_{2}$ (cm$^{-1}$) & \phantom{0}420 & \phantom{0}424 & \phantom{0}404 & \phantom{0}423 & \phantom{0}450 \\
$\omega_{3}$ (cm$^{-1}$) & \phantom{0}765 & \phantom{0}763 & \phantom{0}731 & \phantom{0}763 & \phantom{0}797 \\
\hline
$\varepsilon_{\parallel}^{\infty}$ & 2.41 & 2.43 & 2.64 & 2.41 & 2.42 \\
$\varepsilon_{\perp}^{\infty}$     & 2.37 & 2.38 & 2.58 & 2.38 & 2.39 \\
$\varepsilon_{\parallel}^{0}$      & 4.73 & 5.02 & 5.61 & 4.74 & 4.64 \\
$\varepsilon_{\perp}^{0}$          & 4.51 & 4.65 & 5.20 & 4.52 & 4.43 \\
\hline\hline
\end{tabular}
}
\caption{Comparison of $\alpha$-\ce{SiO2} spectroscopic quantities at 300~\si{K}: main IR peak positions $\omega_i$ (cm$^{-1}$), high-frequency dielectric constants $\varepsilon_{\parallel,\perp}^{\infty}$, and static dielectric constants $\varepsilon_{\parallel,\perp}^{0}$, parallel ($\parallel$) and perpendicular ($\perp$) to the polar axis. ``\texttt{Allegro-pol}'' values are taken from Ref.~\cite{allegro-pol-2025}; ``Direct'' and ``MFMH-0'' denote, respectively, the directly trained \texttt{MACE-Field} model and the fine-tuned \texttt{MACE-Field-MH-0} foundation model, all analysed using the same time-domain protocol (Hann window, Gaussian broadening $\sigma = 20$~\si{cm^{-1}}, branch-invariant wrapping of $\mathbf{P}(t)$). ``DFPT (PBE)'' and ``Expt.'' are representative quartz dielectric benchmarks compiled by He \emph{et al.}~\cite{He2014DFPTDielectrics} (experiment at 298~K). Peak positions are estimated from the plotted spectra with an uncertainty of $\pm 10$-$20$~\si{cm^{-1}}. The directly trained \texttt{MACE-Field} remains close to the \texttt{Allegro-pol} and DFPT references. At the same time, the foundation model is still somewhat softer and more polarisable, especially in the static dielectric response.}
\label{tab:quartz_compare}
\end{table}

For $\alpha$-quartz, we probe the linear response at 300~\si{K}. Spectra are obtained from derivative observables of the learned electric enthalpy evaluated on stored frames from 200~ps trajectories: IR from $\dot{\mathbf{P}}$-$\dot{\mathbf{P}}$ and Raman from $\boldsymbol{\alpha}$-$\boldsymbol{\alpha}$. As in Allegro-pol, we employ a Hann window, Gaussian broadening with $\sigma = 20$~\si{cm^{-1}}, and branch-invariant wrapping of the Berry-phase polarisation $\mathbf{P}(t)$. Figure~\ref{fig:sio2_spectra} combines the time-domain IR/Raman and dielectric comparison with a mode-resolved Raman diagnostic, and Table~\ref{tab:quartz_compare} collects the main quantitative descriptors.

The directly trained \texttt{MACE-Field} model reproduces the three dominant IR peaks at $\omega_1 \approx 1042$~\si{cm^{-1}}, $\omega_2 \approx 424$~\si{cm^{-1}} and $\omega_3 \approx 763$~\si{cm^{-1}}, in excellent agreement (within $\sim 5$~\si{cm^{-1}}) with DFPT and \texttt{Allegro-pol} values~\cite{allegro-pol-2025}. The corresponding high-frequency dielectric constants $\varepsilon_{\parallel,\perp}^{\infty}$ and static values $\varepsilon_{\parallel,\perp}^{0}$ also remain close to the \texttt{Allegro-pol} reference, and broadly consistent with representative PBE DFPT and room-temperature experimental quartz benchmarks (Table~\ref{tab:quartz_compare}). In the Raman fingerprint region, the directly trained model places most of its mode-resolved activity in the 350--500~\si{cm^{-1}} band (about 65\% of the total common-basis activity), reproducing the dominant quartz feature around 430--475~\si{cm^{-1}} and aligning well with the experimental~\cite{ENSLyonQuartzRaman} markers near 450 and 464~\si{cm^{-1}} in Fig.~\ref{fig:sio2_spectra}. The direct model does retain weaker high-frequency Si--O stretching activity, but the mode-resolved panel shows that this activity remains relatively discrete; the broader high-frequency background in the MLMD Raman envelope is therefore likely to reflect finite-trajectory and spectral-estimator effects in addition to the underlying response model.

The \texttt{MACE-Field-MH-0} foundation model retains not only the qualitative band structure but also a realistic quantitative IR/dielectric spectrum. The main IR peaks shift only moderately relative to the directly trained specialist, to $\omega_1 \approx 1007$~\si{cm^{-1}}, $\omega_2 \approx 404$~\si{cm^{-1}} and $\omega_3 \approx 731$~\si{cm^{-1}}, while the dielectric constants move systematically high: $\varepsilon_{\parallel,\perp}^{\infty} \approx 2.64$ and $2.58$, and $\varepsilon_{\parallel,\perp}^{0} \approx 5.61$ and $5.20$. \texttt{MACE-Field-MH-0} is therefore somewhat softer and more polarisable than the directly trained specialist. The mode-resolved analysis shows that this Raman discrepancy is not primarily a phonon-basis failure: the native direct/foundation mode overlap remains high (mean overlap $\approx 0.83$, mean frequency ratio $\approx 0.96$), yet the native Raman-activity correlation is only about 0.14 and remains low even when the foundation response is evaluated on the direct normal-mode basis (common-basis Raman-activity correlation $\approx 0.26$). In that common-basis comparison, the foundation suppresses the dominant 350--500~\si{cm^{-1}} quartz Raman band from about 65\% of the total activity in the direct model to about 14\%, while over-weighting the 650--850~\si{cm^{-1}} and 950--1250~\si{cm^{-1}} bands to about 48\% and 34\%, respectively. The excess high-frequency Raman intensity is therefore primarily a response-head error, most plausibly an overestimate of the relevant polarisability derivatives, rather than merely a consequence of slightly shifted phonon frequencies.

Overall, these \ce{BaTiO3} and $\alpha$-quartz case studies demonstrate that an explicitly field-aware \texttt{MACE-Field} model can be used to drive finite-field MD and to extract non-trivial time-domain observables. The directly trained single-material models provide the most quantitative agreement with DFPT and the \texttt{Allegro-pol} reference. The OMAT-based foundation model still produces a credible intrinsic \ce{BaTiO3} hysteresis loop and a realistic quartz IR/dielectric response. Still, the new mode-resolved Raman analysis indicates that its Raman activity distribution is not yet quantitatively reliable. From a computational-science perspective, the framework is therefore not only a property regressor but also a stable dynamical engine for comparative finite-field simulations, with the Raman case providing a concrete example of where targeted response supervision remains important.

\section{Discussion}\label{sec:discussion}

We introduced \texttt{MACE-Field}, a field-aware, $O(3)$-equivariant interatomic potential that learns a \emph{single}
electric enthalpy $\mathcal F(\{\mathbf R\},\mathbf E)$ and obtains polarisation $\mathbf P$, Born effective charge $Z^*$, and polarisability $\boldsymbol\alpha$ by exact differentiation.
A uniform field couples to learned latent spherical-tensor features via Clebsch-Gordan products, while the final readout
remains a scalar. Benchmarked against semilocal DFT/DFPT reference data, a directly trained cross-chemistry model reproduces Berry-phase
polarisations (including $P_s$) for $\sim$250 ferroelectric materials spanning 61 elements. Separately, a fine-tuned foundation model, \texttt{MACE-Field-MH-0}, jointly trained on $\sim$5.3k DFPT
dielectric/BEC entries, $\sim$2.5k ferroelectric distortion-path structures, and an OMAT-PBE replay set predicts $Z^*$ and $\boldsymbol\alpha$ with high fidelity
(Figs.~\ref{fig:dielectric_parity} and~\ref{fig:refractive_index_parity}), recovers both electronic and
ionic dielectric trends through analytic Hessians, and captures same-branch polarisation and spontaneous-polarisation trends across chemistry. Finally, single-material models for \ce{BaTiO3} and $\alpha$-\ce{SiO2} demonstrate that the same
framework can drive finite-field MD and reproduce hysteresis loops, IR/Raman spectra, and dielectric functions with DFPT accuracy, whilst the \texttt{MACE-Field-MH-0} foundation model demonstrates comparative accuracy with a systematic softening of the resulting dynamics and a less reliable Raman activity distribution (Figs.~\ref{fig:bto_hyst_ft} and~\ref{fig:sio2_spectra}). 
\\

Conceptually, \texttt{MACE-Field} is therefore a prototype of a physics-informed, differentiable foundation model for materials~\cite{ChenOng2022,batatia2025foundationmodelatomisticmaterials}: it combines strong inductive biases (symmetries, enthalpy derivatives) with data-driven learning. It can be retrofitted onto existing energy/force models. Because all observables share one scalar origin, Maxwell reciprocity, the acoustic sum
rule, point-group tensor forms, and branch-invariant learning of Berry-phase $\mathbf P$ hold \emph{by construction}
(Sec.~\ref{sec:identities}).
\\

\paragraph*{Key advances.}
\begin{enumerate}
  \item \textbf{Plug-in field coupling that \emph{inherits \texttt{MACE} foundations}.}
  The field enters only through irrep-wise latent tensor products and equivariant mixing; the \texttt{ACE} graph
  construction and scalar readout are unchanged. Consequently, existing \texttt{MACE} foundation weights
  can be dropped in and fine-tuned to become field-aware. Our \texttt{MACE-Field-MH-0} model shows that this
  inheritance preserves the accuracy of the OMAT-based multihead \texttt{mace-mp-mh-0} prior on energies, forces and stresses (Fig.~\ref{fig:replay_parity}),
  while adding transferable predictions of $Z^*$, $\boldsymbol\alpha$, and derived optical properties.

  \item \textbf{One model across diverse chemistry.}
  Unlike prior unified-enthalpy demonstrations restricted to single materials, \texttt{MACE-Field} is trained
  once across thousands of crystals covering $\gtrsim 80$ elements. The \texttt{MACE-Field-MH-0} dielectric foundation
  model reproduces DFPT BECs and electronic polarisabilities across the MP-Dielectric set (Fig.~\ref{fig:dielectric_parity}),
  and yields realistic distributions of response quantities on unseen replay structures from the OMAT-PBE pool
  (Fig.~\ref{fig:bec_alpha_kde}). It also provides reasonable refractive-index parity on the Matbench dielectric task, with performance in the same broad regime as strong direct-property predictors. However, the substantial overlap with Materials-Project-derived data means that this comparison should be interpreted with caution rather than as a strict out-of-distribution benchmark (Fig.~\ref{fig:refractive_index_parity}).

  \item \textbf{Time-domain finite-field validation.}
  Single-material \texttt{MACE-Field} models for \ce{BaTiO3} and $\alpha$-\ce{SiO2} reproduce DFPT and Allegro-pol benchmarks for polarisation hysteresis and IR/Raman/dielectric spectra, with \texttt{MACE-Field-MH-0} maintaining comparative accuracy despite not being trained on the same MD trajectories. These \texttt{MACE-Field} models match IR peak positions and $\varepsilon^{\infty,0}_{\parallel,\perp}$ for $\alpha$-quartz to within the uncertainty
  of the spectral analysis (Table~\ref{tab:quartz_compare}) and generate \ce{BaTiO3} loops with the expected intrinsic,
  uniform-field coercive scales for a small periodic cell (Fig.~\ref{fig:bto_hyst_ft}).
\end{enumerate}
\par\addvspace{.75\baselineskip} 

\paragraph*{What we learn from the fine-tuned foundation model.}
To our knowledge, \texttt{MACE-Field-MH-0} offers a first demonstration of a field-aware \emph{foundation} model built from a recent multihead OMAT-initialised prior: starting from \texttt{mace-mp-mh-0} and its OMAT-PBE head, we add field-coupling blocks and perform joint multi-headed fine-tuning on DFPT BECs, polarisabilities, Berry-phase polarisations, and a replay set of energies/forces/stresses. This succeeds remarkably well for cross-chemistry dielectric response and also transfers substantial ferroelectric information. The model captures $Z^*$ and $\boldsymbol\alpha$ across chemistry, reconstructs
realistic refractive-index distributions, and (combined with analytic Hessians) yields reasonable electronic
and ionic dielectric constants without direct supervision on $\varepsilon$ (Fig.~\ref{fig:eps_parity}). On the folded same-branch and spontaneous-polarisation benchmarks, it performs as well as a directly trained cross-chemistry ferroelectric model. Outlier analysis reveals that the largest discrepancies in $Z^*$ often correspond to clear pathologies in the DFPT
labels (violations of the acoustic sum rule). In contrast, the learned enthalpy automatically enforces ASR and produces physically sensible charges (Fig.~\ref{fig:bec_error_dist} and Table~\ref{tab:bec-outliers}).

At the same time, \texttt{MACE-Field-MH-0} exposes important limitations of fine-tuning a broadly transferable force field for field response. First, the ionic dielectric constants show significantly larger scatter than the electronic ones. This reflects the fact that $\boldsymbol{\varepsilon}_{\mathrm{ion}}$ depends on both $Z^*$ and the phonon spectrum via the inverse Hessian [Eq.~\eqref{eq:eps_ion_static}], and hence amplifies any residual bias in the
underlying foundation-model Hessians. The softened phonons and overestimated static permittivities in our quartz tests are consistent with a slightly over-smoothed potential-energy surface learned for broad transferability: total energies can remain usable while curvature-sensitive observables such as phonons, elastic constants, Raman derivatives, and ionic dielectric constants drift systematically. In our case, this manifests as mildly red-shifted IR peaks and elevated permittivities, both in the analytic $\varepsilon_0$ and in the time-domain spectra of $\alpha$-quartz (Fig.~\ref{fig:sio2_spectra} and Table~\ref{tab:quartz_compare}).

Second, the cross-chemistry polarisation benchmarks show that the fine-tuned \texttt{MACE-Field-MH-0} foundation can do more than merely approach a directly trained specialist: in the present folded parity metrics, it slightly surpasses the direct model for both same-branch Berry-phase polarisation and spontaneous $P_s$ (Fig.~\ref{fig:pol-parity}). This is instructive. Learning polarisation modulo the polarisation quantum via a folded loss is intrinsically challenging, particularly when combined with heterogeneous semilocal workflows and multitask supervision. Joint training
on BECs provides strong local information, but the branch structure of $\mathbf P$ is a global, path-dependent object; here, however, the combination of foundation pretraining, replay stabilisation, and simultaneous dielectric supervision appears to provide a useful inductive bias rather than a detriment. In addition, MP-Ferroelectric and
MP-Dielectric labels come from independent DFT calculations, so even subtle inconsistencies in geometry, functional, or convergence criteria effectively act as label noise when a single model is asked to reconcile
both.

Third, Raman line shapes remain more model-dependent than either IR peak positions or static dielectric constants. The mode-resolved quartz diagnostic sharpens this conclusion. The foundation model retains phonons broadly similar to those of the direct model (mean native mode overlap $\approx 0.83$ and mean frequency ratio $\approx 0.96$). Still, the Raman-activity parity remains poor both in the native comparison and when the foundation response is evaluated on the direct normal-mode basis. In other words, the excess foundation intensity in the 650--850 and 950--1250~\si{cm^{-1}} bands survives even when the vibrational basis is held fixed. This indicates that the dominant error is in the learned polarisability response, especially the mode-resolved derivatives $\partial\boldsymbol\alpha/\partial Q_m$, rather than primarily in the phonon frequencies themselves.

Taken together, these observations suggest that \emph{field-aware foundations} such as \texttt{MACE-Field-MH-0} can provide not only broadly accurate $Z^*$, $\boldsymbol\alpha$, and derived dielectric constants across chemistry, but also the strongest cross-chemistry ferroelectric parity among the models compared here, together with credible finite-field switching behaviour. Targeted, single-material (or small-chemistry) fine-tuning is still beneficial when the goal is the most quantitative Raman spectra or material-specific finite-field dynamics.
\\

\paragraph*{Position within the broader ferroelectrics landscape.}
The \texttt{MACE-Field} architecture should be viewed as a derivative-consistent model for bulk periodic response under a homogeneous external field. In particular, because the field input is global and no explicit self-consistent long-range Coulomb term is solved, spatially varying depolarising fields, charged defects, and interface electrostatics are outside the present model class. Likewise, bulk phonon effects that depend on explicit long-range Coulomb physics, such as LO-TO splitting, are only learned implicitly from the training data rather than enforced analytically. Addressing those regimes will likely require either site-dependent field conditioning or hybridisation with explicit long-range electrostatics, and will be the subject of future work.
\\

\paragraph*{Outlook.}
There are several directions to extend and exploit the \texttt{MACE-Field} framework:
\begin{itemize}
  \item \textbf{Better foundations and phonons.}
  The behaviour of \texttt{MACE-Field-MH-0} underscores the importance of high-quality forces and Hessians.
  Retraining foundations with tighter DFT settings, explicit phonon or elastic-constant targets, or
  incorporating DFPT force-constant or higher-order derivative information into the loss~\cite{Gonnheimer2025} could harden the phonon spectrum and
  improve $\varepsilon_0$, hysteresis, and Raman response.

  \item \textbf{Mode-resolved Raman supervision.}
  The quartz diagnostics suggest that matching equilibrium $\boldsymbol\alpha$ is not sufficient for quantitatively correct Raman spectra. A more targeted route is to generate DFPT or finite-field polarisabilities for $\pm Q_m$ displacements along Raman-relevant normal modes and to fine-tune directly on $d\alpha_{ij}/dQ_m$, together with isotropic, anisotropic, or activity-based Raman invariants. Mode-space validation should then become a standard check before trusting MLMD Raman envelopes.

  \item \textbf{Hybrid long-range, higher-order, and higher-fidelity response.}
  Combining \texttt{MACE-Field} with site-dependent auxiliary internal-field modules, latent/Ewald terms, or charge equilibration, while preserving derivative consistency, would extend its reach to highly ionic or low-dimensional systems. Nothing in the architecture ties it specifically to semilocal DFT: in principle, one could fine-tune on r$^2$SCAN, hybrid-DFT, or multi-fidelity response datasets where such labels are available. Likewise, selected second- and third-order susceptibilities $\chi^{(2)}$/$\chi^{(3)}$ and hyperpolarisabilities are formally accessible as higher derivatives of $\mathcal F$. Still, quantitative nonlinear responses will require dedicated training labels, as the current datasets primarily constrain the linear, near-zero-field regime.

  \item \textbf{From foundations to targeted specialists.}
  Field-aware foundations such as \texttt{MACE-Field-MH-0} can serve as starting points for lightweight,
  material- or family-specific fine-tuning that sharpens polarisation branches, $P_s$, and Raman
  intensities without retraining from scratch. This hierarchical strategy mirrors current practice in
  energy/force foundation models.

  \item \textbf{High-throughput dielectric and ferroelectric discovery.}
  A \texttt{MACE-Field} foundation fine-tuned on MP-Dielectric and MP-Ferroelectric can be deployed in
  large-scale screening campaigns for ferroelectrics, multiferroics, hyperferroelectrics and high-$\kappa$
  dielectrics, including finite-field MD to probe switching pathways, leakage, and finite-temperature
  response.

  \item \textbf{Broader applications.}
  Beyond inorganic dielectrics, the same idea, a single enthalpy functional that is differentiable with respect to fields and structural degrees of freedom, could be applied to a wide range of systems where field response matters: electrolytes and solid-electrolyte interfaces in batteries, polar liquids and biomolecules in strong fields, heterogeneous catalysis under bias, and geophysical materials in the deep Earth. In each case, a reusable, field-aware force field would enable computational scientists across communities to ask new questions about how materials behave under realistic operating conditions.
\end{itemize}

Methodologically, our work aligns with the broader push towards physics-informed machine learning in computational science, where enforcing physical structure in ML models improves robustness, interpretability and transfer across tasks. To conclude, by learning a single electric enthalpy and differentiating it, \texttt{MACE-Field} unifies $\mathbf P$, $Z^*$, and $\boldsymbol\alpha$ in a symmetry-consistent way that can inherit weights from existing \texttt{MACE} foundations. The present results show that simple coupling of a uniform field to learned equivariant latent features is already sufficient to endow a pretrained foundation with broadly accurate dielectric response, very strong cross-chemistry ferroelectric transfer capability, and credible finite-field dynamics. At the same time, the single-material studies show that the most quantitative Raman intensities and material-specific dynamical response still benefit from dedicated supervision. We therefore view \texttt{MACE-Field} as a practical route to field-aware atomistic modelling, whose natural next step is targeted fine-tuning on higher-fidelity or more locally field-rich data.

\section*{Code and Data Availability}

The \texttt{MACE-Field} source code is available at \url{https://github.com/mdi-group/mace-field}.
\\

The manuscript-specific repository contains processed train/validation/test split files, training shell scripts and configuration files, replay-set selection inputs, and the analysis code used for the figures reported here--including polarisation-branch folding, dielectric post-processing, Matbench-overlap analysis, and hysteresis/spectroscopy plotting--at \url{https://github.com/mdi-group/2025-04-mace-field}. Large binary artefacts, such as trained checkpoints and serialised models, are distributed via repository releases when file sizes permit, rather than tracked directly in git.
\\

The same repository also provides the data-generation scripts used to build the cross-chemistry datasets from public upstream sources that assemble MP-Dielectric from the Materials Project API, and MP-Ferroelectric from the \texttt{MPContribs} API. Together with the processed split files, replay-set selection inputs for the OMAT-PBE multihead fine-tune, and the public \texttt{mace-mp-mh-0} foundation checkpoint with its OMAT-PBE head, these artefacts make the data-preparation, training, and analysis workflow described in this paper publicly reproducible from the repository.
\\

\section*{Acknowledgments}

KTB and BAAM acknowledge support from EPSRC funding (EP/Y014405/1). Via our membership of the UK's HEC Materials Chemistry Consortium, which is funded by EPSRC (EP/L000202), this work used the UK Materials and Molecular Modelling Hub for computational resources, the MMM Hub, which is partially funded by EPSRC (EP/T022213/1, EP/W032260/1 and EP/P020194/1).

\bibliographystyle{apsrev4-2}
\bibliography{bibliography}


\end{document}


\maketitle

\section*{Supplementary Information (SI)}

\subsection*{S1. Training configurations for \texttt{MACE-Field}}

\begin{table*}[t]
  \centering
  \small
  \caption{Training setups and loss weights for the four experiments.}
  \label{tab:training_setups}
  \begin{tabular}{@{}l l l c c c c c c c c c c c@{}}
  \toprule
  Experiment & Data split & Channels &
  \multicolumn{6}{c}{Loss weights} & \multicolumn{4}{c}{Optimiser / schedule} & Seed \\
   & & & $w_E$ & $w_F$ & $w_\sigma$ & $w_P$ & $w_{Z^*}$ & $w_{\alpha}$ & LR & Epochs$_\text{max}$ & Patience & Batch &  \\
  \midrule
  Fine-tuned & 80/10/10 &
  128 & 1.0 & 100.0 & 1.0 & 100.0 & 100.0 & 100.0 & 0.0001 & 300 & 50 & 1 & 23 \\
  Ferroelectrics & 80/10/10 &
  128 & 1.0 & 10.0 & 0.0 & 100.0 & 0.0 & 0.0 & 0.005 & 1000 & 100 & 2 & 23 \\
  BaTiO$_3$ MD & valid 20\% &
  64 & 1.0 & 100.0 & 1.0 & 1.0 & 50.0 & 10.0 & 0.005 & 1000 & 100 & 4 & 23 \\
  $\alpha$-SiO$_2$ MD & valid 20\% &
  64 & 1.0 & 50.0 & 1.0 & 10.0 & 100.0 & 200.0 & 0.01 & 1000 & 100 & 8 & 23 \\
  \bottomrule
  \end{tabular}
\end{table*}

We trained four models with the \texttt{run\_train.py} CLI (DDP via \texttt{torchrun}; double precision). All runs used the
\texttt{MACEField} backbone with field injection at each interaction block (\texttt{--enable\_cueq True}),
branch-invariant polarisation loss enabled (\texttt{--compute\_* True}), and Adam with EMA (\texttt{--ema} with
decay~0.995), cosine/plateau schedule (\texttt{--scheduler\_patience} as listed), weight decay $10^{-8}$, and checkpoints on CPU.
Atomic baselines were set to the dataset average (\texttt{--E0s average}). Errors are reported with the
\texttt{PerAtomRMSEstressvirialsfield} table. Learning rates and early-stopping patience are per-experiment.
\\

\paragraph*{Shared architecture.} We use: 2 interaction layers; correlation order~3; cutoff $r_\text{max}=5$~\AA; $L_{\max}=1$; $\ell_{\max}=3$; 10 radial basis; MLP readout \texttt{16x0e}; mixed residual field coupling; scalar ($L{=}0$) readout. Ferroelectric and dielectric models use 128 channels; single-material MD models (BaTiO$_3$, $\alpha$-SiO$_2$) use 64 channels to compare to \texttt{Allegro-pol}, which also used 64 channels.
\\

\paragraph*{Optimisation and precision.}
Default dtype \texttt{float64}; AMSGrad enabled; gradient EMA; evaluation each epoch
(\texttt{--eval\_interval 1}). Seeds are recorded per run below. Batch sizes are chosen to saturate GPU memory.
\\

\paragraph*{Command lines.}
For reproducibility, the exact CLI flags (data files, weights, and optimiser settings) used in this work match those listed in
Table~\ref{tab:training_setups}; full commands are included in the code repository under \texttt{scripts/}.
\\

\paragraph*{Representative runtime comparison to ordinary \texttt{MACE}.}
Table~\ref{tab:runtime_compare} summarises representative timing information to illustrate the relative cost to base \texttt{MACE}. For training, we compare representative single-material runs of similar depth and width (2 interaction layers, 64 channels) executed with DDP on 4 Nvidia L4 GPUs. We estimate the epoch cost from differences between successive timestamped epoch lines. Base \texttt{MACE} supervises only $E,\mathbf F,\boldsymbol\sigma$, whereas \texttt{MACE-Field} additionally supervises $\mathbf P$, $Z^*$, and $\boldsymbol\alpha$.

For production MLMD, the picture is different. Force-only \texttt{LAMMPS}/ML-IAP loops query only energies and forces inside the time-integration step, so their per-step cost is expected to remain close to that of an architecture-matched ordinary \texttt{MACE} potential. The representative \texttt{MACE-Field} production throughputs in Table~\ref{tab:runtime_compare} therefore characterise the core MD cost. By contrast, when derivative response observables are needed for analysis, the workflow used here commonly evaluates $\mathbf P$, $Z^*$, and $\boldsymbol\alpha$ afterwards in a separate pass over stored frames. This post-processing introduces an additional per-frame inference cost on top of the force-only MD loop.

\begin{table*}[t]
  \centering
  \small
  \caption{Representative runtime comparison between ordinary \texttt{MACE} and \texttt{MACE-Field}.}
  \label{tab:runtime_compare}
  \begin{tabular*}{\textwidth}{@{\extracolsep{\fill}}l l l l@{}}
  \toprule
  \parbox[t]{2.8cm}{\raggedright Workflow} &
  \parbox[t]{3.6cm}{\raggedright \texttt{MACE}} &
  \parbox[t]{3.6cm}{\raggedright \texttt{MACE-Field}} &
  \parbox[t]{5.4cm}{\raggedright Explanation} \\
  \midrule
  \parbox[t]{2.8cm}{\raggedright 135 atom \ce{BaTiO3} training (4 GPUs; 2 layers; 64 channels)} &
  \parbox[t]{3.6cm}{\raggedright $\approx 0.6$~s median epoch interval; larger batch; targets $E,\mathbf F,\boldsymbol\sigma$} &
  \parbox[t]{3.6cm}{\raggedright $\approx 3$~s median epoch interval; smaller batch; targets $E,\mathbf F,\boldsymbol\sigma,\mathbf P,Z^*,\boldsymbol\alpha$} &
  \parbox[t]{5.4cm}{\raggedright Roughly $\times 5$ times slower due to the extra retained/create-graph autograd needed for response-property supervision.} \\ \\
  \parbox[t]{2.8cm}{\raggedright Force-only production MLMD on $\alpha$-\ce{SiO2} supercells of $10^2$--$10^3$ atoms} &
  \parbox[t]{3.6cm}{\raggedright Similar to \texttt{MACE-Field}} &
  \parbox[t]{3.6cm}{\raggedright $\approx 6$ steps/s / $\approx 1$ k-atom-step/s throughput} &
  \parbox[t]{5.4cm}{\raggedright No additional cost for MLMD when response properties are not computed.} \\ \\
  \parbox[t]{2.8cm}{\raggedright Trajectory annotation of response properties} &
  \parbox[t]{3.6cm}{\raggedright Not applicable} &
  \parbox[t]{3.6cm}{\raggedright Inference rates $\approx 1--1.5$~s per stored frame when evaluating $\mathbf P$, $Z^*$, and $\boldsymbol\alpha$} &
  \parbox[t]{5.4cm}{\raggedright This overhead is separate from the force-only MD loop. It arises from reloading each stored structure and running additional response-property inference because the \texttt{LAMMPS}/ML-IAP exports used here do not return these observables directly.} \\
  \bottomrule
  \end{tabular*}
\end{table*}

\subsection*{S2. Data curation and splits}

\paragraph*{Ferroelectric paths.} For each material, we use the two endpoint structures plus 8 evenly spaced interpolates (fixed cell, fractional-coordinate interpolation), giving 10 frames per path in total. 
\\

\paragraph*{Dielectric/BEC set.} We restrict to insulating DFPT entries (GGA-PBE). Forces/stresses from this dataset are not used for training. Splits are made at the MP identifier level, so supercells, symmetry equivalents, or small perturbations of the same MP ID do not cross splits.
\\

\begin{table*}[t]
  \centering
  \small
  \caption{Dataset provenance and scope. Reported ML errors throughout the manuscript are with respect to the underlying DFT/DFPT labels used for each dataset; they therefore do not include the intrinsic error of those reference workflows relative to experiment or higher-level electronic-structure methods.}
  \label{tab:dataset_provenance}
  \begin{tabular*}{\textwidth}{@{\extracolsep{\fill}}l l l l@{}}
  \toprule
  \parbox[t]{2.2cm}{\raggedright Dataset} &
  \parbox[t]{2.6cm}{\raggedright Labels used here} &
  \parbox[t]{4.0cm}{\raggedright Reference workflow} &
  \parbox[t]{6.0cm}{\raggedright Notes} \\
  \midrule
  \parbox[t]{2.2cm}{\raggedright MP-Dielectric} &
  \parbox[t]{2.6cm}{\raggedright $Z^*$, $\boldsymbol\alpha$} &
  \parbox[t]{4.0cm}{\raggedright Materials Project DFPT data assembled via public API scripts (VASP; GGA-PBE family)} &
  \parbox[t]{6.0cm}{\raggedright Used only for dielectric-response heads; energy/forces/stresses from this dataset are not used.} \\ \\
  \parbox[t]{2.2cm}{\raggedright MP-Ferroelectric} &
  \parbox[t]{2.6cm}{\raggedright $E,\mathbf F,\mathbf P$ along distortion paths} &
  \parbox[t]{4.0cm}{\raggedright MP-Ferroelectric workflow of Smidt \emph{et al.}~\cite{smidt-2020-ferrodb}; Berry-phase polarisation with GGA-PBE(+U) calculations and path validation} &
  \parbox[t]{6.0cm}{\raggedright These labels come from a different workflow from MP-Dielectric, so small inter-dataset inconsistencies act as multi-task label noise.} \\ \\
  \parbox[t]{2.2cm}{\raggedright OMAT-PBE replay subset} &
  \parbox[t]{2.6cm}{\raggedright $E,\mathbf F,\boldsymbol\sigma$} &
  \parbox[t]{4.0cm}{\raggedright OMAT-PBE replay data associated with the multihead \texttt{mace-mp-mh-0/1} foundation family~\cite{batatia2025foundationmodelatomisticmaterials}} &
  \parbox[t]{6.0cm}{\raggedright Provides force-field replay regularisation during foundation fine-tuning; contains no field-response labels.} \\ \\
  \parbox[t]{2.2cm}{\raggedright BaTiO$_3$ and $\alpha$-SiO$_2$ MD sets} &
  \parbox[t]{2.6cm}{\raggedright $E,\mathbf F,\boldsymbol\sigma,\mathbf P,Z^*,\boldsymbol\alpha$} &
  \parbox[t]{4.0cm}{\raggedright Single-material datasets released with \texttt{Allegro-pol}~\cite{allegro-pol-2025}} &
  \parbox[t]{6.0cm}{\raggedright Used only for direct finite-field validation and comparison to \texttt{MACE-Field-MH-0} foundation model.} \\
  \bottomrule
  \end{tabular*}
\end{table*}

\paragraph*{Consistency across datasets.} The cross-chemistry heads are intentionally multi-source rather than one perfectly uniform DFT workflow. This improves chemical coverage, but it also introduces a small amount of inter-workflow label noise when a single multi-head model is asked to reconcile MP-Dielectric, MP-Ferroelectric, and replay-set supervision.
\\

\paragraph*{Public artefacts and regeneration.} The public repositories expose the training configurations, split definitions, processed MP-Dielectric and MP-Ferroelectric files, replay-set selection inputs, and the full analysis workflow used in this paper. They also provide the scripts used to regenerate MP-Dielectric from the Materials Project API and MP-Ferroelectric from the \texttt{MPContribs} API, alongside the public \texttt{mace-mp-mh-0} foundation checkpoint and the OMAT-PBE-head fine-tuning setup. In this sense, the data-preparation, training, and analysis pipeline described here is publicly reproducible from repository artefacts and public upstream APIs.

\subsection*{S3. Autograd and folding recipes (as used in code)}

All derivatives are computed on the \emph{interaction} energy (atomic baselines are constant):

\begin{align}
\Omega\,\mathbf P &= -\frac{\partial \mathcal F}{\partial \mathbf E},
\quad
Z^*_{\kappa,ij} = \frac{\partial (\Omega P_i)}{\partial u_{\kappa j}},
\quad
\alpha_{ij} = \frac{\partial P_i}{\partial E_j}.
\end{align}

In PyTorch:
\begin{verbatim}
# Polarisation (per graph)
polar = - torch.autograd.grad(
    outputs=[inter_e], 
    inputs=[E], 
    grad_outputs=[torch.ones_like(inter_e)],
    retain_graph=True, 
    create_graph=True
)[0]

# BECs (stack component-wise)
becs = []
for d in range(3):
    comp = polar[:, d]
    g = torch.autograd.grad(
        outputs=[comp], 
        inputs=[positions],
        grad_outputs=[torch.ones_like(comp)],
        retain_graph=True, create_graph=True
    )[0]
    becs.append(g)
# [n_atoms, 3, 3]
becs = torch.stack(becs, dim=1)   

# Polarisability
alphas = []
for d in range(3):
    comp = polar[:, d]
    g = torch.autograd.grad(
        outputs=[comp], 
        inputs=[E],
        grad_outputs=[torch.ones_like(comp)],
        retain_graph=True, 
        create_graph=True
    )[0]
    alphas.append(g)
# [n_graphs, 3, 3]
alpha = torch.stack(alphas, dim=1)  
\end{verbatim}
\par\addvspace{.75\baselineskip}
\paragraph*{Polarisation folding for general cells.}
Differences between reference and predicted polarisations are defined only up to the polarisation lattice. In the internal units used by the code, the lattice basis is built directly from the ASE-style cell matrix,
\[
B=
\begin{bmatrix}
\mathbf a_1^\mathsf{T}\\
\mathbf a_2^\mathsf{T}\\
\mathbf a_3^\mathsf{T}
\end{bmatrix},
\qquad
Q_{\mathrm{pol}} = B / |\Omega|,
\]
with lattice vectors stored as \emph{rows}. Given
\[
\Delta \mathbf P = \mathbf P_{\mathrm{pred}}-\mathbf P_{\mathrm{ref}},
\]
we convert to polarisation-lattice coordinates by solving
\[
Q_{\mathrm{pol}}^\mathsf{T}\mathbf c^\mathsf{T}=\Delta\mathbf P^\mathsf{T},
\]
then solve the exact closest-vector problem on the polarisation lattice,
\[
\mathbf n^\star=\arg\min_{\mathbf n\in\mathbb Z^3}\left\|(\mathbf c-\mathbf n)Q_{\mathrm{pol}}\right\|_2,
\qquad
\mathbf c_{\mathrm{fold}}=\mathbf c-\mathbf n^\star,
\]
and reconstruct
\[
\Delta\mathbf P_{\mathrm{fold}}=\mathbf c_{\mathrm{fold}}Q_{\mathrm{pol}}.
\]
This row-vector formulation matches the actual tensor layout used in the code and remains exact for any full-rank simulation cell, including non-orthogonal cells. The corresponding implementation is:
\begin{verbatim}
B = cell.view(-1, 3, 3)
vol = torch.linalg.det(B).abs()
    .clamp_min(1e-30).view(-1, 1, 1)
Qpol = B / vol
dP = pred_polarisation.view(-1, 3) 
    - ref_polarisation.view(-1, 3)
c = torch.linalg.solve(
    Qpol.transpose(-2, -1),
    dP.unsqueeze(-1),
).squeeze(-1)
integer_shift = _closest_lattice_shift(c, Qpol)
c_fold = c - integer_shift
dP_fold = dP - torch.einsum("bi,bij->bj", 
    integer_shift, Qpol)
\end{verbatim}
In the present implementation, the default polarisation loss is then evaluated on the folded fractional coordinates using the normalised, cell-aware metric
\begin{equation}
\begin{aligned}
    d_{\mathrm{pol}}^2(\mathbf c_{\mathrm{fold}};Q_{\mathrm{pol}})
    &=
    \mathbf c_{\mathrm{fold}}\,M(Q_{\mathrm{pol}})\,\mathbf c_{\mathrm{fold}}^\mathsf{T},
    \\
    M(Q)&=\frac{3QQ^\mathsf{T}}{\mathrm{tr}(QQ^\mathsf{T})}.
\end{aligned}    
\end{equation}
This rescales the metric so that anisotropic or skewed cells are compared in a geometry-aware way while remaining dimensionless in folded polarisation-quantum coordinates. The same closest-vector folding recipe is also used in the ferroelectric analysis scripts that generate the folded parity and branch-distribution figures in the manuscript, ensuring that training-time branch-invariant supervision and post-processing-time folded parity metrics are consistent.

\subsection*{S4. Finite-field MD and spectroscopy protocols}

\paragraph*{General MD settings.}
The production trajectories discussed in the main text were run in \texttt{LAMMPS}/ML-IAP after model-specific fixed-cell relaxation and a short equilibration stage. The quartz trajectories use fixed-cell NVT dynamics at 300~\si{K}, whereas the driven BaTiO$_3$ loops use athermal fixed-cell \texttt{nve} dynamics with viscous damping at 0~\si{K}. Unless stated otherwise, we use periodic boundary conditions and a timestep $\Delta t=2$\,fs. During time integration, the production exports record the total energy per atom, temperature, full stress tensor, lattice lengths/angles, and the (possibly time-dependent) uniform field $\mathbf E(t)$. In the current \texttt{LAMMPS}/ML-IAP workflow, these production loops evaluate only energies and forces inside the MD step; the saved trajectory is then converted and annotated afterwards with polarisation $\mathbf P(t)$, Born effective charges $Z^*(t)$, and polarisability $\boldsymbol\alpha(t)$ in a separate post-processing pass using a \texttt{MACE-Field} model.
\\

\paragraph*{Autocorrelations and spectra.}
Given $\Delta t$ and $N$ frames ($t_n=n\Delta t$), we form the normalised autocorrelations
\[
C_P(t)=\frac{\sum_i\langle P_i(t)P_i(0)\rangle}{\sum_i\mathrm{Var}[P_i]},\qquad
C_\alpha(t)=\frac{\sum_{ij}\langle \alpha_{ij}(t)\alpha_{ij}(0)\rangle}{\sum_i\mathrm{Var}[\alpha_{ii}]},
\]
obtain the one-sided spectra $\mathrm{Re}\,S_P(\omega)$ and $\mathrm{Re}\,S_\alpha(\omega)$ by rFFT (Hann window), and
compute
\[
\mathrm{IR}(\omega)\propto \omega^2\,\mathrm{Re}\,S_P(\omega),\qquad
\mathrm{Raman}(\omega)\propto \omega^2\,\mathrm{Re}\,S_\alpha(\omega),
\]
with Gaussian broadening $\sigma=20~\mathrm{cm}^{-1}$ for presentation. During analysis, $\mathbf P(t)$ is folded at each
step onto the nearest image of the polarisation lattice using the same general-cell recipe described above.
\\

\paragraph*{Dielectric constants from fluctuations.}
Directional components are obtained from time averages and fluctuations
\[
\varepsilon_{\infty,i}=1+\frac{4\pi}{\varepsilon_0}\,\langle \alpha_{ii}\rangle,\qquad
\varepsilon_{0,i}=\varepsilon_{\infty,i}+\frac{4\pi}{\varepsilon_0}\,\frac{\Omega\,\mathrm{Var}[P_i]}{k_BT},
\]
and we plot $\bar\varepsilon_\infty=\tfrac13\sum_i\varepsilon_{\infty,i}$ and
$\bar\varepsilon_0=\tfrac13\sum_i\varepsilon_{0,i}$ as horizontal guides in $\mathrm{Re}\,\varepsilon(\omega)$.
\\

\paragraph*{Frequency-dependent $\varepsilon(\omega)$.}
From $S_{P,i}(\omega)$ we construct
\[
\varepsilon_i(\omega)\approx 1+\big(\varepsilon_{0,i}-1\big)
\left[1-i\,\omega\,\frac{S_{P,i}(\omega)}{\mathrm{Var}[P_i]}\right],
\]
and report the Cartesian averages of $\mathrm{Re}\,\varepsilon(\omega)$ and the loss $-\mathrm{Im}\,\varepsilon(\omega)$.

\subsubsection*{S4.1 BaTiO\texorpdfstring{$_3$}{3}: finite-field hysteresis}

\textbf{Structure and calculator.}
For \ce{BaTiO3}, we start from the DFT-relaxed 135-atom supercell from Allegro-pol~\cite{allegro-pol-2025} for the production hysteresis runs discussed here. The directly trained specialist uses the field-aware \texttt{MACECalculator} with \texttt{MACEField}
(\texttt{MACE-Field-BaTiO3.model}; double precision); the OMAT-based foundation comparison uses the corresponding \texttt{MACEField-omat-dielectric.model} export with the same wrapper.

\textbf{Thermostat and timestep.}
After fixed-cell relaxation, the structure is equilibrated for 5000 steps (10~ps) at $T=0$~\si{K} with \texttt{nve} dynamics plus viscous damping. The driven production loop continues at the same timestep, $\Delta t=2$\,fs, with the same athermal dynamics. For quasi-static reference points, we cool from $300$~\si{K} down to $0$~\si{K} in steps of $50$~\si{K} whilst performing ionic relaxations at fixed field values on a grid along the polar axis.

\textbf{Field protocol (dynamic loop).}
We apply a cosine field along $\hat{\mathbf z}$ throughout the production run:
\begin{align*}
E_z(t) &= E_0 \cos\!\left(\frac{2\pi t}{T_\mathrm{per}}\right), \\
E_0&=0.3636~\mathrm{V}\,\text{\AA}^{-1}\;(\approx 36.4~\si{\mega\volt\per\centi\metre}), \\
T_\mathrm{per}&=100000~\text{steps}\;(200~\mathrm{ps}\text{ at }\Delta t=2~\mathrm{fs})
\end{align*}
The loop is sampled by logging $\{E_z(t),P_{x,y,z}(t)\}$ each MD step and
plotting $P$ vs $E$ to extract coercive fields and remanent polarisations. Because the simulation cell is small, periodic, and driven by a homogeneous field, the resulting loop probes intrinsic homogeneous switching rather than domain nucleation and growth. The extracted coercive fields are therefore expected to be much larger than the experimental values, which are governed by defects, domain walls, and nucleation kinetics.

\begin{figure*}[t]
\centering
\includegraphics[width=\textwidth]{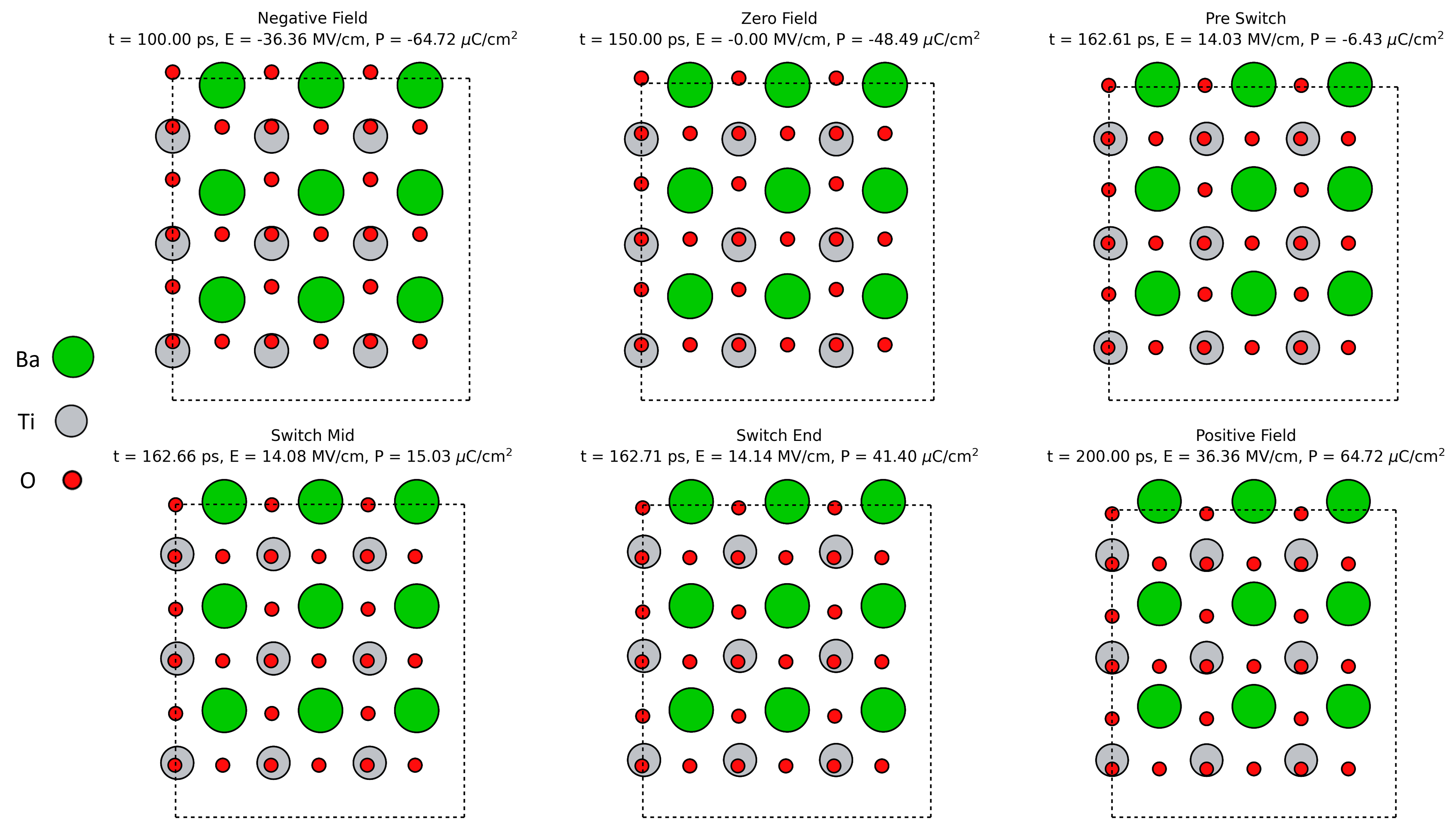}
\caption{\textbf{Representative atomistic snapshots from the \texttt{MACE-Field-MH-0} OMAT-based foundation model \ce{BaTiO3} hysteresis trajectory.} The six frames correspond to the circular markers shown on the main-text switching-trace panel (Figure 11): negative field, zero field, pre-switch, switch-mid, switch-end, and positive field. Titles report the time, instantaneous field, and reconstructed polarisation for each frame. Together, they visualise a coherent reversal of the tetragonal off-centring during intrinsic homogeneous switching in the small periodic cell.}
\label{fig:bto_snapshots_si}
\end{figure*}

\textbf{Outputs.}
The run writes a per-step LAMMPS thermo trace and a per-step atomic trajectory, which are then converted to \texttt{extxyz} and annotated with $\mathbf P$, $Z^*$, and $\boldsymbol\alpha$. The manuscript figures use these annotated trajectories together with a diagnostic plot of field, polarisation, and Ti off-centring over one full cycle.
The six representative structures shown in Fig.~\ref{fig:bto_snapshots_si} are sampled from the same annotated trajectory, and their times coincide with the circular markers overlaid on the main-text switching trace.

\subsubsection*{S4.2 $\alpha$-SiO\texorpdfstring{$_2$}{2}: IR/Raman and $\varepsilon(\omega)$ from MLMD}

\textbf{Structure and calculator.}
$\alpha$-quartz was retrieved as \texttt{mp-7000} and expanded to a 72-atom periodic supercell for the production runs discussed here. We used the
\texttt{MACE-field-SiO2.model} and \texttt{MACEField} for the directly trained specialist, and the corresponding \texttt{MACEField-omat-dielectric.model} export for the foundation comparison.

\textbf{Production MD.}
After fixed-cell relaxation, we equilibrate for 5000 steps (10~ps) in fixed-cell NVT at $T=300$~\si{K}, then run a 100000-step production trajectory (200~ps) at the same temperature with $\Delta t=2$\,fs and zero external field. The LAMMPS production loop writes thermo data and atomic positions every step; $\mathbf P(t)$, $Z^*(t)$, and all nine $\alpha_{ij}(t)$ components are reconstructed afterwards in the annotation pass.

\begin{figure*}[t]
\centering
\subfigure[Representative 300 K snapshots from the \texttt{MACE-Field-MH-0} OMAT-based foundation model $\alpha$-\ce{SiO2} trajectory at 0, 40, 80, 120, 160, and 200 ps.]{%
\includegraphics[width=.60\linewidth]{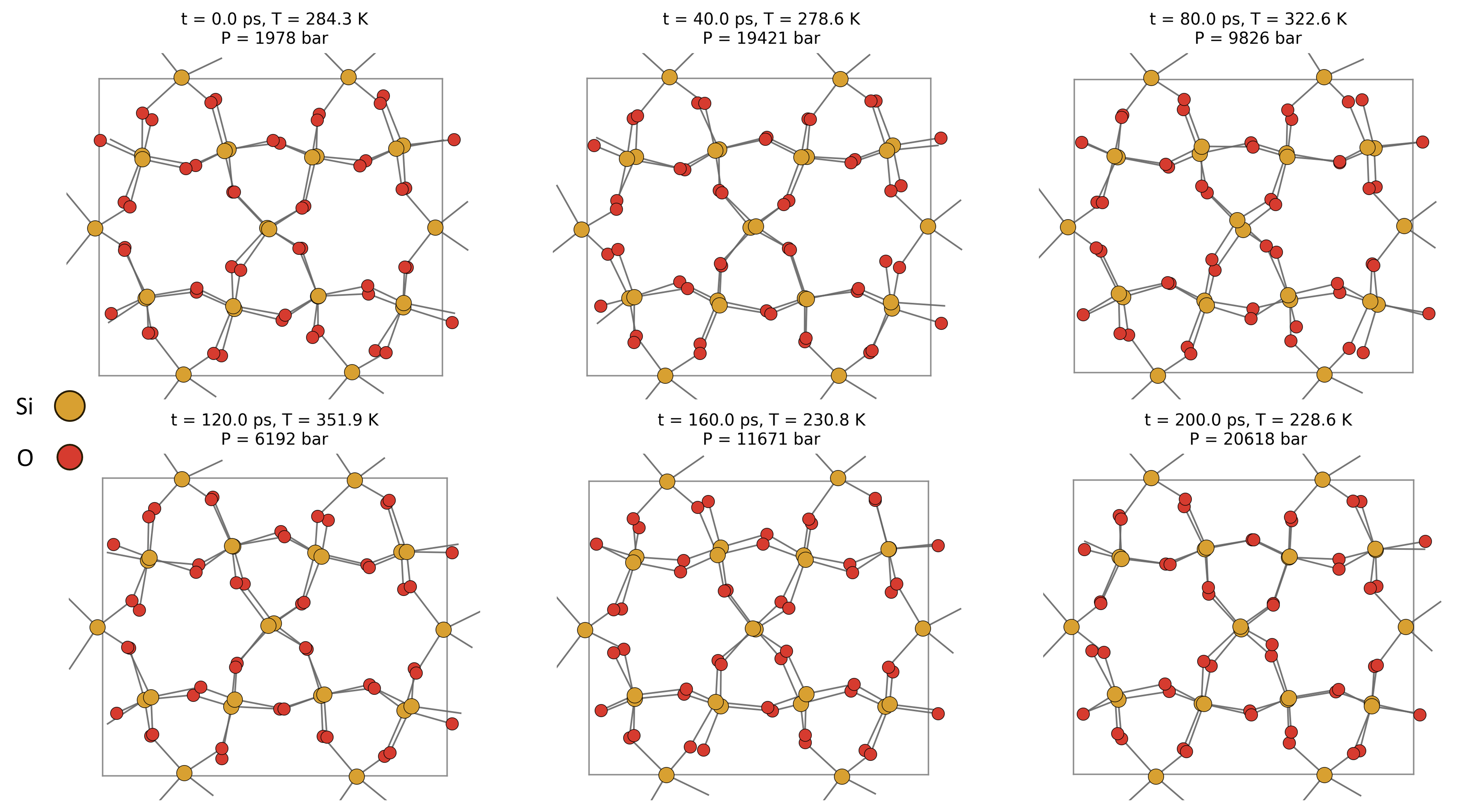}}
\hfill
\subfigure[Thermal diagnostics for the same trajectory. The marker positions correspond to the six frames shown at left.]{%
\includegraphics[width=.38\linewidth]{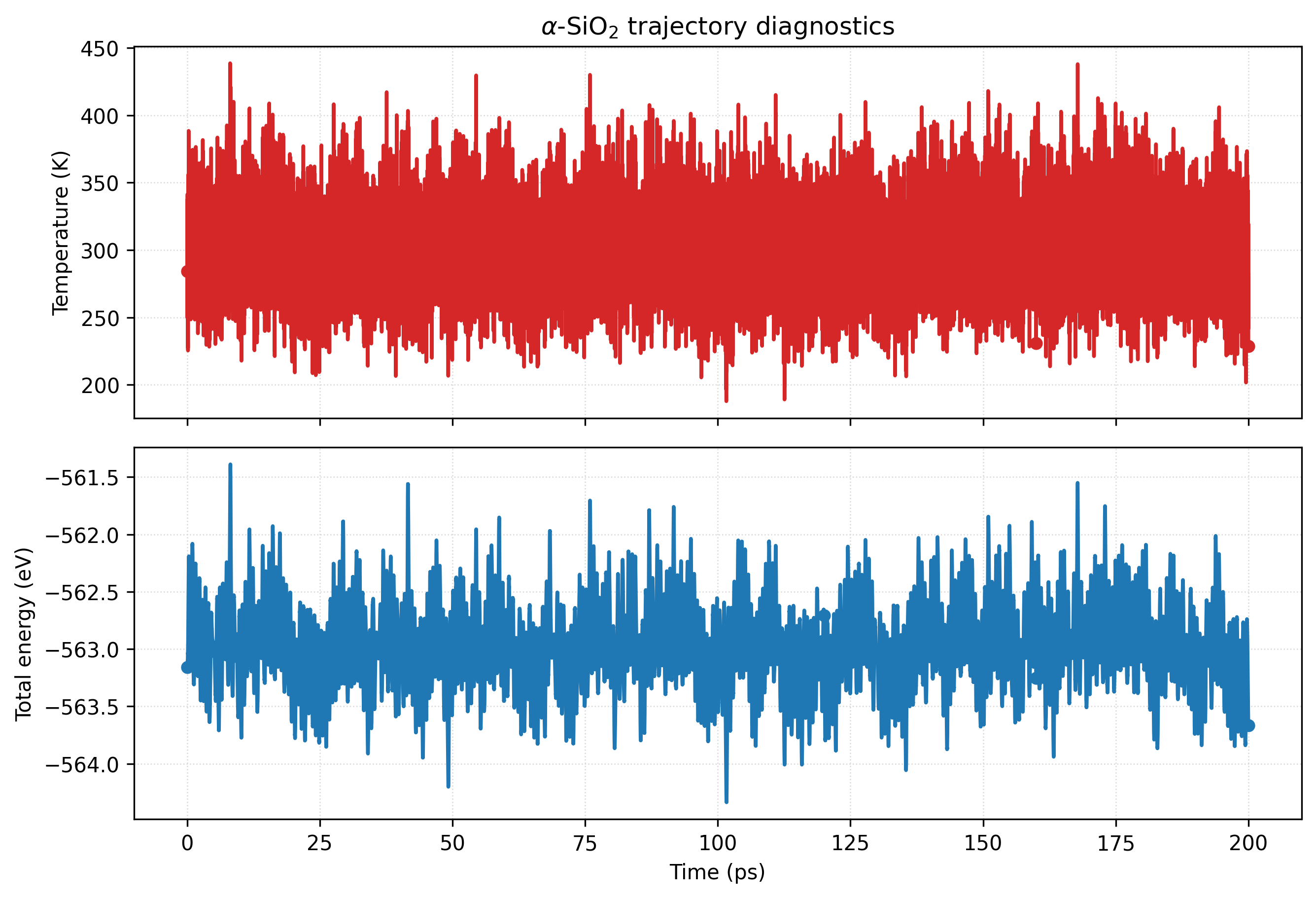}}
\caption{\textbf{Representative structures and trajectory diagnostics for the \texttt{MACE-Field-MH-0} OMAT-based foundation model $\alpha$-\ce{SiO2} MLMD run.} The snapshot panel shows six evenly spaced frames from the 200 ps production trajectory used for the IR/Raman/$\varepsilon(\omega)$ analysis in the main text. The accompanying thermo trace shows that the trajectory remains in the intended finite-temperature regime without obvious long-time drift, supporting the interpretation of the time-domain spectroscopic observables.}
\label{fig:sio2_snapshots_si}
\end{figure*}

\textbf{Spectral analysis.}
From the annotated trajectory, we compute:
(i) IR spectrum from the normalised $\dot{\mathbf P}$--$\dot{\mathbf P}$ (equivalently $P$--$P$) autocorrelation,
(ii) Raman spectrum from the $\boldsymbol\alpha$--$\boldsymbol\alpha$ autocorrelation,
(iii) $\varepsilon_\infty$ and $\varepsilon_0$ from $\langle\alpha_{ii}\rangle$ and $\mathrm{Var}[P_i]$,
and (iv) the frequency-dependent dielectric function $\varepsilon(\omega)$ using the expression above.
For presentation we apply Gaussian broadening ($\sigma=20~\mathrm{cm}^{-1}$) and plot IR, Raman, $\mathrm{Re}\,\varepsilon$,
and the loss $-\mathrm{Im}\,\varepsilon$ on a shared $\omega$ axis. Separate small-field relax-and-analyse calculations on the same trajectories reproduce the same qualitative ordering as the time-domain analysis: the \texttt{MACE-Field-MH-0} OMAT-based foundation model gives systematically larger $\varepsilon_\infty$ and $\varepsilon_0$ than the directly trained specialist. Representative frames from the trajectory obtained from \texttt{MACE-Field-MH-0}, together with temperature and total-energy traces, are shown in Fig.~\ref{fig:sio2_snapshots_si}.

\subsection*{S5. Parities and training curves}

\begin{figure*}[t]
\centering
\subfigure[OMAT-PBE replay head]{%
\includegraphics[width=.8\linewidth]{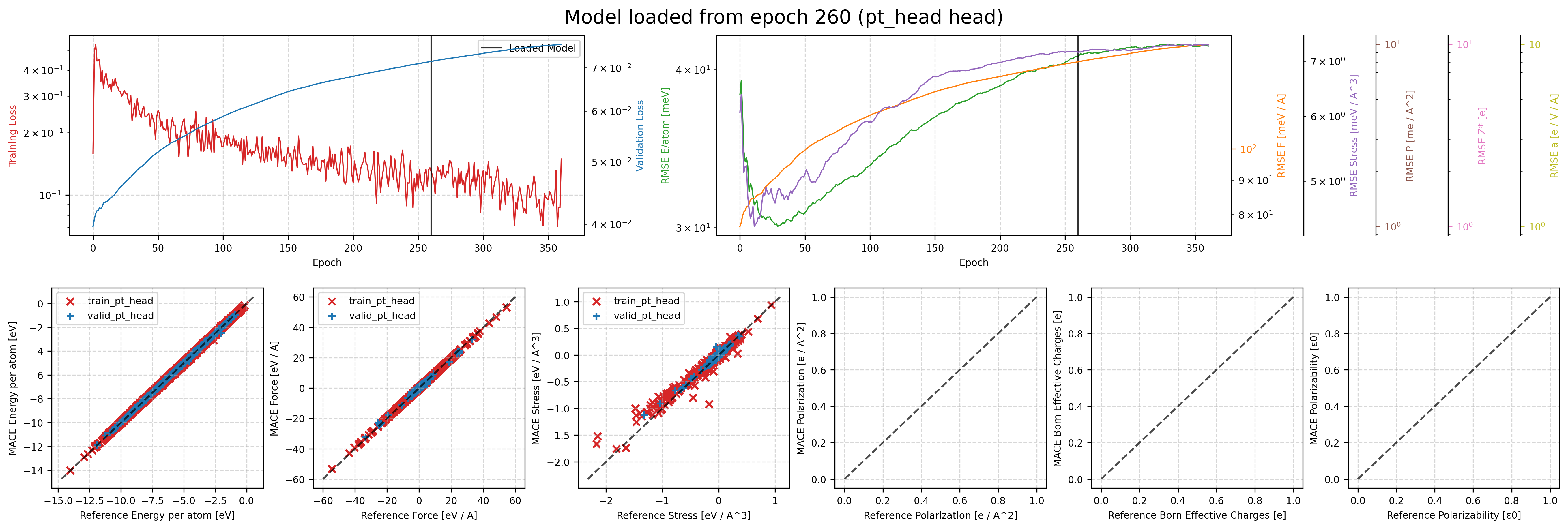}}
\subfigure[MP-Dielectric head]{%
\includegraphics[width=.8\linewidth]{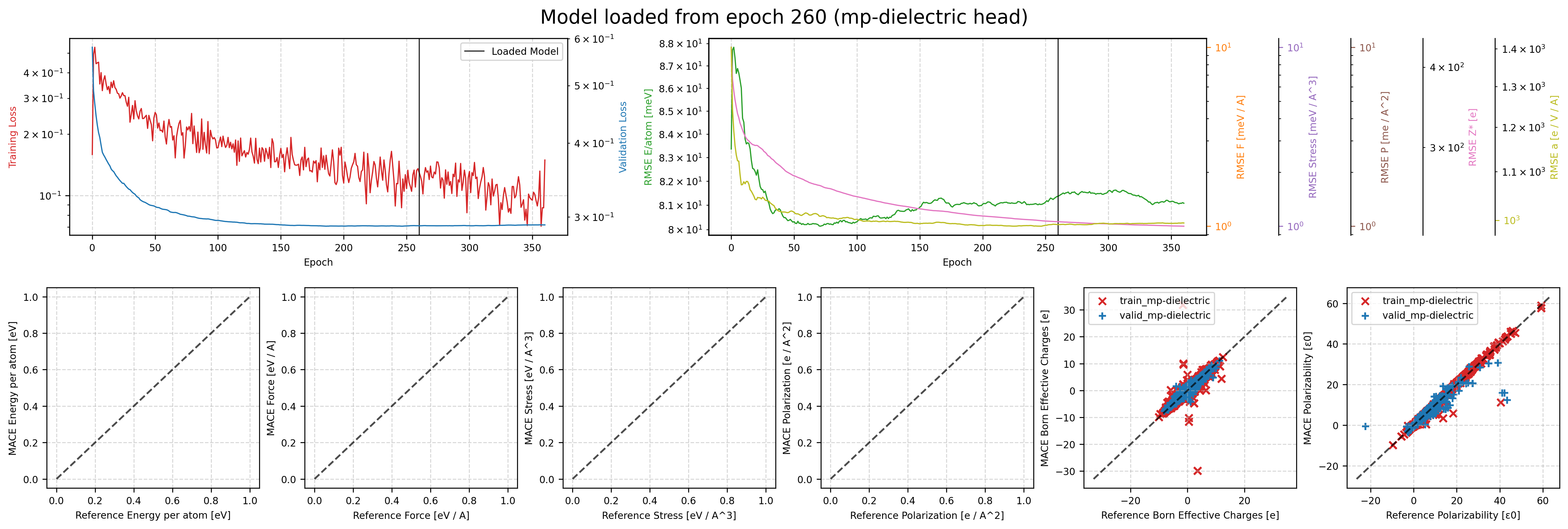}}
\subfigure[MP-Ferroelectric head]{%
\includegraphics[width=.8\linewidth]{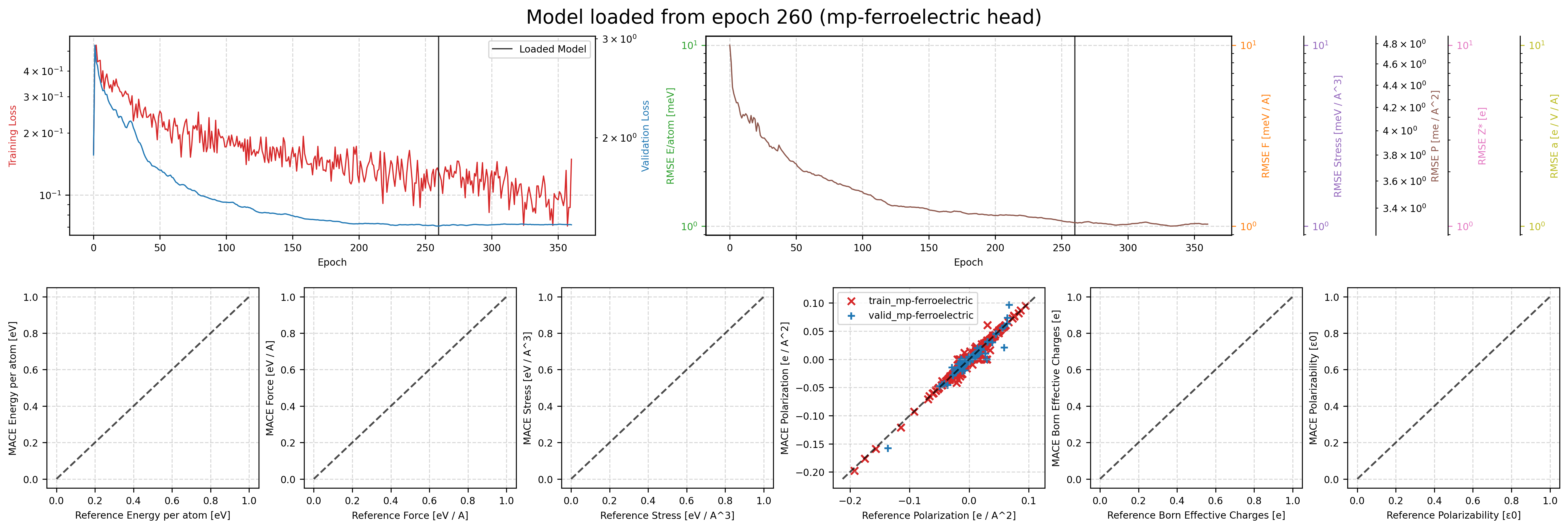}}
\caption{\textbf{Head-resolved training dynamics and parities for the OMAT-based \texttt{MACE-Field-MH-0} fine-tune.} Each subfigure corresponds to one head of the multihead optimisation: OMAT-PBE replay ($E,\mathbf F,\boldsymbol\sigma$), MP-Dielectric ($Z^*,\boldsymbol\alpha$), and MP-Ferroelectric ($\mathbf P$ together with the underlying $E,\mathbf F$ targets present in that dataset). The top row in each panel shows total training/validation loss and the per-target RMSE traces versus epoch; the vertical black line marks the selected checkpoint. The lower parity plots show only the observables defined for that head, so unused targets appear as blank axes.}
\label{fig:foundation-curves}
\end{figure*}

\begin{figure*}[t]
\centering
\subfigure[Ferroelectric (cross-chemistry) model]{%
\includegraphics[width=.8\linewidth]{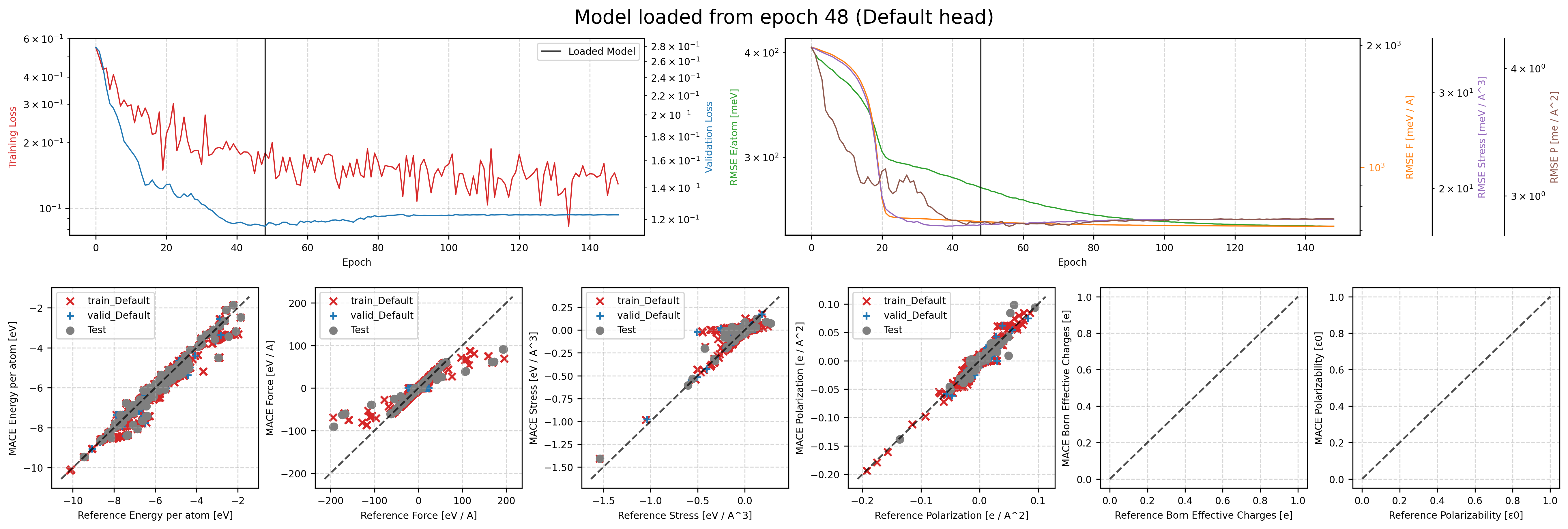}}
\subfigure[BaTiO$_3$ model]{%
\includegraphics[width=.8\linewidth]{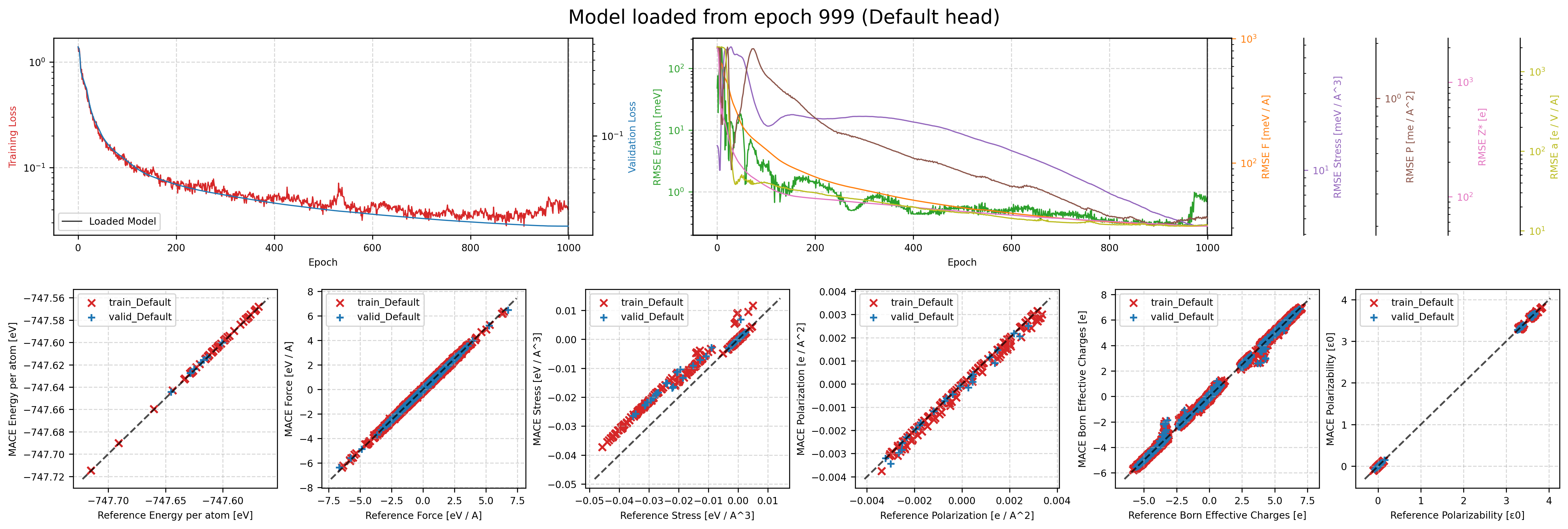}}
\subfigure[$\alpha$-SiO$_2$ model]{%
\includegraphics[width=.8\linewidth]{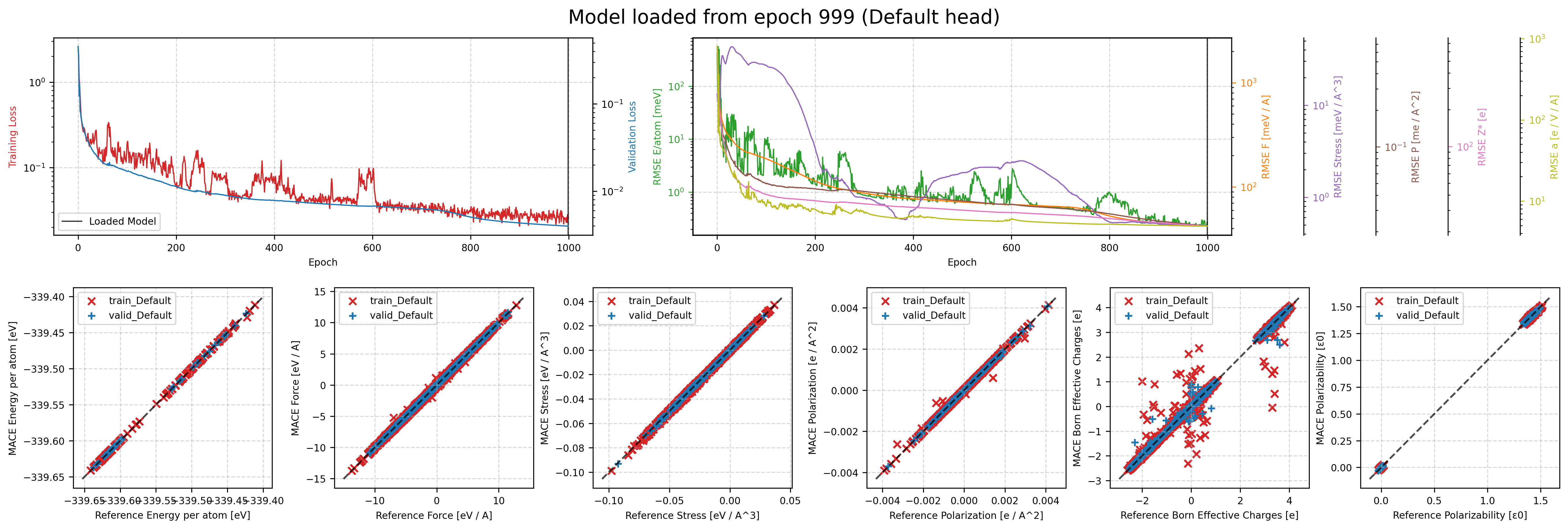}}
\caption{\textbf{Training dynamics and parities for the directly trained models.} \emph{Top of each subfigure:} total training/validation loss and per-target RMSE versus epoch; the vertical black line marks the selected checkpoint. \emph{Bottom:} parity plots for the observables present in each training setup (train/validation/test splits as indicated in the legends; dashed line is $y{=}x$). The cross-chemistry ferroelectric model is trained on $E,\mathbf F,\mathbf P$ along MP-Ferroelectric distortion paths, so its unused $Z^*$ and $\boldsymbol\alpha$ panels are blank. The \ce{BaTiO3} and $\alpha$-\ce{SiO2} specialists use the full $(E,\mathbf F,\boldsymbol\sigma,\mathbf P,Z^*,\boldsymbol\alpha)$ supervision required for the finite-field MLMD validation in the main text.}
\label{fig:parities_and_curves}
\end{figure*}

Figures~\ref{fig:foundation-curves} and~\ref{fig:parities_and_curves} compile the learning curves and parity plots for the models used throughout the paper. Figure~\ref{fig:foundation-curves} resolves the three heads of the OMAT-based \texttt{MACE-Field-MH-0} fine-tune separately: OMAT-PBE replay, MP-Dielectric, and MP-Ferroelectric. Figure~\ref{fig:parities_and_curves} then shows the directly trained cross-chemistry ferroelectric model together with the directly trained single-material \ce{BaTiO3} and $\alpha$-\ce{SiO2} specialists.
For each panel, the top row shows the training/validation loss versus epoch together with the per-target RMSE traces (energy, forces, stress, polarisation $\mathbf P$, Born effective charges $Z^*$, and polarisability $\boldsymbol\alpha$ where present; units follow the axes). The vertical black line marks the checkpoint used elsewhere in the manuscript. The bottom row displays predicted versus reference values on the train/validation (and, where applicable, test) splits with the $y{=}x$ guide. Tight clustering around the diagonal indicates low bias and good calibration, while empty parity panels indicate targets that are not present in that particular head or training setup.

\textbf{Fine-tuned (cross-chemistry) foundation model.} This model is fine-tuned using multiple heads starting from the multihead \texttt{mace-mp-mh-0} foundation model with its OMAT-PBE head. Here, the loss targets are energies, forces and stresses from a 10000-structure OMAT-PBE replay subset associated with the \texttt{mace-mp-mh-0/1} family, Born effective charges and polarisabilities from MP-Dielectric and polarisations from MP-Ferroelectric.

\textbf{Ferroelectric (cross-chemistry) model.} This model is trained on distortion-path structures with supervision on $(E,\mathbf F,\mathbf P)$ only. Parities for these quantities are tight across materials; $Z^*$ and $\boldsymbol\alpha$ (not included in the loss) show larger scatter, as expected, but remain physically reasonable due to derivative consistency of the learned enthalpy.

\textbf{BaTiO\textsubscript{3}.} Trained on \emph{ab initio} MD frames with supervision on $(E,\mathbf F,\boldsymbol\sigma,\mathbf P, Z^*,\boldsymbol\alpha)$, the model converges smoothly and attains near-linear parities for all observables, enabling the finite-field hysteresis simulation in the main text.

\textbf{$\alpha$-SiO\textsubscript{2}.} Trained analogously on $\alpha$-quartz trajectories, the model shows similarly steady convergence and diagonal parities for $\mathbf P$, $Z^*$ and $\boldsymbol\alpha$, supporting the IR/Raman and $\varepsilon(\omega)$ comparisons discussed in the main text.

\bibliography{bibliography}